\documentclass[aps,prb,times,twocolumn,amsmath,amssymb,
superscriptaddress,floatfix,showpacs]{revtex4}

\usepackage{color} 
\usepackage{graphicx}
\usepackage{subfigure}
\usepackage{bbold} 
\usepackage{dsfont}
\usepackage{multirow}
\usepackage[section]{placeins}

\DeclareMathOperator{\Tr}{Tr}

\begin{document}


\title{Living on the edge : ground-state selection in ``quantum spin-ice'' pyrochlores}


\author{Han Yan}
%
\affiliation{Okinawa Institute of Science and Technology, Onna-son,
Okinawa 904-0412, Japan}
\affiliation{Clarendon Laboratory, University of Oxford, Parks Rd.,
Oxford OX1 3PU, UK} 

\author{Owen Benton}
%
\affiliation{Okinawa Institute of Science and Technology, Onna-son,
Okinawa 904-0412, Japan} 
\affiliation{H.\ H.\ Wills Physics Laboratory,
University of Bristol,  Tyndall Av, Bristol BS8--1TL, UK}

\author{Ludovic Jaubert}
%
\affiliation{Okinawa Institute of Science and Technology, Onna-son, 
Okinawa 904-0412, Japan}
\affiliation{Rudolf Peierls Centre for Theoretical Physics, University
of Oxford, 1--6 Keeble Rd, Oxford OX1 3NP, UK} 

\author{Nic Shannon}
%
\affiliation{Okinawa Institute of Science and Technology, Onna-son,
Okinawa 904-0412, Japan}
\affiliation{Clarendon Laboratory, University of Oxford, Parks Rd.,
Oxford OX1 3PU, UK} 
\affiliation{H.\ H.\ Wills Physics Laboratory,
University of Bristol, Tyndall Av, Bristol BS8--1TL, UK}


\date{\today}


\begin{abstract}
The search for new quantum phases, especially in frustrated magnets, is central
to modern condensed matter physics.
One of the most promising places to look is in
rare-earth pyrochlore magnets with highly-anisotropic exchange interactions, 
materials closely related to the spin ices Ho$_2$Ti$_2$O$_7$ and Dy$_2$Ti$_2$O$_7$.
Here we establish a general theory of magnetic order in 
these materials.
We find that many of their most interesting properties can be traced back to the 
``accidental'' degeneracies where phases with different symmetry meet.
These include the ordered ground state selection by fluctuations 
in Er$_2$Ti$_2$O$_7$, the ``dimensional-reduction'' observed in
Yb$_2$Ti$_2$O$_7$, and the absence of magnetic order in Er$_2$Sn$_2$O$_7$.
\end{abstract}




\maketitle


Like high-energy physics, condensed matter physics 
is dominated by the idea of symmetry.
Any physical property which {\it cannot} be traced back to a broken 
symmetry is therefore of enormous fundamental interest. 
In this context, the spin liquid phases found in frustrated magnets 
are a rich source of inspiration~\cite{balents10}.
Perhaps the most widely studied examples are the  ``spin ice'' states in  
Ho$_{2}$Ti$_{2}$O$_{7}$ and Dy$_{2}$Ti$_{2}$O$_{7}$, classical 
spin-liquids famous for their magnetic monopole excitations~\cite{castelnovo12}.
And there is now good reason to believe that a {\it quantum} 
spin-liquid phase, in which the magnetic monopoles are elevated to the role 
of ``elementary'' particles, could exist in spin-ice like materials where quantum 
effects play a larger 
role~\mbox{\cite{hermele04,banerjee08,savary12-PRL108,shannon12,lee12,benton12,savary13}}.


\begin{figure}[ht!]
\centering\includegraphics[width=\columnwidth]{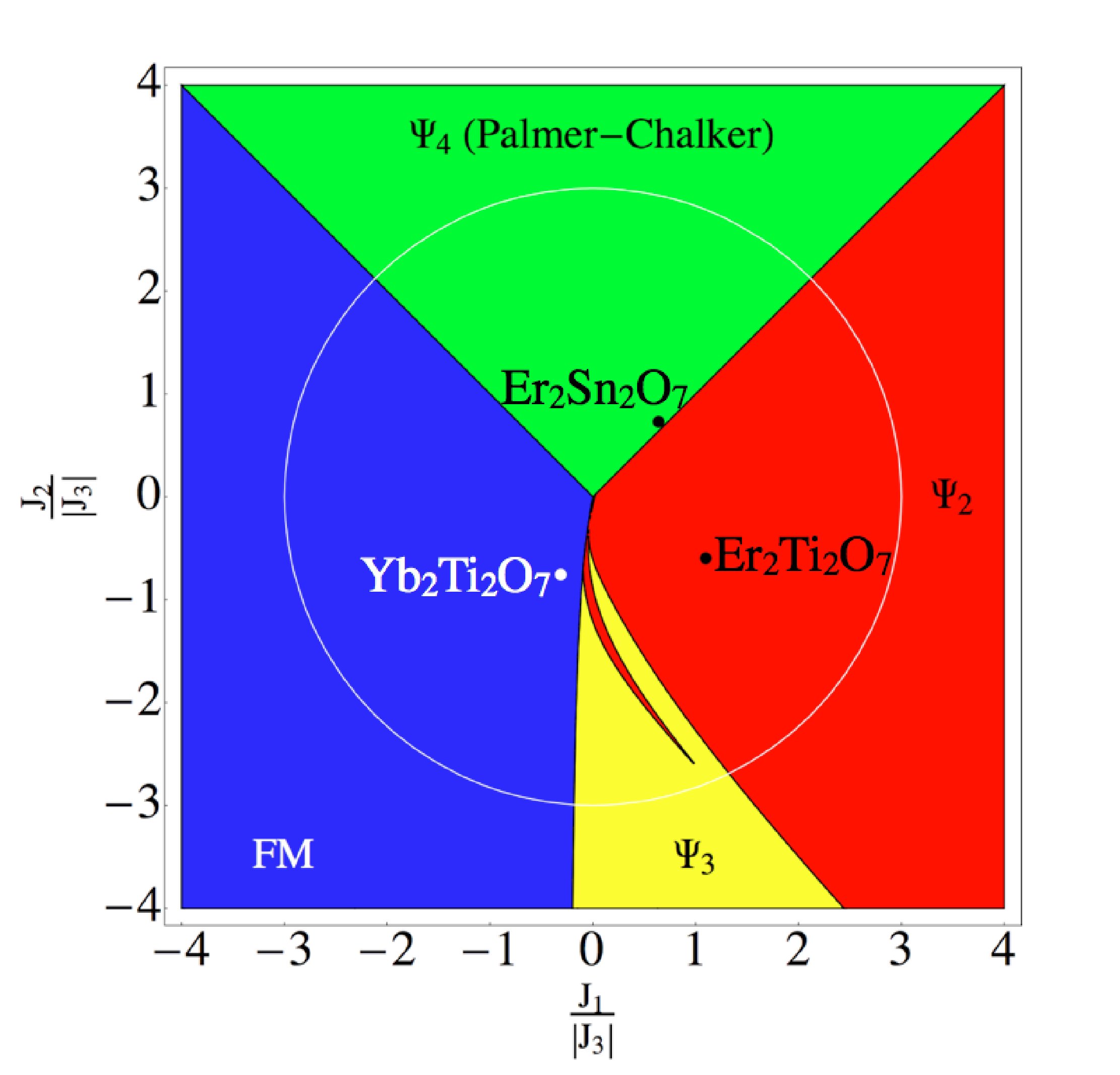} 
\caption{
Classical ground state phase diagram for a pyrochlore magnet with 
anisotropic exchange interactions.
The model considered is the most general nearest-neighbour 
exchange Hamiltonian on the pyrochlore lattice 
${\mathcal H}_{\sf ex}$~[Eq.~(\ref{eq:Hex1})],  
with ferromagnetic ``pseudo-dipolar'' interaction ($J_3 < 0$), 
and vanishing Dzyaloshinskii-Moriya interactions ($J_4 = 0$).
There are four distinct ordered phases, illustrated in the insets to
Fig.~\ref{fig:finite-temperature-phases}.
Points correspond to known parameters for 
Yb$_2$Ti$_2$O$_7$ [\onlinecite{ross11-PRX1}],
Er$_2$Ti$_2$O$_7$ [\onlinecite{savary12-PRL109}],
and Er$_2$Sn$_2$O$_7$ [\onlinecite{guitteny13}],
setting $J_4 = 0$.
} 
\label{fig:classical-phase-diagram}
\end{figure}


Fortunately there are a wide range of materials in which to look for such a state.
The best candidates for a ``quantum spin ice'' are rare-earth pyrochlore 
oxides ${\sf R}_2$${\sf A}_2$O$_7$ in which the magnetic ions have a doublet 
ground state, and highly-anisotropic exchange interactions.
The physical properties of these materials depend on the choice of rare-earth
${\sf R}^{3+}$  and transition metal ${\sf A}^{4+}$, and are fabulously 
diverse~\cite{bloete69,gardner10}.
In addition to spin ices, this family includes a wide range of systems that order
magnetically, spin glasses and systems where local moments couple to itinerant
electrons.
And significantly, a number of materials, including 
Tb$_2$Ti$_2$O$_7$~[\onlinecite{gardner99,mirebeau02}] and 
Er$_2$Sn$_2$O$_7$~[\onlinecite{matsuhira02,lago05,shirai07,sarte11,guitteny13}] 
have never been seen to order at {\it any} temperature.


\begin{figure*}[htpb]
\centering
\hspace{3cm}
\centering
\includegraphics[width=18cm]{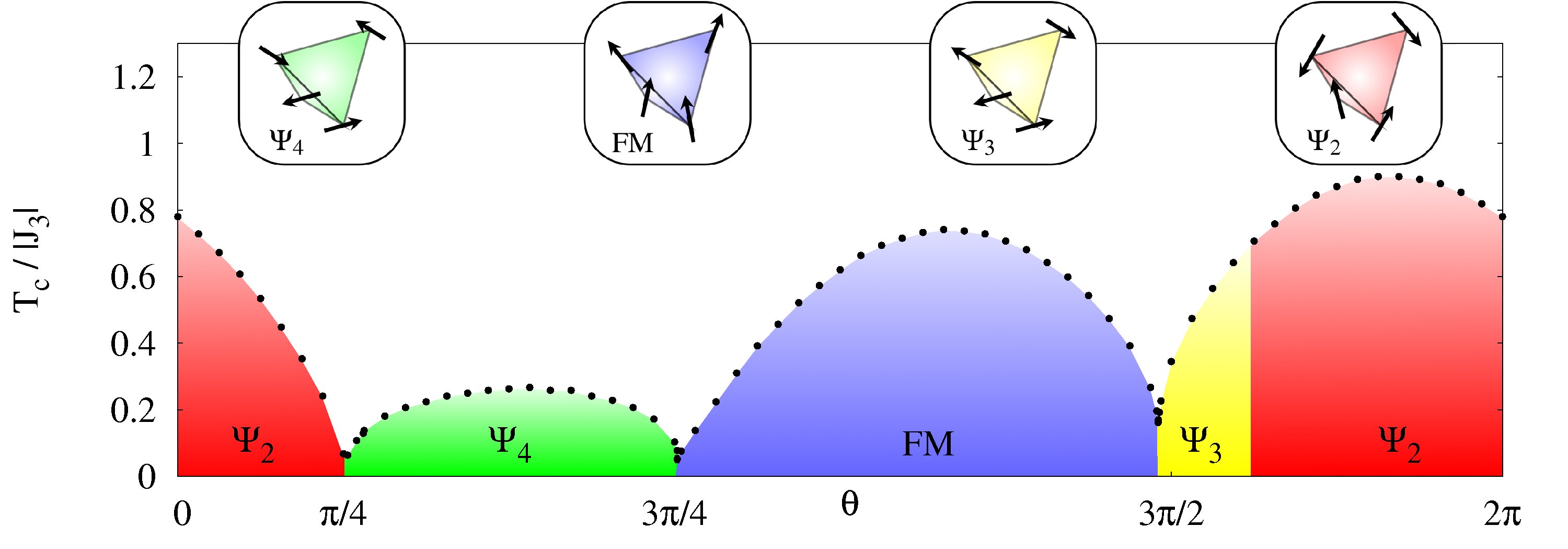}
\caption{
Finite-temperature phase diagram for pyrochlore magnet with 
anisotropic exchange interactions.
The model considered is ${\mathcal H}_{\sf Ex}$ [Eq.~(\ref{eq:Hex1})], 
with $J_1 = 3|J_3| \cos\theta$, $J_2 = 3|J_3| \sin\theta$, $J_3 < 0$, 
and $J_4 \equiv 0$, corresponding to the white circle in 
Fig.~\ref{fig:classical-phase-diagram}.  
Points show finite temperature phase transitions found from 
classical Monte Carlo simulation, as described in the supplementary 
information.  
The four ordered phases, 
Palmer-Chalker ($\Psi_4$), 
non-collinear ferromagnetic (FM), 
coplanar antiferromagnetic ($\Psi_3$) 
and non-coplanar antiferromagnetic ($\Psi_2$), 
are illustrated at the top of the figure.
Each of these phases is six-fold degenerate, with zero crystal momentum, 
and is completely specified by the spin configuration in a single tetrahedron.  
} 
\label{fig:finite-temperature-phases}
\end{figure*} 


In this article we single out two of the best-characterised examples of rare-earth
pyrochlore magnets, Er$_2$Ti$_2$O$_7$ and Yb$_2$Ti$_2$O$_7$, and explore how 
their properties fit into the ``bigger picture'' of magnetism on the pyrochlore lattice.
We find that both the ``order by disorder'' ground-state selection in 
Er$_2$Ti$_2$O$_7$~\cite{champion03}, and the ``dimensional reduction'' observed in 
Yb$_2$Ti$_2$O$_7$~\cite{ross09}, can be understood in terms of proximity to nearby 
zero-temperature phase transitions.
In the process, we establish a general phase diagram for magnetic order in pyrochlore 
magnets with anisotropic exchange interactions, and identify where these ordered states 
might give way to unconventionally ordered or spin liquid phases.
These results place a third material, Er$_2$Sn$_2$O$_7$~[\onlinecite{guitteny13}],  
tantalizingly close to a region of quantum disorder.


The physics driving the spin-ice state in Ho$_2$Ti$_2$O$_7$ and Dy$_2$Ti$_2$O$_7$ 
is predominantly classical.
In these materials, the magnetic ions have Ising moments of $\sim 10\mu_b$, which couple 
through dipolar interactions \cite{siddharthan99,den_hertog00}.
In contrast, the magnetic ions in Yb$_2$Ti$_2$O$_7$ and Er$_2$Ti$_2$O$_7$ 
have a doublet ground state with XY character, and relatively small effective 
moment~\cite{bloete69,cao09}.
In this case, quantum effects play a much larger role, and interactions between spins 
can be described by an 
anisotropic nearest-neighbour exchange
Hamiltonian~\cite{curnoe07,mcclarty09,ross11-PRX1,applegate12,zhitomirsky12,savary12-PRL109,oitmaa-arXiv}
\begin{eqnarray}
\mathcal{H}_{\sf ex} = \sum_{\langle ij \rangle}
J^{\mu\nu}_{ij}S^\mu_i S^\nu_j
\label{eq:Hex1}
\end{eqnarray}
where the sum on $\langle ij \rangle$ runs over the bonds of the
pyrochlore lattice.
The cubic symmetry of the pyrochlore lattice permits only four
independent parameters to enter the matrix $J^{\mu\nu}_{ij}$ \cite{curnoe07}, 
and for a bond directed along $(0,-1,-1)$ this is given by 
\begin{eqnarray}
{\bf J}_{01} 
  = \begin{pmatrix}
    J_{2} & J_{4} &J_{4}\\
   -J_{4 } & J_{1} &J_{3}\\
   -J_{4} & J_{3} &J_{1}
   \end{pmatrix}
\label{eq:J01}
\end{eqnarray}
with the corresponding matrix for all other bonds 
found by applying lattice symmetry operations.  


\begin{figure*}[htpb] 
\includegraphics[width=18cm]{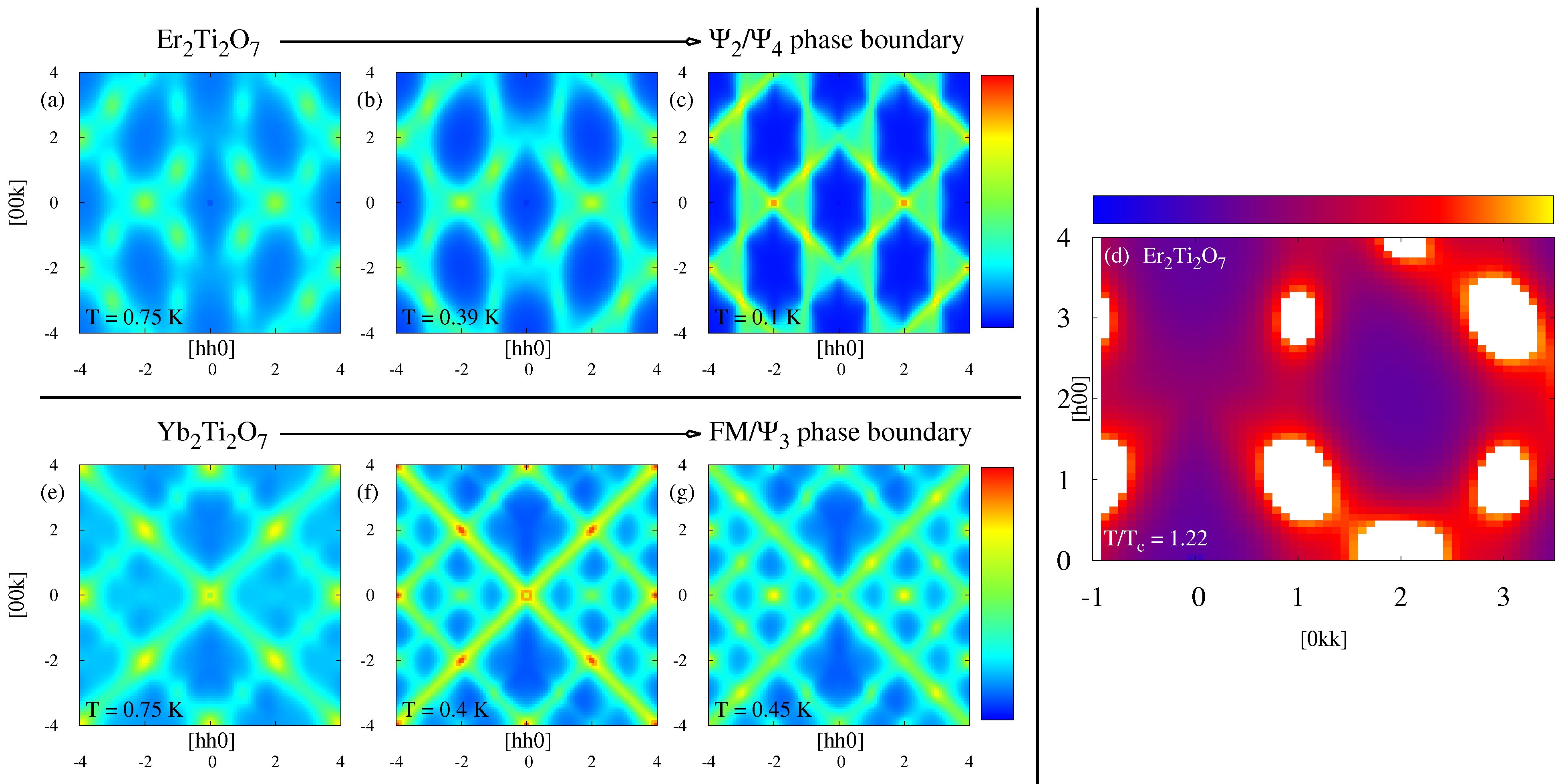}
\caption{
Correlations in the high-temperature paramagnetic phase, as revealed by the quasi-elastic 
structure factor $S({\bf q})$.
\mbox{(a-c)} results for parameters interpolating from (a) Er$_2$Ti$_2$O$_7$ [\onlinecite{savary12-PRL109}]
to (c) the boundary of the Palmer-Chalker phase ($\Psi_4$).
The diffuse scattering characteristic of the $\Psi_2$ phase
evolve into sharp features reminiscent of pinch points
when bordering the $\Psi_4$ phase.
(d) detail of S(q) for parameters appropriate to Er$_2$Ti$_2$O$_7$, 
plotted with a colour scale chosen 
to match Fig. 14 of [\onlinecite{dalmas12}].
Results are taken from classical Monte Carlo simulations carried out 
for 
(a) $J_2=-0.06$~\textrm{mK}, $ T=750$~\textrm{mK}; 
(b) $J_2=0.06$~\textrm{mK}, $ T=390$~\textrm{mK}; 
(c) $J_2=0.11$~\textrm{mK}, $ T=100$~\textrm{mK}.  
(d) $J_2=-0.06$~\textrm{mK}, $ T=616$~\textrm{mK}. 
%
In all cases, 
$J_1=-0.11$~\textrm{mK}, 
$J_3=-0.1$~\textrm{mK},  
$J_4\equiv 0$, 
and $S({\bf q})$ has been calculated using g-tensor appropriate 
to Er$_2$Ti$_2$O$_7$~[\onlinecite{savary12-PRL109}].
\mbox{(e-g)} results for parameters interpolating from   
Yb$_2$Ti$_2$O$_7$ [cf. Ref.~(\onlinecite{ross11-PRX1})],
to the border of the $\Psi_3$ phase.
The rods of scattering along $[111]$ directions, interpreted    
as evidence of dimensional reduction in Yb$_2$Ti$_2$O$_7$ 
[\onlinecite{ross11-PRB84}], evolve into weakly-dispersing, 
low-energy excitations in the neighbouring $\Psi_3$ phase.
Results are taken from classical Monte Carlo simulations of 
${\mathcal H}_{\sf ex}$ [Eq.~(\ref{eq:Hex1})] for 
(e) $J_1=-0.09$~\textrm{meV}, $ T=750$~\textrm{mK}; 
(f) $J_1=-0.04$~\textrm{meV}, $ T=400$~\textrm{mK}; 
(g) $J_1=-0.0288$~\textrm{meV}, $ T=450$~\textrm{mK}.  
In all cases, 
$J_2=-0.22$~\textrm{meV}, 
$J_3=-0.29$~\textrm{meV}, 
$J_4\equiv 0$, 
and $S({\bf q})$ has been calculated using g-tensor appropriate 
to Yb$_2$Ti$_2$O$_7$~[\onlinecite{hodges01}].
}
\label{fig:Sq} 
\end{figure*}


This model supports an extremely rich variety of different ground states.  
For $J_1$=$-J_2$=$J_3$=$J_4$=$-1/3$, $\mathcal{H}_{\sf ex}$~[Eq.~(\ref{eq:Hex1})] 
favours the ``ice rules'' states found in Ho$_2$Ti$_2$O$_7$ and 
Dy$_2$Ti$_2$O$_7$~[\onlinecite{harris97,ramirez99,pomaranski13}].  
Quantum tunnelling between these states gives rise to a ``${\sf U}(1)$'' quantum spin liquid with 
photon-like excitations~\cite{hermele04,banerjee08,shannon12,benton12}, and  
at a mean-field level, this is the ground state of $\mathcal{H}_{\sf ex}$  
for a moderately wide range of ($J_1$, $J_2$, $J_3$, $J_4$)~[\onlinecite{savary12-PRL108,lee12,savary13}].  
Meanwhile, for  $J_1$=$J_2$=$J$, and $J_3$=$J_4$=$0$, 
$\mathcal{H}_{\sf ex}$ reduces to the Heisenberg model on a pyrochlore lattice, also studied 
as a quantum spin-liquid~\cite{canals98,conlon09}.  
Nonetheless, materials such as Er$_2$Ti$_2$O$_7$ --- which is extremely well-described by 
$\mathcal{H}_{\sf ex}$ \cite{savary12-PRL109,zhitomirsky12,oitmaa-arXiv} ---  {\it do} order 
magnetically~\cite{champion03}.


Since $\mathcal{H}_{\sf ex}$~[Eq.~(\ref{eq:Hex1})] does not, in general, 
possess {\it any} continuous spin-rotation symmetry, different ordered phases 
can be characterised by the lattice symmetries which they break.
Moreover, the model $\mathcal{H}_{\sf ex}$~[Eq.~(\ref{eq:Hex1})] 
has the remarkable property that it is {\it always} 
possible to find a classical ground state with ${\bf q}=0$, i.e. 
one in which the spin-configuration in a single four-site unit cell 
is repeated across the entire lattice (a proof of this statement is given 
in the supplementary information).  
The great richness of the problem stems from the fact that 
this does {\it not} preclude the existence of other, degenerate, ground 
states at finite ${\bf q}$, or of a continuous ground-state manifold 
at ${\bf q}=0$. 


In fact, where the different ground states of a single tetrahedron
are linked by a symmetry which leaves at least one spin unchanged, we find that it is {\it always} 
possible to construct alternative ground states with finite ${\bf q}$.
Such ``accidental'' degeneracies are common, and we will 
argue below that they drive not only the spin-liquid phases of  
$\mathcal{H}_{\sf ex}$~[Eq.~(\ref{eq:Hex1})], but also many 
of the interesting phenomena associated with ordered phases.


Considering first what happens for ${\bf q}=0$, we find that 
$\mathcal{H}_{\sf ex}$~[Eq.~(\ref{eq:Hex1})] 
supports four distinct classes of magnetically ordered 
ground state, transforming like the ${\sf A_2}$, 
${\sf E}$, ${\sf T_2}$, and ${\sf T_1}$ irreducible representations of the point group 
of a tetrahedron, ${\sf T_d}$.  
The Hamiltonian $\mathcal{H}_{\sf ex}$~[Eq.~(\ref{eq:Hex1})] can be 
expressed in terms of the associated order parameters ${\bf m}_\lambda$
\begin{eqnarray}
{\mathcal H}_{\sf ex}^{[{\sf T_d}]}
  &=&  \frac{1}{2}
    \left[ 
       a_{\sf A_2} \, m_{\sf A_2}^2 
       + a_{\sf E}\, {\bf m}^2_{\sf E} 
       + a_{\sf T_2}\, {\bf m}^2_{\sf T_2} 
       + a_{\sf T_{1, A}}\, {\bf m}^2_{\sf T_{1, A}} 
    \right.
    \nonumber\\
    && \quad \left. 
       + a_{\sf T_{1, B}}\, {\bf m}^2_{\sf T_{1, B}} 
       + a_{\sf T_{1, AB}}\, {\bf m}_{\sf T_{1, A}} \cdot {\bf m}_{\sf T_{1, B}}
    \right].
\label{eq:HTd1}
\end{eqnarray}
where the order parameters ${\bf m}_\lambda$
and coefficients 
\mbox{$a_\lambda 
= a_{\lambda, 1} J_1 
+ a_{\lambda, 2} J_2 
+ a_{\lambda, 3} J_3 
+ a_{\lambda, 4} J_4$}   
are defined in the supplementary information.
For classical spins this model must be solved subject to the constraints 
that each spin is separately normalised to ${\bf S}^2 = S^2$.
We note that it is also possible to derive ${\mathcal H}_{\sf ex}^{[{\sf T_d}]}$ 
as the ${\bf q} = 0$ limit of a classical field theory reminiscent of electromagnetism.
This approach will be developed further elsewhere.


Starting from ${\mathcal H}^{[{\sf T_d}]}_{\sf ex}$ [Eq.~(\ref{eq:HTd1})],  it is possible to find which 
form of 4-sublattice order has the lowest energy for {\it any} given set of parameters 
$(J_1,J_2,J_3,J_4)$. 
However 
Yb$_2$Ti$_2$O$_7$~[\onlinecite{ross11-PRX1}], 
Er$_2$Ti$_2$O$_7$~[\onlinecite{savary12-PRL109}]
and Er$_2$Sn$_2$O$_7$~[\onlinecite{guitteny13}] 
all have ferromagnetic pseudodipolar interaction \mbox{$J_3 < 0$}, 
and small Dzyaloshinskii-Moriya 
interaction \mbox{$J_4 \ll |J_1|, |J_2|,|J_3|$}.  
In what follows, we therefore concentrate on the most general form of ground 
state possible for $J_3 < 0$ and $J_4 \equiv 0$.
Our results are summarised in Fig.~\ref{fig:classical-phase-diagram}.


We find four distinct ordered phases, which we label following the 
conventions of [\onlinecite{poole07,kovalev93}]:
i) $\Psi_4$ --- a coplanar antiferromagnetic ``Palmer-Chalker''  [\onlinecite{palmer00}]
phase, transforming with ${\sf T}_2$; 
ii) FM --- a non-collinear phase with finite magnetisation, 
transforming with ${\sf T}_1$.  
This is the ground state for parameters appropriate to 
Yb$_2$Ti$_2$O$_7$~[\onlinecite{ross11-PRX1}] or 
Yb$_2$Sn$_2$O$_7$~[\onlinecite{yaouanc13}]; 
iii) $\Psi_3$ --- a coplanar antiferromagnetic phase, selected by 
fluctuations from a one-dimensional manifold of states transforming 
with ${\sf E}$.  
This state was amoung those enumerated by 
Bramwell, Gingras and Reimers \cite{bramwell94};
iv) $\Psi_2$ --- a non-coplanar antiferromagnetic phase, selected by fluctuations 
from the same one-dimensional manifold of states as $\Psi_3$. 
This phase is the known ground state in Er$_2$Ti$_2$O$_7$ [\onlinecite{champion03}], 
originally studied by Champion and Holdsworth in the context of a Heisenberg 
model with single-ion anisotropy~\cite{champion04}.
Each of these phases singles out a unique [100] axis, and is six-fold degenerate.
The $T \to 0$ phase boundary between the $\Psi_2$ and $\Psi_3$ phases in 
Fig.~\ref{fig:classical-phase-diagram} was determined using classical spin-wave theory; 
analytic expressions for all the other phase boundaries are given in the supplementary 
information.
Ground state spin configurations for each of these phases are shown in Fig.~\ref{fig:finite-temperature-phases}.


\begin{figure}[htpb] 
\includegraphics[width=0.9\columnwidth]{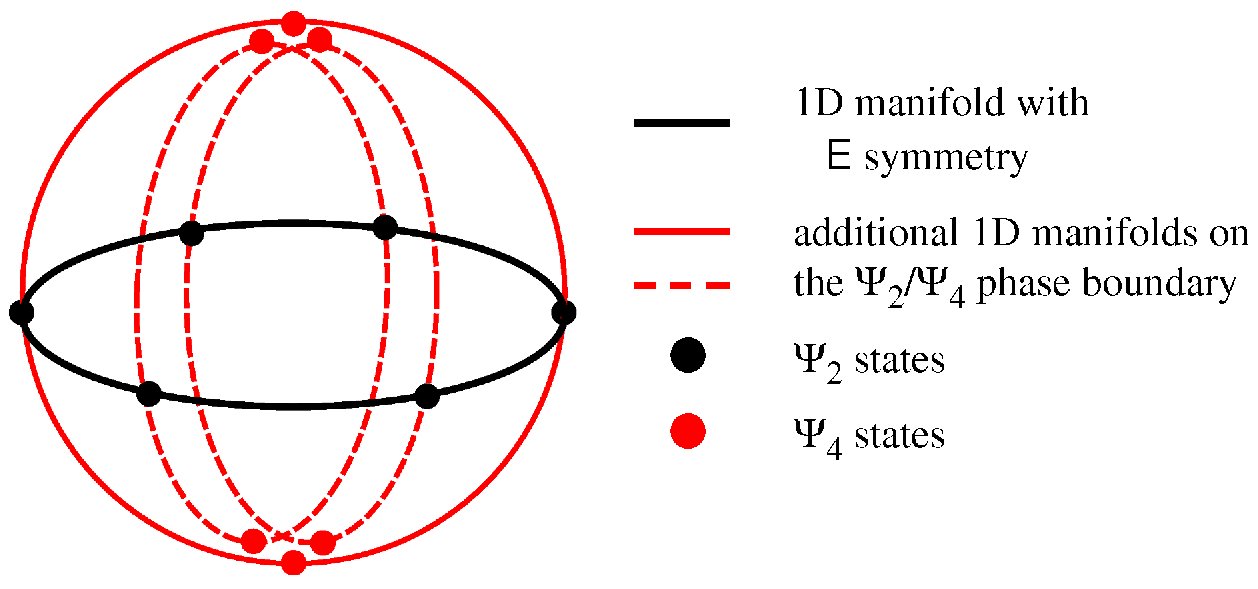}
\caption{
Structure of the ${\bf q}=0$ classical ground-state manifold at the boundary 
between the Palmer-Chalker phase ($\Psi_4$), and the one-dimensional manifold of 
states with ${\sf E}$ symmetry. 
The black circle denotes the manifold of ${\sf E}$--symmetry ground 
states, including the six $\Psi_2$ ground states (black dots).
These are connected to the six $\Psi_4$ ground states with ${\sf T_2}$ 
symmetry (red dots), by three, additional, one-dimensional manifolds 
(solid and dashed red lines).
} 
\label{fig:ground-state-manifold} 
\end{figure}


We have also explored the finite-temperature evolution of these phases 
using classical Monte Carlo simulation.
These results are summarised in Fig.~\ref{fig:finite-temperature-phases}.
The most striking feature of simulations is the complex evolution of 
correlations in the paramagnetic phase, highlighted in 
Fig.~\ref{fig:Sq}.


While these results provide an important context for experiment, 
they leave unanswered the important question of {\it why} fluctuations favour 
$\Psi_2$ for parameters appropriate to 
Er$_2$Ti$_2$O$_7$~\cite{savary12-PRL109,zhitomirsky12}, 
and $\Psi_3$ elsewhere.
Similarly, knowing that Yb$_2$Ti$_2$O$_7$  might be expected to order
ferromagnetically does little to explain the apparent dimensional-reduction 
seen in its paramagnetic phase~\cite{ross09,ross11-PRB84}.
And the absence of magnetic order in Er$_2$Sn$_2$O$_7$
remains mysterious.
The key to understanding all of these problems lies in the structure 
of the ground state manifold where phases with different symmetry meet.


\begin{figure}[htpb] 
\includegraphics[width=9cm]{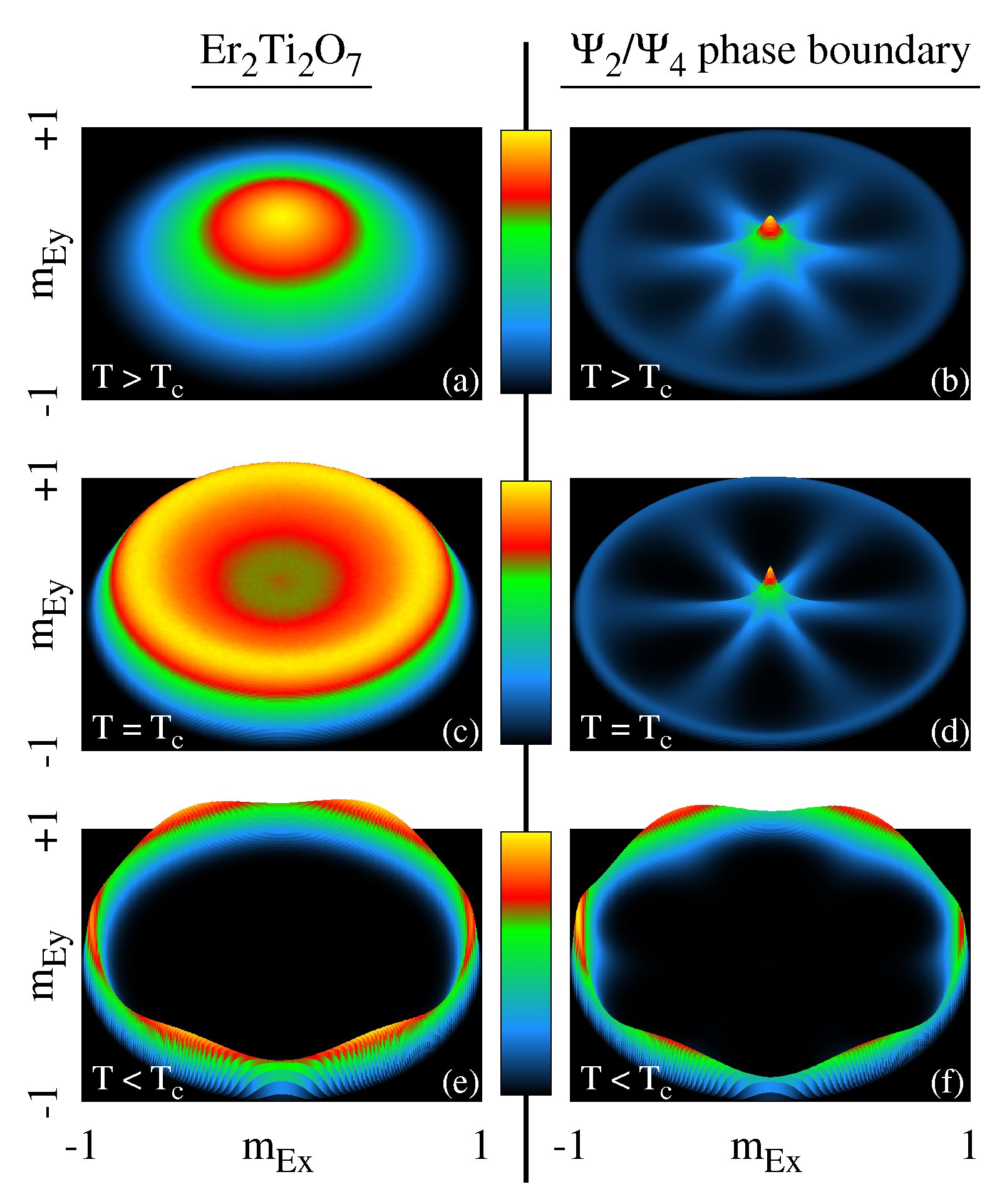}
\caption{
Selection of the $\Psi_2$ ground state by thermal fluctuations, 
as revealed by the probability density $P({\bf m}_{\sf E})$.
\mbox{(a, c, e)} results for parameters appropriate to Er$_2$Ti$_2$O$_7$ 
[\onlinecite{savary12-PRL109}], setting $J_4 = 0$.
\mbox{(b, d, f)} results for parameters bordering the Palmer-Chalker ($\Psi_4$)
ground state.
(a) broad distribution of fluctuations for $T > T_c$.
(b)  ``spoked wheel'' pattern, characteristic of the enlarged ground-state manifold 
[Fig.~\ref{fig:ground-state-manifold}], visible even for $T > T_c$.
(c) emergence of a one-dimensional manifold of states at $T = T_c$.
(d) emergence of a one-dimensional manifold of states at $T = T_c$, 
showing branching at the six $\Psi_2$ states.   
(e) selection of six $\Psi_2$ ground states within the manifold for $T < T_c$.
(f) selection of six $\Psi_2$ ground states within the manifold for $T < T_c$.
Results are taken from classical Monte Carlo simulations of 
${\mathcal H}_{\sf ex}$~[Eq.~(\ref{eq:Hex1})], with parameters
described in the supplementary information.  
%
} 
\label{fig:probability-distribution} 
\end{figure}


Let us first consider Er$_2$Ti$_2$O$_7$. 
Early heat capacity measurement of Er$_2$Ti$_2$O$_7$ 
revealed a transition at $T_c=1.25$ K, releasing an entropy 
$\Delta s \approx 0.97 k_B \ln 2$ per spin, consistent 
with the ordering of the ground state doublet of Er \cite{bloete69}. 
Later work identified this transition as a textbook example of ``order by disorder'',
with fluctuations selecting a $\Psi_2$ ground state from a continuous manifold of
states with local easy-plane character~\cite{champion03,mcclarty09,zhitomirsky12,savary12-PRL109}.
Using parameters taken from [\onlinecite{savary12-PRL109}], we have 
confirmed that Er$_2$Ti$_2$O$_7$ lies within a $\Psi_2$ phase.
Estimates from our classical Monte Carlo simulations give $T_c \approx 500 \text{mK}$, 
somewhat lower than in experiment, but with excellent agreement with experimental
measurements~\cite{dalmas12} of $S({\bf q})$ within the paramagnetic state 
[Fig.~\ref{fig:Sq}(d)].
However these parameters also place Er$_2$Ti$_2$O$_7$ relatively close to the 
boundary with the neighbouring Palmer-Chalker phase 
\mbox{[cf. Fig.~\ref{fig:classical-phase-diagram}]} 


On this boundary, we find that it is possible to deform the ground 
state continuously from 
$\Psi_2$ to a corresponding Palmer-Chalker state.
In more formal terms, the ground state of ${\mathcal H}_{\sf ex}^{[{\sf T_d}]}$ [Eq.~(\ref{eq:HTd1})] 
is enlarged from a single one-dimensional manifold connecting the $\Psi_2$ and $\Psi_3$ 
states, to a set of connected one-dimensional manifolds, which also interpolate 
to the Palmer-Chalker states \mbox{[Fig.~\ref{fig:ground-state-manifold}]}.
Since these manifolds branch at $\Psi_2$, $\Psi_2$ gains an additional soft 
set of excitations, and therefore has a lower free energy than $\Psi_3$.
These arguments remain valid at finite temperature [Fig.~\ref{fig:probability-distribution}],
and in the presence of quantum fluctuations, which are already 
known to favour $\Psi_2$ order [\onlinecite{mcclarty09,savary12-PRL109,zhitomirsky12}].  
We can therefore understand the ``order by disorder'' selection $\Psi_2$ 
in Er$_2$Ti$_2$O$_7$ from its proximity to the Palmer-Chalker phase.


We note that an exactly parallel argument predicts that $\Psi_3$ should be 
favoured approaching the boundary with the ferromagnet 
\mbox{[cf. Fig.~\ref{fig:classical-phase-diagram}]}.  
Where these two phase boundaries approach one another, the soft modes 
associated with the two different sets of manifolds compete.  
This leads to the complicated, re-entrant 
behaviour seen in Fig.~\ref{fig:classical-phase-diagram}, and studied for quantum 
spins in [\onlinecite{wong13}].   


We now turn to Yb$_2$Ti$_2$O$_7$.
Quasi-elastic neutron scattering from the paramagnetic phase of Yb$_2$Ti$_2$O$_7$ 
is dominated by dramatic ``rod''-like features in the [111] directions of reciprocal 
space.
First observed almost ten years ago~\cite{bonville04}, these rods of scattering
have since been interpreted as evidence of dimensional reduction~\cite{ross11-PRB84,ross09}
and, in the context of $\mathcal{H}_{\sf ex}$~[Eq.~(\ref{eq:Hex1})], as evidence 
of significant anisotropic exchange interactions~\cite{ross11-PRX1,thompson11}.
They are a robust feature of $S({\bf q})$, as calculated from $\mathcal{H}_{\sf ex}$~[Eq.~(\ref{eq:Hex1})] 
within both the (semi-classical) random phase approximation~\cite{thompson11,chang12}, 
and classical Monte Carlo simulation [Fig.~\ref{fig:Sq}].
However, despite their ubiquity, the origin of these rods of scattering remains mysterious.


Dimensional reduction is a well-defined feature of
$\mathcal{H}_{\sf ex}$~[Eq.~(\ref{eq:Hex1})].
The classical ground states of $\mathcal{H}_{\sf ex}$ reduce to a set of 
independent kagome planes on the boundary between FM and Palmer-Chalker 
phases, and to a set of independent chains on the boundary between the 
$\Psi_2$ and Palmer-Chalker phases.
However the rods of scattering seen in Yb$_2$Ti$_2$O$_7$ occur for parameters
where the ground state of $\mathcal{H}_{\sf ex}$ is expected to 
be ordered and fully three-dimensional.
They can instead be traced back to dimensionally-reduced excitations --- quasi-degenerate 
lines of low-lying spin wave excitations, which evolve into low-lying excitations of $\Psi_3$
on the boundary between the FM and the $\Psi_3$ phases.
This progression is clear in the evolution of $S({\bf q})$ from parameters 
appropriate to Yb$_2$Ti$_2$O$_7$ [Fig.~\ref{fig:Sq}(e)]
to the border of the $\Psi_3$ phase [Fig.~\ref{fig:Sq}(g)].


Our classical Monte Carlo simulations predict that Yb$_2$Ti$_2$O$_7$ orders 
at $450\ \text{mK}$, a little higher than the $T_c \approx 250\ \text{mK}$ found in those 
samples which show a phase transition~\cite{bloete69,ross11-PRB84,chang12}.
We note that the detailed form of $S({\bf q})$ observed in those 
samples of Yb$_2$Ti$_2$O$_7$ which do {\it not} exhibit a phase 
transition~\cite{ross09,ross11-PRB84,ortenzio-arXiv} is closer to predictions 
bordering on the $\Psi_3$ phase [Fig.~\ref{fig:Sq}(f)--(g)], than to 
those for the parameters given in \cite{ross11-PRX1} [Fig.~\ref{fig:Sq}(e)].
In particular, the noticeable enhancement of scattering at [220] 
offers strong evidence for the proximity of a $\Psi_3$ phase.
It is therefore interesting to ask what effect disorder~\cite{ross12,ortenzio-arXiv}, 
and quantum fluctuations, have on the 
ground state of $\mathcal{H}_{\sf ex}$~[Eq.~(\ref{eq:Hex1})] ?

\begin{figure}
\centering
\includegraphics[width=0.91\columnwidth]{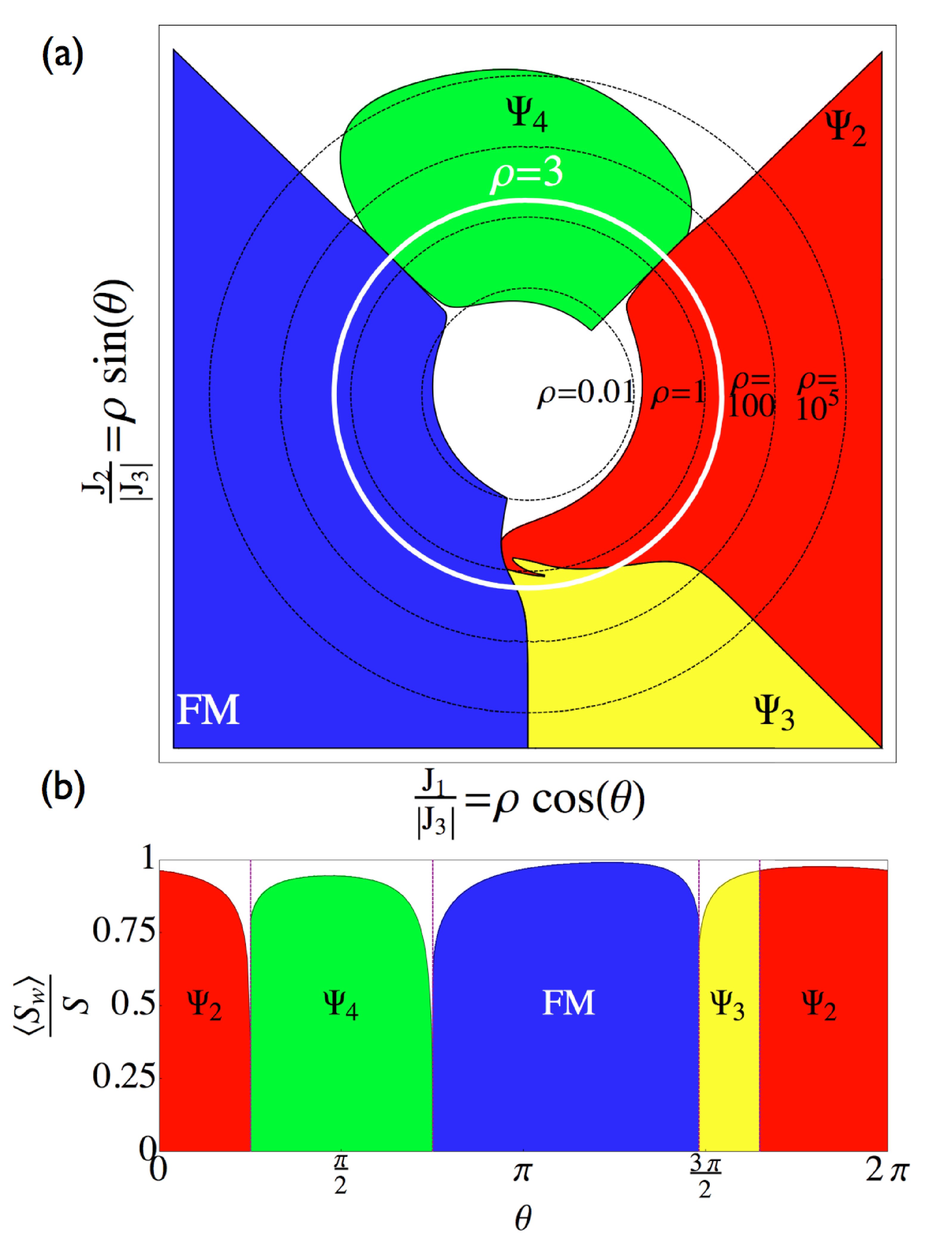}
\caption{
Effect of quantum fluctuations on magnetic order.   
(a) Ground-state phase diagram for a pyrochlore magnet with 
anisotropic exchange interactions, 
plotted in log-polar coordinates.
Blank regions indicate where magnetic order is entirely eliminated
by quantum fluctuations.
(b) Fraction of full classical moment achieved in ordered phases,
for $J_1 = 3|J_3| \cos\theta$, 
$J_2 = 3|J_3| \sin\theta$.  
Away from phase boundaries, the ordered moment is close to 
its full classical value.
All results are obtained within linear spin-wave theory for 
${\mathcal H}_{\sf ex}$~[Eq.~(\ref{eq:Hex1})], setting $J_3 < 0$ 
and $J_4 \equiv 0$.   
}
\label{fig:quantum-phase-diagram}
\end{figure}

We have explored the effect of quantum fluctuations on each of the four ground states
of $\mathcal{H}_{\sf ex}$~[Eq.~(\ref{eq:Hex1})], within linear spin wave theory.   
These results are summarised in Fig.~\ref{fig:quantum-phase-diagram}.
As expected, the enlargement of the ground state manifold at phase boundaries has 
a profound influence on fluctuations, eliminating order entirely for a wide range 
of parameters.
This tendency is most pronounced where the symmetry of the model is highest, e.g.
near the high symmetry point $|(J_1,J_2)|/|J_3|\to 0$, and approaching the 
Heisenberg line $J_1 = J_2$, $J_3 = J_4 = 0$.
However, deep within the ordered phases, all excitations are gapped, 
and the ordered moment approaches its full, classical, value.


The implications of these results for Er$_2$Sn$_2$O$_7$ are striking.
Like Er$_2$Ti$_2$O$_7$ and Yb$_2$Ti$_2$O$_7$, the magnetic ions 
in Er$_2$Sn$_2$O$_7$ have a Kramers doublet ground state \cite{matsuhira02}, 
and are believed to be well-described by $\mathcal{H}_{\sf ex}$~[Eq.~(\ref{eq:Hex1})] 
[\onlinecite{guitteny13}].
Correlations reminiscent of the Palmer-Chalker phase have been observed
in neutron scattering~\cite{guitteny13}, and magnetization measurements
show some evidence of spin-freezing at low temperatures~\cite{guitteny13}. 
Nonetheless, Er$_2$Sn$_2$O$_7$ shows {\it no} evidence of 
magnetic order, in thermodynamic measurements \cite{sarte11,guitteny13}, 
$\mu$SR \cite{lago05}, or neutron scattering \cite{sarte11,guitteny13}, 
down to a temperature of $20\ \text{mK}$ \cite{lago05}.


Our ground state analysis places Er$_2$Sn$_2$O$_7$ extremely 
close to the boundary between $\Psi_4$ and $\Psi_2$  [Fig.~\ref{fig:classical-phase-diagram}].  
Classical Monte Carlo simulations, using parameters taken 
from \cite{guitteny13}, predict a transition into Palmer-Chalker ($\Psi_4$) state at 
\mbox{$T_c = 204 \pm 5\ \text{mK}$}.
However linear spin-wave theory --- which typically {\it underestimates} quantum 
effects --- predicts that there is a narrow region of disorder between the 
$\Psi_4$ and $\Psi_2$ phases.
Er$_2$Sn$_2$O$_7$ lies on the edge of this disordered region, and is therefore a strong candidate 
for a quantum spin liquid.


In conclusion, rare-earth pyrochlore oxides offer a treasure-trove of different 
magnetic phases, including both classical and quantum spin liquids.
In this article we have established a general theory of magnetic order 
in materials where exchange interactions are limited to nearest-neighbour bonds.
We find that ordered phases of different symmetry are connected by enlarged 
ground-state manifolds.  
These ``accidental'' degeneracies have a profound effect, driving both ground state 
selection in Er$_2$Ti$_2$O$_7$, 
and the ``dimensional-reduction'' seen in Yb$_2$Ti$_2$O$_7$.
They also open the door to quantum spin liquids  --- a situation which 
may be realised in Er$_2$Sn$_2$O$_7$.
In this context, it could be extremely interesting to explore the effect of 
pressure and chemical substitution on systems like Yb$_2$Ti$_2$O$_7$ 
and Er$_2$Sn$_2$O$_7$, living on the edge of conventional magnetic order.\\


{\it Acknowledgments:} 
%
The authors are pleased to acknowledge helpful conversations with 
Bruce Gaulin, 
Michel Gingras,
Edwin Kermarrec, 
Isabelle Mirebeau,
Sylvain Petit,
Karlo Penc,
and Kate Ross, 
and a critical reading of the manuscript by Mathieu Taillefumier.
This work was supported by OIST.

\clearpage

\appendix

\section{Definition of model}

\begin{figure}
\centering\includegraphics[width=0.7\columnwidth]{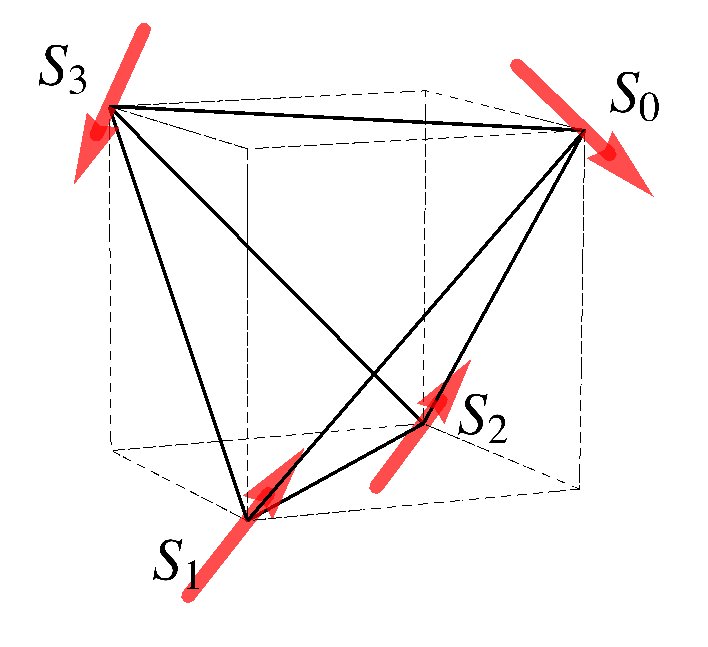}
\caption{
\label{fig:spindef}
A single tetrahedron within the pyrochlore lattice, 
showing the convention used in labelling sites.  
The positions of the magnetic sites relative to the centre of the 
tetrahedron are defined in Eq.~\ref{eq:r}. 
}
\end{figure}

The pyrochlore lattice is a corner-sharing network of tetrahedra.   
The lattice has overall cubic symmetry $Fd\overline{3}m$, and is bipartite in tetrahedra, 
with the centres of the tetrahedra forming a diamond lattice.
In what follows, we adopt the convention of numbering the spins in a 
tetrahedron $0$, $1$, $2$, and $3$, as shown in Fig.~\ref{fig:spindef}, 
with sites at positions 
\begin{eqnarray}
{\bf r}_0 & = & \left( \frac{{1}}{2},\frac{{1}}{2},\frac{{1}}{2} \right),\nonumber \\
{\bf r}_1 & = & \left( \frac{{1}}{2},-\frac{{1}}{2},-\frac{{1}}{2} \right),\nonumber \\
{\bf r}_2 & = & \left(-\frac{{1}}{2},\frac{{1}}{2},-\frac{{1}}{2} \right),\nonumber \\
{\bf r}_3 & = & \left(-\frac{{1}}{2},-\frac{{1}}{2},\frac{{1}}{2} \right).
\label{eq:r}
\end{eqnarray}
relative to the centre of the tetrahedron.
The spins at the four sites are denoted as ${\bf S}_{i=0,1,2,3}$.


The most general nearest-neighbour exchange Hamiltonian,
 respecting the symmetry of the pyrochlore lattice,  for
pseudospin-$1/2$ variables ${\bf S}_i$ representing
a Kramers doublet on each site,  can be written as \cite{curnoe07} 
%
\begin{eqnarray}
\mathcal{H}_{\sf ex} 
   = \sum_{\langle ij \rangle}
            J^{\mu\nu}_{ij} S^\mu_i S^\nu_j
\protect\label{eq:Hex}
\end{eqnarray}
where the sum on $\langle ij \rangle$ runs over the bonds of the
pyrochlore lattice and
\begin{eqnarray}
&{\bf J}_{01} 
  = \begin{pmatrix}
    J_2 & J_4 &J_4 \\
   -J_4 & J_1 &J_3 \\
   -J_4 & J_3 &J_1
   \end{pmatrix} 
   \quad
&{\bf J}_{02} 
  = \begin{pmatrix}
    J_1 & -J_4 & J_3 \\
    J_4 & J_2 & J_4 \\
    J_3 & -J_4 & J_1
   \end{pmatrix}    \nonumber\\
&{\bf J}_{03} 
  = \begin{pmatrix}
    J_1 & J_3 & -J_4 \\
    J_3 & J_1 & -J_4 \\
    J_4 & J_4 & J_2
   \end{pmatrix} 
   \quad
&{\bf J}_{12} 
  = \begin{pmatrix}
    J_1 & -J_3 & J_4 \\
    -J_3 & J_1 & -J_4 \\
    -J_4 & J_4 & J_2
   \end{pmatrix}    \nonumber\\
&{\bf J}_{13} 
  = \begin{pmatrix}
     J_1 & J_4 & -J_3 \\
     -J_4 & J_2 & J_4 \\
     -J_3 & J_4 & J_1
   \end{pmatrix} 
   \quad
&{\bf J}_{23} 
  = \begin{pmatrix}
    J_2 & -J_4 & J_4 \\
    J_4 & J_1 & -J_3 \\
    -J_4 & -J_3 & J_1
   \end{pmatrix} \nonumber\\ 
\end{eqnarray}
We can characterize the four different exchange parameters 
$J_{i=1,2,3,4}$  as
$J_1$ --- $XY$ interaction; 
$J_2$ --- Ising interaction; 
$J_3$ --- ``pseudo-dipolar'' interaction, and 
$J_4$ --- Dzyaloshinskii-Moriya interaction.
Values of $J_{i=1,2,3,4}$ for a given material can be determined from, e.g., 
inelastic neutron scattering experiments carried out in magnetic field 
\cite{ross11-PRX1,savary12-PRL109}


\section{Conditions for the existence of 4-sublattice order}

Here we prove the assertion that the most general possible Hamiltonian for nearest-neighbour 
exchange interactions on the pyrochlore lattice --- $\mathcal{H}_{\sf ex}$ [Eq.~(\ref{eq:Hex})] --- 
{\it always} possesses a classical ground state with vanishing crystal momentum ${\bf q}=0$, 
and 4-sublattice long-range order.


Since $\mathcal{H}_{\sf ex}$ includes only nearest-neighbour bonds, all of which belong uniquely
to a single tetrahedron, $\mathcal{H}_{\sf ex}$ can be written in terms of a sum over individual 
tetrahedra $l$
\begin{eqnarray}
\mathcal{H}_{\sf ex} 
  = \sum_{l \in \sf A} \mathcal{H}_{\sf ex}^{\sf A}[l]
  + \sum_{l \in \sf B} \mathcal{H}_{\sf ex}^{\sf B}[l]
\label{eq:HAandB}
\end{eqnarray}
%


\begin{table*}
\begin{tabular}{ | c | c |  c | }
\hline
\multirow{2}{*}{}
   order  & 
   definition in terms   & 
   associated
\\
   parameter & 
   of spin components & 
   ordered phases 
\\
\hline
\multirow{1}{*}{}
   $m_{\sf A_2}$ & 
   $\frac{1}{2 \sqrt{3} } 
     \left(S_0^x+S_0^y+S_0^z+S_1^x-S_1^y-S_1^z-S_2^x+S_2^y-S_2^z-S_3^x-S_3^y+S_3^z
     \right)$ & 
    ``all in-all out'' 
\\   
\hline
\multirow{1}{*}{} 
   ${\bf m}_{\sf E}$ & 
   $\begin{pmatrix}
         \frac{1}{2 \sqrt{6} } \left( -2 S_0^x + S_0^y + S_0^z - 2 S_1^x - S_1^y-S_1^z+2 S_2^x + S_2^y-
              S_2^z +2 S_3^x-S_3^y +S_3^z \right) \\
         \frac{1}{2 \sqrt{2}} \left( -S_0^y+S_0^z+S_1^y-S_1^z-S_2^y-S_2^z+S_3^y+S_3^z \right)
     \end{pmatrix}$ &
     $\Psi_2$ and $\Psi_3$ \\ 
\hline
\multirow{1}{*}{}
   ${\bf m}_{\sf T_{1, A}}$  & 
   $\begin{pmatrix}
        \frac{1}{2} (S_0^x+S_1^x+S_2^x+S_3^x) \\
        \frac{1}{2} (S_0^y+S_1^y+S_2^y+S_3^y) \\
        \frac{1}{2} (S_0^z+S_1^z+S_2^z+S_3^z)
   \end{pmatrix} $ &
   non-collinear FM
\\
\hline
\multirow{1}{*}{}
   ${\bf m}_{\sf T_{1, B}}$  & 
   $\begin{pmatrix}
          \frac{-1}{2\sqrt{2}} (S_0^y+S_0^z-S^1_y-S^1_z-S_2^y+S_2^z+S_3^y-S_3^z)  \\
          \frac{-1}{2\sqrt{2}} (S_0^x+S_0^z-S_1^x+S_1^z-S_2^x-S_2^z+S_3^x-S_3^z)  \\
          \frac{-1}{2\sqrt{2}} ( S_0^x+S_0^y-S_1^x+S_1^y+S_2^x-S_2^y-S_3^x-S_3^y) 
   \end{pmatrix}$ &
    non-collinear FM 
\\
\hline
\multirow{1}{*}{}
   ${\bf m}_{\sf T_2} $ & 
   $\begin{pmatrix}
        \frac{1}{2 \sqrt{2}} 
        \left(
         -S_0^y+S_0^z+S_1^y-S_1^z+S_2^y+S_2^z-S_3^y-S_3^z
        \right) 
        \\
        \frac{1}{2 \sqrt{2}} 
        \left(
        S_0^x-S_0^z-S_1^x-S^1_z-S_2^x+S_2^z+S_3^x+S_3^z
        \right) \\
        \frac{1}{2 \sqrt{2} }
        \left(
        -S_0^x+S_0^y+S_1^x+S_1^y-S_2^x-S_2^y+S_3^x-S_3^y
        \right)
      \end{pmatrix} $  &
       Palmer-Chalker ($\Psi_4$)     
\\ 
\hline
\end{tabular}
\caption{
Order parameters ${\bf m}_\lambda$, describing how the point group symmetry of a 
4-site tetrahedral unit cell, $T_d$, is broken by conventional magnetically-ordered 
phases on the pyrochlore lattice. The $\Psi_i$ notations are taken from~\cite{poole07}.
}
\label{table:m_lambda}
\end{table*}


\noindent where $\sf A$ and $\sf B$ refer to the two distinct sublattices of tetrahedra.
It follows that any state which minimises the energy of each individual tetrahedron 
must therefore be a ground state.
Tetrahedra on the $\sf A$ and $\sf B$ sublattices are related through inversion
about a single site $\mathcal{I}$.
Since $\mathcal{I}^2=1$,
\begin{eqnarray}
{\bf S}_i \cdot {\bf J}_{ij} \cdot {\bf S}_j
   &=& {\bf S}_i \cdot \mathcal{I}^2 \cdot {\bf J}_{ij} \cdot \mathcal{I}^2  \cdot {\bf S}_j \nonumber  \\
   &=&  {\bf S}_i \cdot \mathcal{I} \cdot {\bf J}_{ij} \cdot \mathcal{I} \cdot {\bf S}_j \nonumber  \\
   \implies   {\bf J}_{ij}  &=&  \mathcal{I} \cdot {\bf J}_{ij} \cdot \mathcal{I} 
\end{eqnarray}
where we have used the fact that ${\bf S}_i$ is invariant under inversion.
We infer that 
\begin{eqnarray}
\mathcal{H}_{\sf ex}^{\sf A}[l] &=& \mathcal{H}_{\sf ex}^{\sf B}[l'] = \mathcal{H}_{\sf ex}^{\sf tet}
\label{eq:inversionsymmetry}
\end{eqnarray}
and the Hamiltonian for {\it any} tetrahedron $l$ is the same, 
regardless of which sublattice it belongs to.


In the case of classical spins, $[\mathcal{H}_{\sf ex}^{\sf A} , \mathcal{H}_{\sf ex}^{\sf B}] = 0$,
and we can construct a ground state of $\mathcal{H}_{\sf ex}$ by choosing {\it any} state which
minimises the energy of a single tetrahedron, and repeating it across all ${\sf A}$-sublattice
(or ${\sf B}$-sublattice) tetrahedra. 
The equivalence of Hamiltonians for ${\sf A}$ or ${\sf B}$ 
sublattices [Eq.~(\ref{eq:inversionsymmetry})] {\it guarantees} that all ${\sf B}$-sublattice
(or ${\sf A}$-sublattice) will automatically have the minimum energy. Therefore there always exists
a ${\mathbf q} = 0$ classical ground state with 4-sublattice long-range order,
even in the presence of finite Dzyaloshinskii-Moriya interaction $J_4$.


This ${\mathbf q} = 0$, 4-sublattice order is unique --- up to the degeneracy of the  
ground state for a single tetrahedron --- {\it provided} that the spin on every site 
of the tetrahedron points in a different direction in each of these  ground states. 
Away from phase boundaries, this is true for {\it all} of the 4-sublattice ordered phases 
discussed in the main text, each of which is 6-fold degenerate.
However if two of the ground states of a single tetrahedron share a common spin --- i.e.  
the spin on a given site points in the same direction in more than one ground state --- then 
it is {\it always} possible to construct other ground states with finite ${\mathbf q}$.


Let us suppose, for example, that two different ground states for a single tetrahedron 
have identical orientation of the spin on site $0$, but different orientation of the spins on sites
$1$, $2$ and $3$.   
In this case it is possible to divide the pyrochlore lattice into a set of parallel kagome planes,
containing spins associated with sites $1$,$2$ and $3$ of a tetrahedron, separated by 
triangular-lattice planes associated with site $0$.   
Since each successive kagome plane can take on one of two different spin 
configurations, the number of such ground states grows as $2^{\sf N_K}$, 
where ${\sf N_K}$ is the number of kagome planes, 
and encompasses all possible ${\mathbf q} \parallel [111]$.
Dimensional reduction of this type occurring for example, on the boundary between
the FM and Palmer-Chalker phases, is described below.


An even larger degeneracy occurs for the 
``two in, two out"  states, made famous by the spin ice problem.
In this case there are a total of $6$ possible ground states for a single 
tetrahedron, but each possible spin orientation, on each site, belongs 
to $3$ different ground states.
The total number of possible ground states on the lattice is then 
extensive, $\Omega_{\sf ice} \sim (3/2)^{N/2}$, 
where $N$ is the total number of sites in the lattice.
This manifold of ``ice'' states includes ground states with 
all possible ${\mathbf q}$.

\section{Symmetry classification of ordered phases}

\subsection{Definition of order parameters ${\bf m}_\lambda$}
\label{def_m_lambda}

The symmetry operations of a tetrahedron form a 24-element group ${\sf T}_d$, 
with elements~:
$8 \times C_3$ --- $\frac{2 \pi}{3}$ rotation around a $[111]$ axis;   
$3 \times C_2$ --- $\pi$ rotation around $[100]$ axis;
$6 \times S_4$ --- $\frac{\pi}{2}$ rotation around a $[100]$ axis followed by reflection 
in the same $[100]$ plane; 
$6 \times \sigma_d$ --- reflection in $[011]$ plane; 
$\epsilon$ --- the identity~\cite{kovalev93}.


It is possible to define order parameters ${\bf m}_\lambda$, transforming with the 
non-trivial
irreducible representations 
\mbox{$\lambda = \{\ {\sf A}_2$, ${\sf E}$, ${\sf T}_1$, ${\sf T}_2\ \}$} 
of ${\sf T}_d$, which fully characterise all possible 4-sublattice ordered states 
on a pyrochlore lattice.
These are listed in Table~\ref{table:m_lambda}.


The order-parameter susceptibly 
\begin{eqnarray}
\chi_\lambda (T)
   = \frac{ \langle | {\bf m}_\lambda |^2 \rangle 
              - \langle {\bf m}_\lambda \rangle^2}{T}
\label{eq:chi_lambda}
\end{eqnarray}
associated with each ${\bf m}_\lambda$ is a useful tool 
for determining phase transitions in finite-temperature simulations.
We note that in the case of the two, coupled $\sf T_1$ order parameters, 
${\bf m}_{\sf T_{1,A}}$
and ${\bf m}_{\sf T_{1,B}}$ (see Table \ref{table:m_lambda}), 
it is convenient to group both order parameters into
a single susceptibility.
%
%

\subsection{Expression of Hamiltonian in terms of order parameters}

For classical spins undergoing 4-sublattice order, the Hamiltonian $\mathcal{H}_{\sf ex}$ 
[Eq.~(\protect\ref{eq:Hex})] can be rewritten in terms of these order parameters, to give
\begin{eqnarray}
{\mathcal H}_{\sf ex}^{[{\sf T_d}]}
  &=&  \frac{1}{2}
    \left[ 
       a_{\sf A_2} \, m_{\sf A_2}^2 
       + a_{\sf E}\, {\bf m}^2_{\sf E} 
       + a_{\sf T_2}\, {\bf m}^2_{\sf T_2} 
       + a_{\sf T_{1, A}}\, {\bf m}^2_{\sf T_{1, A}} 
    \right.
    \nonumber\\
    && \quad \left. 
       + a_{\sf T_{1, B}}\, {\bf m}^2_{\sf T_{1, B}} 
       + a_{\sf T_{1, AB}}\, {\bf m}_{\sf T_{1, A}} \cdot {\bf m}_{\sf T_{1, B}}
    \right].
\label{eq:HTd}
\end{eqnarray}
with coefficients 
\begin{eqnarray}
a_{\sf A_2} &=& -2 J_1+J_2-2(J_3+2J_4)\nonumber\\
a_{\sf E} &=& -2 J_1+J_2+J_3+2J_4\nonumber\\ 
a_{\sf T_2} &=& -J_2+J_3-2J_4\nonumber\\
a_{\sf T_{1,A}} &=& 2 J_1+J_2\nonumber\\ 
a_{\sf T_{1, B}} &=& -J_2-J_3+2J_4\nonumber\\ 
a_{\sf T_{1, AB}} &=& -\sqrt{8} J_3
\label{eq:owencoeff}
\end{eqnarray}


For the purpose of finding the ground states, the Hamiltonian ${\mathcal H}_{\sf ex}^{[{\sf T_d}]}$ 
can be reduced to a form quadratic in ${\bf m}_\lambda$ by a  
coordinate transformation 
\begin{eqnarray}
{\bf m}_{\sf T_{1, A'}} &=& \cos \theta_{\sf T_1} \ {\bf m}_{\sf T_{1, A}} 
        - \sin \theta_{\sf T_1} \ {\bf m}_{\sf T_{1, B}} \nonumber\\
{\bf m}_{\sf T_{1, B'}} &=& \sin \theta_{\sf T_1} \ {\bf m}_{\sf T_{1, A}} 
        + \cos \theta_{\sf T_1}\ {\bf m}_{\sf T_{1, B}} 
 \label{eq:rotatedT1}
\end{eqnarray}
where 
\begin{eqnarray}
\theta_{\sf T_1} = \frac{1}{2}\arctan{ \left( \frac{\sqrt{8} J_3}{2J_1+2J_2+J_3-2J_4} \right)}.
\label{eq:FMangle}
\end{eqnarray}
is the canting angle between spins and the relevant [100] axis. 
The Hamiltonian then becomes
\begin{eqnarray}
{\mathcal H'}_{\sf ex}^{[{\sf T_d}]} 
   &=& \frac{1}{2} \big[ a_{\sf A_2} m_{\sf A_2}^2+ a_{\sf E} {\bf m}^2_{\sf E}
           + a_{\sf T_2}{\bf m}^2_{\sf T_2}  \nonumber \\
     && + a_{\sf T_{1 A^{\prime}}} {\bf m}^2_{T_{1 A^{\prime}}}
           + a_{\sf T_{1 B^{\prime}}} {\bf m}^2_{\sf T_{1 B^{\prime}}} \big].
\label{eq:H'Td}
\end{eqnarray}
with coefficients  given in Table~\ref{table:coefficients}.


The minimisation of the energy is always subjected to the 
constraint that every spin has fixed length $S^2=1/4$.
It is convenient to express this as
\begin{eqnarray}
{\bf S}_0^2+{\bf S}_1^2+{\bf S}_2^2+{\bf S}_3^2&=&1 \nonumber \\
{\bf S}_0^2+{\bf S}_1^2-{\bf S}_2^2-{\bf S}_3^2&=&0 \nonumber \\
{\bf S}_0^2-{\bf S}_1^2+{\bf S}_2^2-{\bf S}_3^2&=&0 \nonumber \\
{\bf S}_0^2-{\bf S}_1^2-{\bf S}_2^2+{\bf S}_3^2&=&0
\label{eq:spinconstraints}
\end{eqnarray}
for our further calculation. We note that the addition of single-ion anisotropy only changes the
coefficient $a_\lambda$ in equation~(\ref{eq:H'Td}) and so could be easily included in our analysis.

%
%

\begin{table}
\begin{tabular}{ | c | c  | }
\hline
coefficient   & definition in terms of \\
of $|{\bf m}_\lambda|^2$ &  exchange parameters \\
\hline
$a_{\sf A_2}$ & $-2J_1+J_2-2(J_3+2J_4)$ \\
\hline
$a_{\sf E}$ & $-2 J_1+J_2+J_3+2J_4$\\
\hline
$a_{\sf T_2}$ & $-J_2+J_3-2J_4$ \\
\hline
$a_{\sf T_{1, A'}}$ & $(2 J_1+J_2)\cos^2(\theta_{\sf T_1})$ \\
& $-(J_2+J_3-2J_4)\sin^2(\theta_{\sf T_1}) +\sqrt{2} J_3 \sin(2 \theta_{\sf T_1})$ \\
\hline
$a_{\sf T_{1, B'}}$ &$(2 J_1+J_2)\sin^2(\theta_{\sf T_1})-$ \\
 & $(J_2+J_3-2J_4)\cos^2(\theta_{\sf T_1})
-\sqrt{2} J_3 \sin(2 \theta_{\sf T_1})$ \\
\hline
\end{tabular}
\caption{Coefficients $a_\lambda$ of the scalar invariants $|{\bf m}_\lambda|^2$ 
appearing in ${\mathcal H'}_{\sf ex}^{[{\sf T_d}]}$ ~[Eq.~(\ref{eq:H'Td})].
The classical ground states of $\mathcal{H}_{\sf ex}$ [Eq.~(\protect\ref{eq:Hex})] for a given 
set of parameters ($J_1$, $J_2$, $J_3$, $J_4$) can be found by identifying the coefficient(s) 
$a_\lambda$ with the lowest value, and imposing the constraint Eq.~(\ref{eq:spinconstraints}) 
on the associated ${\bf m}_\lambda$.
The canting angle $\theta_{\sf T_1}$ is defined in Eq.~(\ref{eq:FMangle}).
}
\label{table:coefficients}
\end{table}

\subsection{Phases and phase transitions predicted by ${\mathcal H'}_{\sf ex}^{[{\sf T_d}]}$}

The Hamiltonian ${\mathcal H'}_{\sf ex}^{[{\sf T_d}]}$ ~[Eq.~(\ref{eq:H'Td})] leads
very directly to a classical ground state phase diagram.
The sum of the squares of the order parameters are constrained
via Eq.~(\ref{eq:spinconstraints})
\begin{equation}
 m_{\sf A_2}^2 
       + {\bf m}^2_{\sf E} 
       + {\bf m}^2_{\sf T_2} 
       + {\bf m}^2_{\sf T_{1, A'}} 
       + {\bf m}^2_{\sf T_{1, B'}} 
        \equiv \sum_{\lambda}m_{\lambda}^2 =1,
\label{eq:constraint}
\end{equation}
and each individual order parameter is constructed to have a maximal 
magnitude of unity
\begin{eqnarray}
\max {\bf m}_{\lambda}^2=1.
\end{eqnarray}
Taken together, these facts imply that the classical ground state of 
$\mathcal{H}_{\sf ex}$~[Eq.~(\ref{eq:Hex})] can be found by first 
identifying the coefficient $a_\lambda$ of 
${\mathcal H'}_{\sf ex}^{[{\sf T_d}]}$~[Eq.~(\ref{eq:HTd})] with the 
minimum value, and then imposing the constraint Eq.~(\ref{eq:spinconstraints}) 
on ${\bf m}_\lambda$.

\begin{figure}
\centering\includegraphics[width=0.9\columnwidth]{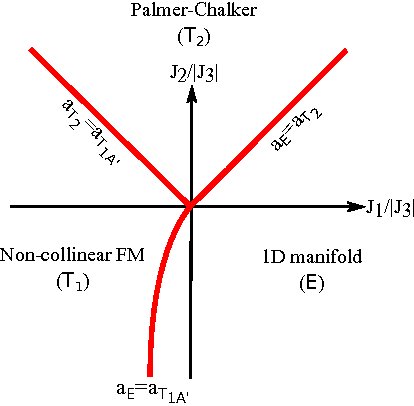}
\caption{
Classical ground state phase diagram of $\mathcal{H}_{\sf ex}$ [Eq.~(\ref{eq:Hex})]
for $J_3<0$, $J_4=0$, as a function of $(J_1,J_2)/|J_3|$.
In the absence of fluctuations, the ground states are 
a non-collinear FM transforming with the ${\sf T_1}$ irrep of ${\sf T_d}$; 
a one-dimensional manifold of states transforming with the ${\sf E}$ irrep of ${\sf T_d}$; 
and the Palmer-Chalker phase, a coplanar antiferromagnet transforming with 
the ${\sf T_2}$ irrep of ${\sf T_d}$.
All three phases have long-range 4-sublattice order.
Analytical expressions for the boundaries between phases are given in 
Eq.~(\ref{eq:boundary-T2-E}--\ref{eq:boundary-E-T1}), with coefficients
$a_\lambda$ defined in Table~\ref{table:coefficients}.  
\label{fig:phase}}
\end{figure}

For $J_3<0$ and $J_4\equiv0$, the coefficients $a_\lambda$ with the  
lowest values can be $a_{\sf E}$, $a_{\sf T_{1 A'}}$, and $a_{\sf T_2}$, depending on
the values of $J_1$ and $J_2$, 
and the ground states found have ${\sf E}$, ${\sf T}_1$ and ${\sf T}_2$ symmetry.
The boundaries between these phases occur where
\mbox{$a_{\sf T_2} = a_{\sf E}$}, 
\mbox{$a_{\sf T_2} = a_{\sf T_{1, A^{\prime}}}$},
and 
\mbox{$a_{\sf E} = a_{\sf T_{1, A^{\prime}}}$}.  
%
In the present case, these expressions reduce to
\begin{eqnarray}
   a_{\sf T_2} = a_{\sf E \phantom{, A^{\prime}}}  &<&
     a_{\sf T_2}, a_{\sf T_{1 B^{\prime}}}, a_{\sf A_2} 
        \Rightarrow   
       J_2 = \phantom{-}J_1 > 0
        \label{eq:boundary-T2-E} \\
   a_{\sf T_2} = a_{\sf T_{1, A^{\prime}}}  &<&
    a_{\sf E}, a_{\sf T_{1 B^{\prime}}}, a_{\sf A_2} 
       \  \Rightarrow
       J_2 = -J_1 > 0
        \label{eq:boundary-T2-T1} 
        \\
   a_{\sf E} = a_{\sf T_{1, A^{\prime}}}   &<&
    a_{\sf T_2}, a_{\sf T_{1 B^{\prime}}}, a_{\sf A_2}  
         \Rightarrow
      J_2 = \frac{J_1 (4 J_1 - 5 J_3)}{4 J_1 - J_3}
                 < 0
             \nonumber\\ 
        \label{eq:boundary-E-T1}
\end{eqnarray}
where $\theta_{\sf T_1}$ is defined in Eq.~(\ref{eq:FMangle}).
The regions bounded by these curves are shown in Fig.~\ref{fig:phase}.  


On the boundaries between phases with different symmetry, the set of possible 
ground states include states with finite values of {\it both} order parameters, 
subject to the constraint Eq.~(\ref{eq:spinconstraints}).
The three distinct $T=0$ ground states for  $J_3<0$ and $J_4\equiv0$, and the associated 
phase boundaries, are discussed in detail below. 


\section{4-sublattice ordered states for \mbox{$J_3 < 0$, $J_4 \equiv 0$}.}


\begin{figure}
\centering
\subfigure[]{%
  \includegraphics[width=.4\linewidth]{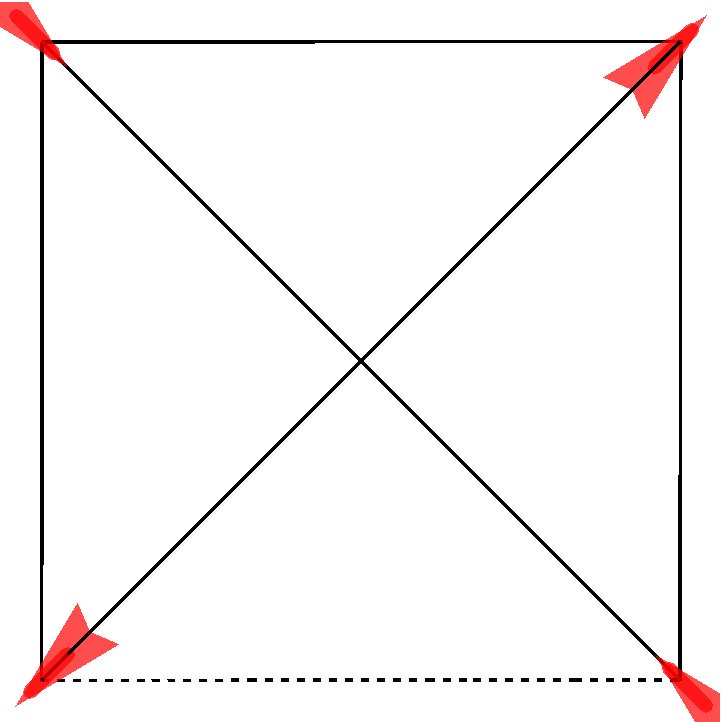}}%
    \qquad
\subfigure[]{%
  \includegraphics[width=.4\linewidth]{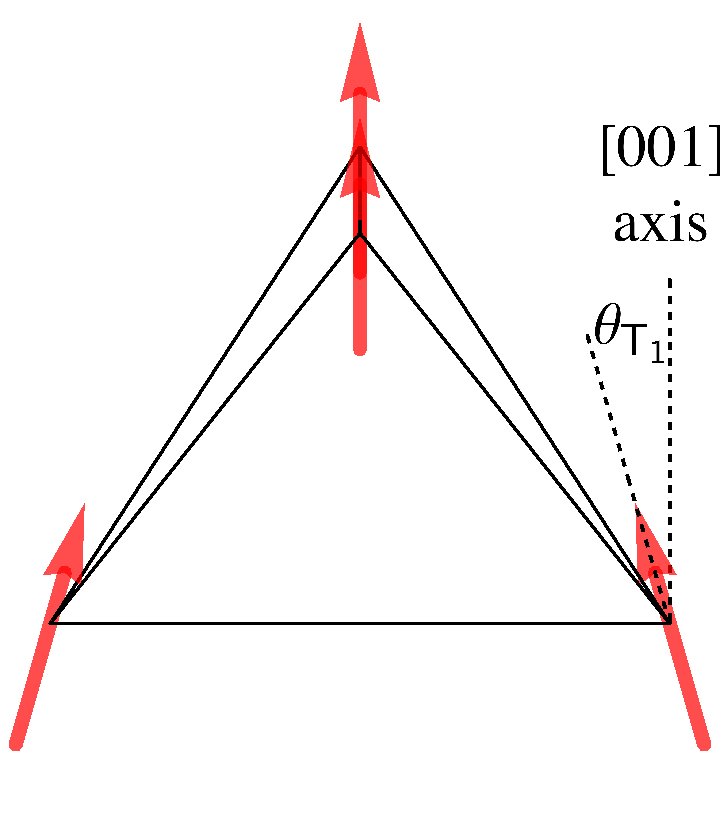}}%
\caption{\label{fig:ferrophase} 
Spin-configuration in the 4-sublattice non-collinear FM phase :
(a) viewed along the $[001]$ axis;   
(b) viewed slightly off the $[110]$ axis. 
The magnetisation is aligned with the $[001]$ axis.
Spins are canted into the plane perpendicular to this, 
with canting angle $\theta_{\sf T_1}$, in an ``ice-like'' manner.
}
\end{figure}

\subsection*{Non-collinear FM with $T_1$ symmetry}
\label{ferromagnet}

In a region bounded by 
\mbox{$a_{\sf T_{1,A^\prime}} = a_{\sf T_2} $} [Eq.~(\ref{eq:boundary-T2-T1})] ,
and \mbox{$a_{\sf T_{1,A^\prime}} = a_{\sf E}$} [Eq.~(\ref{eq:boundary-E-T1})] 
--- cf. Fig.~\ref{fig:phase} --- 
the energy 
is  minimised by setting 
\begin{eqnarray}
   {\mathbf m}^2_{\sf T_{1,A'}} = 1
\end{eqnarray}
and 
\begin{eqnarray}
   m_{\sf A_2} = {\bf m}_{\sf E} = {\bf m}_{\sf T_2} = {\bf m}_{\sf T_{1 B^{\prime}}}=0
\end{eqnarray}
The constraints on the total length of the spin, Eq.~(\ref{eq:spinconstraints}) 
further imply that 
\begin{eqnarray}
   m_{\sf T_ {1 A^{\prime}}}^y m_{\sf T_ {1 A^{\prime}}}^z&=&0 \nonumber \\
   m_{\sf T_ {1 A^{\prime}}}^x m_{\sf T_ {1 A^{\prime}}}^z&=&0 \nonumber \\
   m_{\sf T_ {1 A^{\prime}}}^x m_{\sf T_ {1 A^{\prime}}}^y&=&0.
\end{eqnarray}
It follows that there are 6 possible ground states
\begin{eqnarray}
   {\bf m}_{\sf T_ {1 A^{\prime}}} =
  \begin{pmatrix}
      \pm1\\
       0\\
       0
   \end{pmatrix},
   \begin{pmatrix}
       0\\
       \pm1\\
       0
    \end{pmatrix},
    \begin{pmatrix}
       0\\
       0\\
       \pm1
     \end{pmatrix}.
\end{eqnarray}
Written in terms of spins, these are 6, non-collinear ferromagnetic (FM) 
ground states, with typical spin configuration
\begin{eqnarray}
\mathbf{S}_0 &=& 
  S \left (\sin \theta_{\sf T_1} /\sqrt{2},
   \sin \theta_{\sf T_1} /\sqrt{2},
   \cos\theta_{\sf T_1} \right)
   \nonumber \\
\mathbf{S}_1 &=& 
 S  \left (-\sin \theta_{\sf T_1} /\sqrt{2},
   \sin \theta_{\sf T_1} /\sqrt{2},
   \cos\theta_{\sf T_1} \right)
   \nonumber \\
\mathbf{S}_2 &=& 
  S \left (\sin \theta_{\sf T_1} /\sqrt{2},
   -\sin \theta_{\sf T_1} /\sqrt{2},
   \cos\theta_{\sf T_1} \right)
   \nonumber \\
\mathbf{S}_3 &=& 
  S \left( -\sin \theta_{\sf T_1} /\sqrt{2},
   -\sin \theta_{\sf T_1} /\sqrt{2},
   \cos\theta_{\sf T_1} \right)
\label{eq:fmgs2}
\end{eqnarray}
where $\theta_{\sf T_1}$ is given by Eq. (\ref{eq:FMangle}).


The magnetisation of this FM ground state, illustrated in Fig.~\ref{fig:ferrophase}, 
is parallel to a $[001]$ axis, with spins canted away from this axis, in an ``ice-like'' 
manner.
This state has been identified as the ground state in Yb$_2$Sn$_2$O$_7$, 
where it was referred to as a ``splayed FM''~[\onlinecite{yaouanc13}], and in those 
samples of Yb$_2$Ti$_2$O$_7$ which order at low temperature~\cite{chang12}.


\begin{figure}
\centering
\subfigure[]{%
  \includegraphics[width=.4\linewidth]{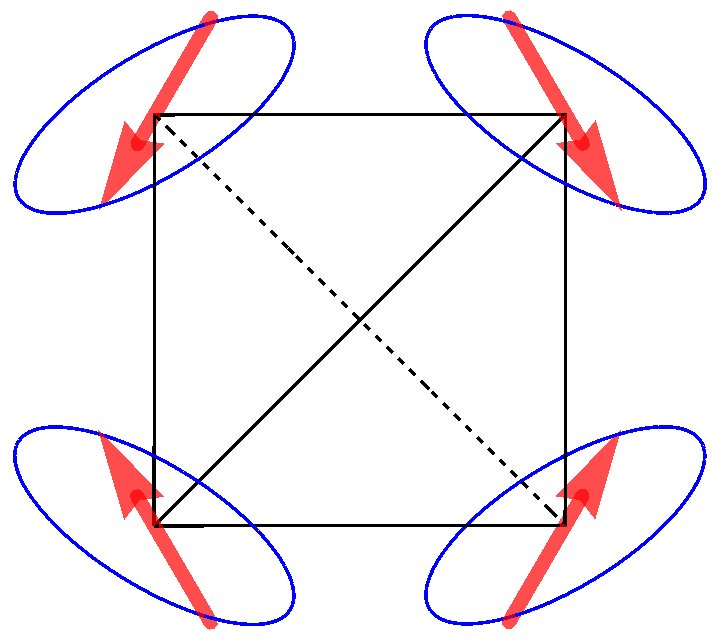}}%
    \qquad
\subfigure[]{%
  \includegraphics[width=.4\linewidth]{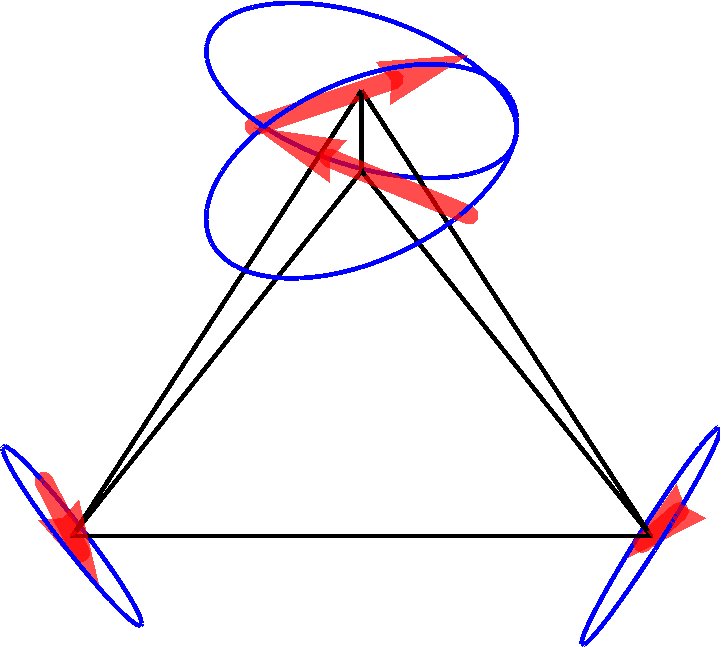}}%
\caption{
Example of a spin configuration within 
the one-dimensional manifold of states 
transforming with the ${\sf E}$ irrep of ${\sf T_d}$~:
(a) viewed along $[001]$ axis; 
(b) viewed slightly off the $[110]$ axis.
The manifold posses 4-sublattice long-range order, with spins
lying in the  ``$XY$'' 
plane perpendicular to the local $[111]$ axis at each site.
The manifold is continuous, and can be parameterised
with a single angle $\theta_{\sf E}$.
\label{fig:u1phase}}
\end{figure}

\subsection*{One-dimensional manifold of states with ${\sf E}$ symmetry}
\label{1Dmanifold}

In a region bounded by 
\mbox{$a_{\sf E} = a_{\sf T_{1,A^{\prime}}} $} [Eq.~(\ref{eq:boundary-E-T1})] 
and \mbox{$a_{\sf E} = a_{\sf T_2}$} [Eq.~(\ref{eq:boundary-T2-E})] 
--- cf.  Fig. \ref{fig:phase} ---
the energy
is minimised by setting
\begin{eqnarray}
   {\bf m}_{\sf E}^2=1
   \label{eq:Enorm}
\end{eqnarray}
and
\begin{eqnarray}
   {m}_{\sf A_2}={\bf m}_{\sf T_2}
   = {\bf m}_{\sf T_{1 A^\prime}}={\bf m}_{\sf T_{1 B^{\prime}}}.
   \label{eq:Ezerofields}
\end{eqnarray}


These solutions {\it automatically} satisfy the constraint on the total length 
of the spin Eq.~(\ref{eq:spinconstraints}), and are conveniently characterised 
by writing 
\begin{eqnarray}
   {\bf m}_{\sf E} = \ (\cos \theta_{\sf E},\  \sin \theta_{\sf E})
   \label{eq:thetaE}
\end{eqnarray}
It follows that the ground state is a continuous, one-dimensional 
manifold of states parameterised by the single angle 
\mbox{$0 \leq \theta_{\sf E} < 2\pi$}.
The spin configuration in this manifold is given by
\begin{eqnarray}
\mathbf{S}_0 & = & S \bigg(
                             \sqrt{\frac{2}{3}}\cos(\theta_{\sf E}),\:
                             \sqrt{\frac{2}{3}}\cos(\theta_{\sf E}+\frac{2 \pi}{3}),\nonumber \\
                             && \qquad \sqrt{\frac{2}{3}}\cos(\theta_{\sf E}-\frac{2\pi}{3})
                             \bigg) \nonumber\\
\mathbf{S}_1 & = & S \bigg(
                             \sqrt{\frac{2}{3}}\cos(\theta_{\sf E}),\:
                             -\sqrt{\frac{2}{3}}\cos(\theta_{\sf E}+\frac{2\pi}{3}), \nonumber \\
                            	&& \qquad -\sqrt{\frac{2}{3}}\cos(\theta_{\sf E}-\frac{2\pi}{3})
                             \bigg) \nonumber\\
\mathbf{S}_2 & = & S \bigg(
                             -\sqrt{\frac{2}{3}}\cos(\theta_{\sf E}),\:
                             \sqrt{\frac{2}{3}}\cos(\theta_{\sf E}+\frac{2\pi}{3}),\nonumber \\
                            && \qquad -\sqrt{\frac{2}{3}}\cos(\theta_{\sf E}-\frac{2\pi}{3})
                             \bigg) \nonumber\\
\mathbf{S}_3 & = & S \bigg(
                             -\sqrt{\frac{2}{3}}\cos(\theta_{\sf E}),\:
                             -\sqrt{\frac{2}{3}}\cos(\theta_{\sf E}+\frac{2\pi}{3}),\: \nonumber \\
                            && \qquad \sqrt{\frac{2}{3}}\cos(\theta_{\sf E}-\frac{2\pi}{3})
                             \bigg).
                             \label{eq:u1gs2} 
\end{eqnarray}
Within this one-dimensional manifold of states, each spin $\mathbf{S}_i$
lies in the local ``$XY$'' plane normal to $\mathbf{\hat{r}}_i$ [cf Eq. (\ref{eq:r})].


\begin{figure}
\centering
\subfigure[]{%
  \includegraphics[width=.4\linewidth]{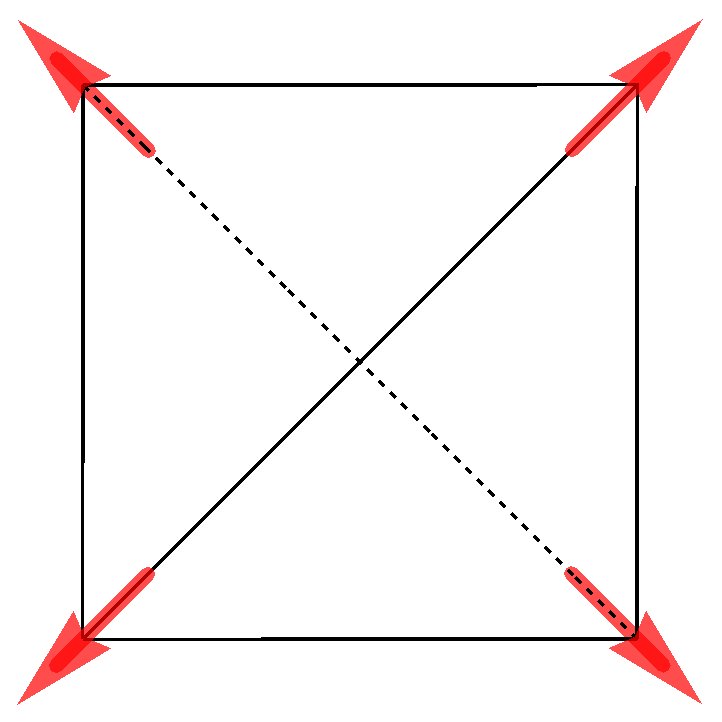}}%
    \qquad
\subfigure[]{%
  \includegraphics[width=.4\linewidth]{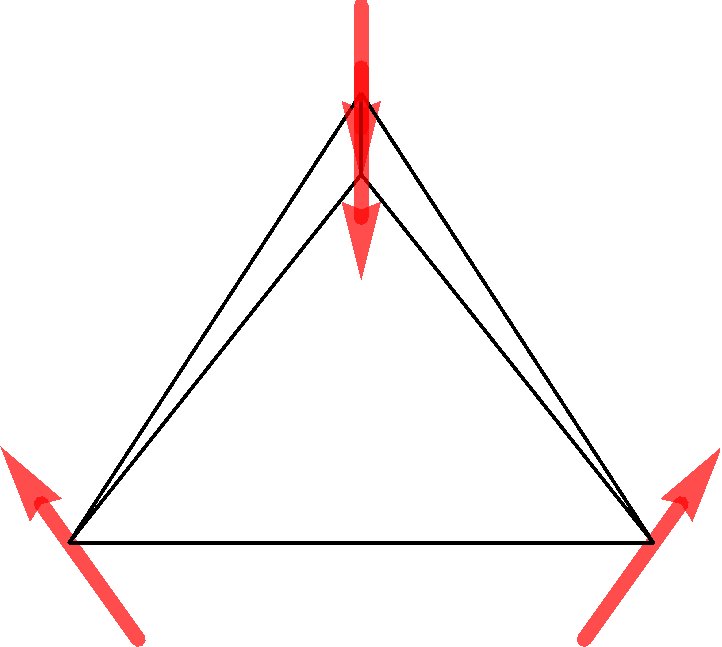}}%
\caption{
Spin configuration in the 4-sublattice non-coplanar antiferromagnet, $\Psi_2$, 
selected by fluctuations from the one-dimensional manifold of states 
transforming with ${\sf E}$~:
(a) viewed along $[001]$ axis; 
(b) viewed slightly off the $[110]$ axis.
At the phase boundary with the Palmer-Chalker phase, 
each of the six $\Psi_2$ ground states can be transformed continuously
into a Palmer-Chalker state.
\label{fig:psi2}}
\end{figure}

\subsection*{Non-coplanar antiferromagnet, $\Psi_2$, with ${\sf E}$ symmetry}

For parameters bordering on the Palmer-Chalker phase [cf Fig. 1, main text], 
fluctuations select a non-coplanar antiferromagnet, $\Psi_2$, from the one-dimensional 
manifold of states transforming with ${\sf E}$.
The $\Psi_2$ ground state is six-fold degenerate, with spins canted symmetrically 
out of the $[100]$ plane.


The six spin configurations for $\Psi_2$ states are given by Eq.~(\ref{eq:u1gs2}) with 
$\theta_{\sf E} = \frac{n \pi}{3}$, $n=0,1,2\ldots 5$.
The $\Psi_2$ state is characterised by the primary order parameter ${\bf m}_{\sf E}$ 
[cf.~Table~\ref{table:m_lambda}], and by $c_{\sf E} > 0$, where 
\begin{eqnarray}
c_{\sf E} = \langle \cos 6\theta_{\sf E} \rangle 
\label{eq:ctheta}
\end{eqnarray}
An example of a typical spin configuration is shown in Fig.~(\ref{fig:psi2}). 

\subsection*{Coplanar antiferromagnet, $\Psi_3$, with ${\sf E}$ symmetry}

For parameters bordering on the non-collinear FM phase, fluctuations
select a coplanar antiferromagnet, $\Psi_3$, from the one-dimensional 
manifold of states transforming with ${\sf E}$.
The $\Psi_3$ ground state is six-fold degenerate, with spins lying in 
a common $[100]$ plane.

\begin{figure}
\centering
\subfigure[]{%
  \includegraphics[width=.4\linewidth]{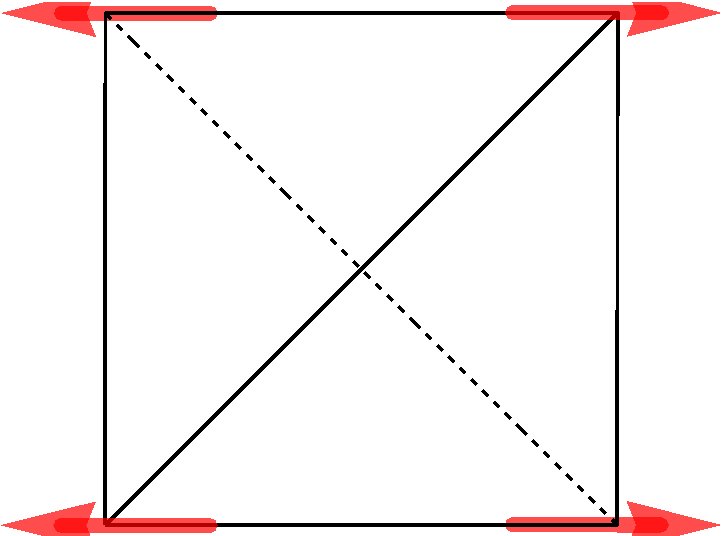}}%
  \qquad
\subfigure[]{%
  \includegraphics[width=.4\linewidth]{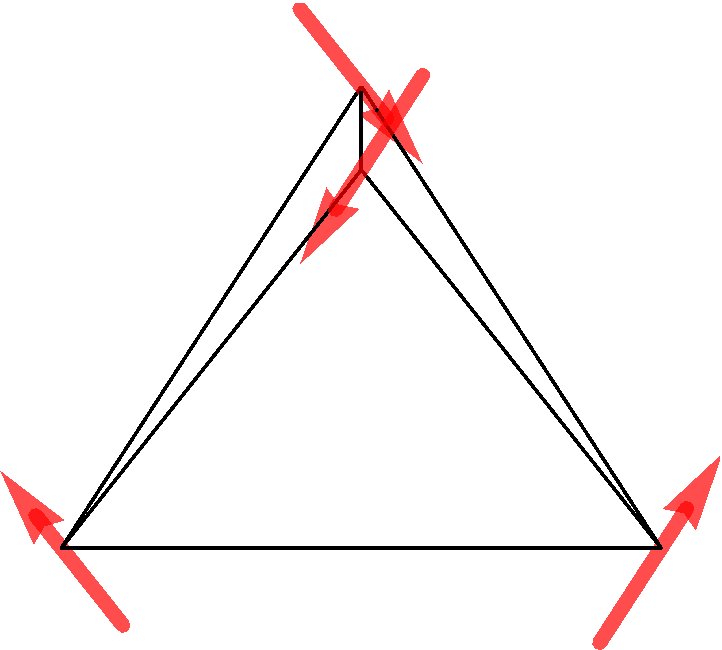}}%
\caption{
Spin configuration in the 4-sublattice coplanar antiferromagnet, $\Psi_3$, 
selected by fluctuations from the one-dimensional manifold of states 
transforming with ${\sf E}$~:
(a) viewed along $[001]$ axis; 
(b) viewed slightly off the $[110]$ axis.
At the phase boundary with the non-collinear FM phase, each of the six $\Psi_3$ 
ground states can be transformed continuously into a non-collinear FM state.
\label{fig:psi3}}
\end{figure}

The six spin configurations for $\Psi_3$ states are given by Eq.~(\ref{eq:u1gs2}) with 
$\theta =\frac{(2 n+1) \pi}{6}$, $n=0,1,2\ldots 5$.    
These states are characterised by a finite value of the order parameter ${\bf m}_{\sf E}$ 
[cf Table~\ref{table:m_lambda}], and by $c_{\sf E} < 0$ [cf. Eq.~(\ref{eq:ctheta})].
An example of a typical spin configuration is shown in Fig.~(\ref{fig:psi3}). 


Taken together $\Psi_2$ and $\Psi_3$ form a complete basis for the 
$\sf E$ irrep of ${\sf T}_d$.


\subsection*{Palmer-Chalker phase [$\Psi_4$] with ${\sf T_2}$ symmetry}

In a region bounded by 
\mbox{$a_{\sf T_2} = a_{\sf T_{1,A^{\prime}}} $} [Eq.~(\ref{eq:boundary-T2-T1})] 
and \mbox{$a_{\sf T_2} = a_{\sf E}$} [Eq.~(\ref{eq:boundary-T2-E})] 
--- cf.  Fig. \ref{fig:phase} ---
the energy
is minimised by setting
\begin{eqnarray}
{\bf m}_{\sf T_2}^2=1
\end{eqnarray}
and
\begin{eqnarray}
m_{\sf A_2}={\bf m}_{\sf E}=
 {\bf m}_{\sf T_{1 A^{\prime}}}={\bf m}_{\sf T_{1 B^{\prime}}}=0
\end{eqnarray}


The constraints on the total length of the spin, Eq.~(\ref{eq:spinconstraints}) 
further imply that 
\begin{eqnarray}
{\bf m}_{\sf T_2}^2=1 \\
m_{\sf T_2}^y m_{\sf T_2}^z=0 \\
m_{\sf T_2}^x m_{\sf T_2}^z=0 \\
m_{\sf T_2}^x m_{\sf T_2}^y=0
\end{eqnarray}
giving us a set of 6 ground states
\begin{eqnarray}
{\bf m}_{\sf T_2}=
\begin{pmatrix}
\pm 1 \\
0 \\
0
\end{pmatrix},
\begin{pmatrix}
0 \\
\pm1 \\
0
\end{pmatrix},
\begin{pmatrix}
0 \\
0 \\
\pm 1
\end{pmatrix}.
\end{eqnarray}

\begin{figure}
\centering
\subfigure[]{%
  \includegraphics[width=.4\linewidth]{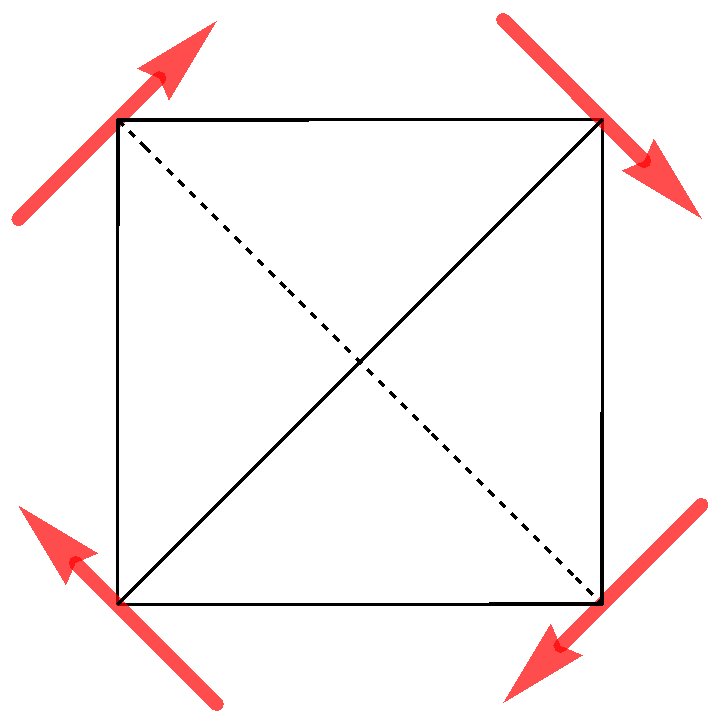}}%
    \qquad
\subfigure[]{%
  \includegraphics[width=.4\linewidth]{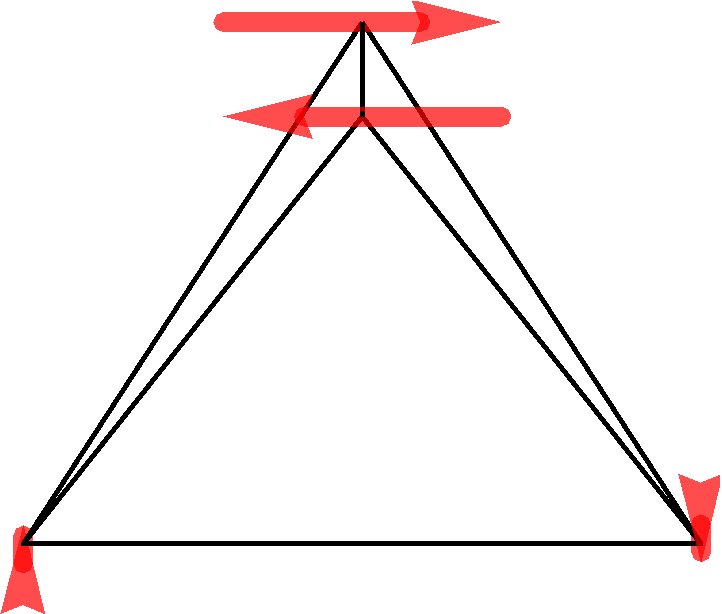}}%
\caption{
Spin configuration in the 4-sublattice Palmer-Chalker phase, $\Psi_4$, 
transforming with the ${\sf T_2}$ irrep of ${\sf T_d}$~:
(a) viewed along $[001]$ axis; 
(b) viewed slightly off the $[110]$ axis.
At the phase boundary with the $\Psi_2$ phase, each of the six Palmer-Chalker 
ground states can be transformed continuously into a $\Psi_2$ state.
\label{fig:PCphase}}
\end{figure}

Within these ground states spins are arranged in helical manner in a 
common $[100]$ plane, with a typical spin configuration given by 
(see Fig.~\ref{fig:PCphase}).
\begin{eqnarray}
\mathbf{S}_0&= & S (\sqrt{2}/2,-\sqrt{2}/2,0)  \nonumber \\
\mathbf{S}_1&= & S (-\sqrt{2}/2,-\sqrt{2}/2,0) \nonumber \\
\mathbf{S}_2&= & S (\sqrt{2}/2,\sqrt{2}/2,0)   \nonumber \\
\mathbf{S}_3&= & S (-\sqrt{2}/2,\sqrt{2}/2,0)
\label{eq:pcgs-2}
\end{eqnarray}


This phase is the ``Palmer-Chalker" phase, first
identified as the ground state of a model with antiferromagnetic
nearest neighbour Heisenberg interactions and long range
dipolar interactions on the pyrochlore lattice \cite{palmer00}.

\section{Ground-state degeneracy on classical phase boundaries}

\subsection{Boundary between Palmer-Chalker phase and the one-dimensional 
                    manifold of states with ${\sf E}$ symmetry}

The boundary between the Palmer-Chalker phase and the one-dimensional manifold 
of states with ${\sf E}$ symmetry occurs when $a_{\sf E} = a_{\sf T_2}$
[cf. Eq.~(\ref{eq:boundary-T2-E})].  
In this case, ${\mathcal H'}_{\sf ex}^{[{\sf T_d}]}$~[Eq.~(\ref{eq:H'Td})]
is minimised by setting 
\begin{eqnarray}
{\bf m}_{\sf E}^2+{\bf m}_{\sf T_2}^2=1
\end{eqnarray}
and
\begin{eqnarray}
m_{\sf A_2}={\bf m}_{\sf T_{1 A^{\prime}}}={\bf m}_{\sf T_{1 B^{\prime}}}=0 .
\end{eqnarray}
Substituting from Eq.~(\ref{eq:thetaE}), and imposing the constraint 
Eq.~(\ref{eq:spinconstraints}), we find
\begin{eqnarray}
2 m_{\sf E} m_{\sf T_2}^x \sin(\theta_{\sf E})-m_{\sf T_2}^y m_{\sf T_2}^z=0 \nonumber \\
2 m_{\sf E} m_{\sf T_2}^y \sin\left(\theta_{\sf E}-\frac{2 \pi}{3}\right)
-m_{\sf T_2}^x m_{\sf T_2}^z=0 \nonumber \\
2 m_{\sf E} m_{\sf T_2}^z \sin\left(\theta_{\sf E}+\frac{2 \pi}{3}\right)-
m_{\sf T_2}^x m_{\sf T_2}^y=0.
\label{eq:U1PCconstraints}
\end{eqnarray}


\begin{figure}
\begin{centering}
\includegraphics[width=.8\columnwidth]{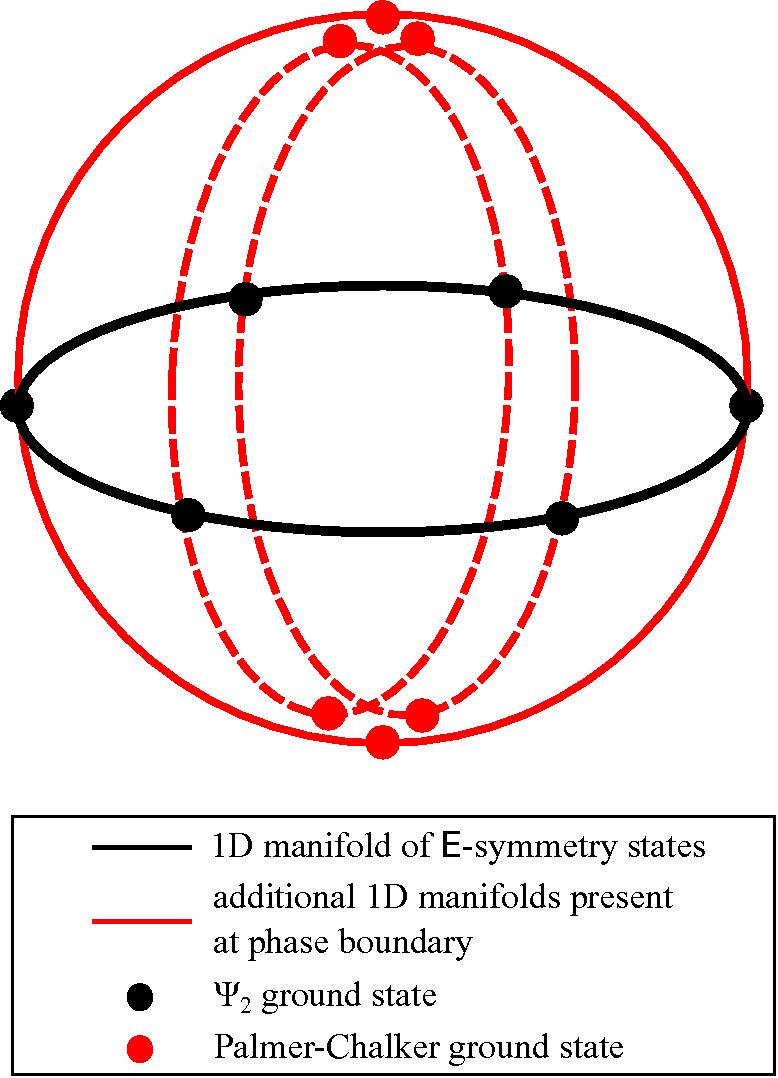}
\par\end{centering}
\caption{
\label{fig:manifold-T2-E} 
Structure of the ground state manifold at the boundary between the 
Palmer-Chalker (PC) phase and the one-dimensional manifold of 
states with ${\sf E}$ symmetry. 
The black circle denotes the manifold of ${\sf E}$--symmetry ground 
states, including the six $\Psi_2$ ground states (black dots).
At the boundary with the PC phase, this manifold branches at 
the $\Psi_2$ states, to connect with three, additional, one-dimensional 
manifolds.
These manifolds in turn interpolate to the six Palmer-Chalker 
ground states with ${\sf T_2}$ symmetry (red dots).
An exactly equivalent picture holds on the boundary between the non-collinear 
ferromagnet (FM), and the one-dimensional manifold 
of states with ${\sf E}$ symmetry. 
However in this case the different manifolds intersect at the $\Psi_3$ states.
}
\end{figure}


It is easy to show that there are no solutions to Eqs.~(\ref{eq:U1PCconstraints})
where more than one component of ${\bf m}_{\sf T_2}$ is finite.
There are, however, three distinct one-dimensional manifolds which connect pairs of 
Palmer-Chalker states to the one-dimensional manifold of $\sf E$-symmetry states:
\begin{eqnarray}
{\bf m}_{\sf E} 
   = \cos(\alpha) 
     \begin{pmatrix}
        1 \\
        0
    \end{pmatrix}, 
    \
{\bf m}_{\sf T_2}
   = \sin(\alpha) 
       \begin{pmatrix}
        1 \\
        0 \\
        0
      \end{pmatrix} 
\label{eq:spoke1}
\\
{\bf m}_{\sf E} 
   = \cos(\beta)
        \begin{pmatrix}
           -\frac{1}{2} \\
           \frac{\sqrt{3}}{2}
      \end{pmatrix}, 
      \
{\bf m}_{\sf T_2}
   = \sin(\beta) 
        \begin{pmatrix}
             0 \\
             1 \\
             0
         \end{pmatrix} 
\label{eq:spoke2}
\\
{\bf m}_{\sf E} 
    = \cos(\gamma)
         \begin{pmatrix}
            -\frac{1}{2} \\
           -\frac{\sqrt{3}}{2}
        \end{pmatrix}, 
        \
{\bf m}_{\sf T_2}
      = \sin(\gamma) \begin{pmatrix}
            0 \\
            0 \\
            1
         \end{pmatrix} 
\label{eq:spoke3}
\end{eqnarray}
where the angles $\alpha$, $\beta$ and $\gamma$ run from $0$ to $2\pi$.  


A typical spin configuration for one of the three connecting manifolds is
\begin{eqnarray}
{\bf S}_{0} & = & S \sqrt{\frac{2}{3}} \left( -\cos(\alpha),
   \:\cos \left( \alpha+\frac{\pi}{3} \right),
   \:\cos \left( \alpha-\frac{\pi}{3} \right) \right) \nonumber\\
{\bf S}_{1} & = & S \sqrt{\frac{2}{3}} \left( -\cos(\alpha),
   \:-\cos \left( \alpha+\frac{\pi}{3} \right),
   \:-\cos \left( \alpha-\frac{\pi}{3} \right) \right) \nonumber\\
{\bf S}_{2} & = & S \sqrt{\frac{2}{3}} \left( \cos(\alpha),
   \:\cos \left( \alpha-\frac{\pi}{3} \right),
   \:-\cos \left( \alpha+\frac{\pi}{3} \right) \right),\nonumber\\
{\bf S}_{3} & = & S \sqrt{\frac{2}{3}} \left( \cos(\alpha),
   \:-\cos \left( \alpha-\frac{\pi}{3} \right),
   \:\cos \left( \alpha+\frac{\pi}{3} \right) \right) .\label{con1}
\end{eqnarray}
where $\alpha = 0$ corresponds to the $\Psi_2$ ground state
with $\theta_{\sf E} = 0$, and $\alpha = \pi/2$ to one of the six 
Palmer-Chalker ground states.
These manifolds are illustrated in Fig.~\ref{fig:manifold-T2-E}.

\subsection{Boundary between the non-collinear ferromagnet and the 
                   one-dimensional manifold of states with ${\sf E}$ symmetry}

The boundary between the Palmer-Chalker phase and the one-dimensional manifold 
of states with ${\sf E}$ symmetry occurs when $a_{\sf E} = a_{\sf T_{1, A^\prime}}$
[cf. Eq.~(\ref{eq:boundary-E-T1})].  
In this case, ${\mathcal H'}_{\sf ex}^{[{\sf T_d}]}$~[Eq.~(\ref{eq:H'Td})]
is minimised by setting 
\begin{eqnarray}
{\bf m}_{\sf T_{1,A'}}^2 + {\bf m}_{\sf E}^2 = 1
\end{eqnarray}
and
\begin{eqnarray}
m_{\sf A_2}  = {\bf m}_{\sf T_{1,B'}}  = {\bf m}_{\sf T_2}  =  0.
\end{eqnarray}
Imposing the constraint Eq. \ref{eq:spinconstraints} we obtain
\begin{eqnarray}
2 m_{\sf E} m_{\sf T_1}^x \cos(\theta_{\sf E})
    &=& - \frac{\mu(\theta_{\sf T_1})}{\nu(\theta_{\sf T_1})}
            m^y_{\sf {T_1 B^{\prime}}} m^z_{\sf {T_1 B^{\prime}}} 
            \nonumber \\
2 m_{\sf E} m_{\sf T_1}^y \cos\left(\theta_{\sf E}-\frac{2\pi}{3}\right)
    &=& - \frac{\mu(\theta_{\sf T_1})}{\nu(\theta_{\sf T_1})}
           m^x_{\sf {T_1 B^{\prime}}} m^z_{\sf {T_1 B^{\prime}}} 
           \nonumber \\
           2 m_{\sf E} m_{\sf T_1}^z \cos\left(\theta_{\sf E}+\frac{2\pi}{3}\right)
   &=& - \frac{\mu(\theta_{\sf T_1})}{\nu(\theta_{\sf T_1})} 
           m^x_{\sf {T_1 B^{\prime}}} m^y_{\sf {T_1 B^{\prime}}} 
           \nonumber \\
\end{eqnarray}
where $\theta_{\sf FM}$ is the (fixed) canting angle [Eq. (\ref{eq:FMangle})],
$\theta_{\sf E}$ is the (variable) angle within the U(1) manifold
[Eq. (\ref{eq:thetaE})].
For the parameters considered here, $\mu$ and $\nu$ are always 
finite and are given by
\begin{eqnarray}
\mu(\theta_{\sf T_1}) 
    &=& (\sqrt{2} \cos(\theta_{\sf T_1})-\sin(\theta_{\sf T_1})) 
            \nonumber  \\
\nu(\theta_{\sf T_1})
     &=& (\sin(\theta_{\sf T_1})^2+\sqrt{2}\sin(2 \theta_{\sf T_1}))
     \label{eq:munu}
\end{eqnarray}


Arguments identical to those developed for the boundary with
the Palmer-Chalker phase, give us three further 1D manifolds
addition to that associated with the $\sf E$ phase.
However the intersections of the manifolds are now located at 
$\theta_{\sf E}=\frac{2 (n+1) \pi}{6}$,  corresponding
to the $\Psi_3$ states.
This explaining the model's general entropic preference 
for $\Psi_3$ states in the region proximate to the 
ferromagnetic phase.


A typical spin configuration for one of the three connecting manifolds,
parameterised by an angle $\eta$ is
\begin{eqnarray}
{\bf S}_{0} & = &  S \bigg( \cos(\theta_{\sf T_1}) \sin(\eta),
   \:\frac{1}{\sqrt{2}}(-\cos(\eta) +\sin(\eta) \sin(\theta_{\sf T_1}) ), \nonumber \\
&&\frac{1}{\sqrt{2}}(\cos(\eta) +\sin(\eta) \sin(\theta_{\sf T_1}) \bigg) \nonumber\\
   {\bf S}_{1} & = &  S \bigg( \cos(\theta_{\sf T_1}) \sin(\eta),
   \:\frac{1}{\sqrt{2}}(\cos(\eta) -\sin(\eta) \sin(\theta_{\sf T_1}) ), \nonumber \\
&&\frac{1}{\sqrt{2}}(-\cos(\eta) -\sin(\eta) \sin(\theta_{\sf T_1}) \bigg)\nonumber\\
   {\bf S}_{2} & = &   S \bigg( \cos(\theta_{\sf T_1}) \sin(\eta),
   \:\frac{1}{\sqrt{2}}(-\cos(\eta) -\sin(\eta) \sin(\theta_{\sf T_1}) ), \nonumber \\
&&\frac{1}{\sqrt{2}}(-\cos(\eta) +\sin(\eta) \sin(\theta_{\sf T_1}) \bigg)\nonumber\\
   {\bf S}_{3} & = & S \bigg( \cos(\theta_{\sf T_1}) \sin(\eta),
   \:\frac{1}{\sqrt{2}}(\cos(\eta) +\sin(\eta) \sin(\theta_{\sf T_1}) ), \nonumber \\
&&\frac{1}{\sqrt{2}}(\cos(\eta)-\sin(\eta) \sin(\theta_{\sf T_1}) \bigg).\label{conFMU1}
\end{eqnarray}
Here $\eta = 0$ corresponds to the $\Psi_3$ ground state with 
$\theta_{\sf E} = \pi/2$, and $\eta = \pi/2$ to one of the six FM ground states.

\subsection{Boundary between the Palmer-Chalker phase and non-collinear ferromagnet}

The boundary between the Palmer-Chalker phase and the non-collinear ferromagnet 
occurs when $a_{\sf T_2} = a_{\sf T_{1, A^{\prime}}}$ 
[cf. Eq.~(\ref{eq:boundary-T2-T1})].  
In this case, ${\mathcal H'}_{\sf ex}^{[{\sf T_d}]}$~[Eq.~(\ref{eq:H'Td})]
is minimised by setting 
\begin{eqnarray}
&&{\bf m}_{\sf T_2}^2+{\bf m}_{\sf T_{1, A^{\prime}}}^2=1 \nonumber \\
\end{eqnarray}
and
\begin{eqnarray}
m_{\sf A_2}={\bf m}_{\sf E}={\bf m}_{\sf T_{1 B'}}=0.
\end{eqnarray}


Imposing the constraint Eq. \ref{eq:spinconstraints} we obtain
\begin{eqnarray}
&&-m_{\sf T_2}^y m_{\sf T_2}^z + 
(\sin( \theta_{\sf T_1})^2
+\sqrt{2}\sin(2 \theta_{\sf T_1})) m_{\sf T_{1 A'}}^y  m_{\sf T_{1 A'}}^z  \nonumber \\
&& \ \ \ + (\sqrt{2} \cos( \theta_{\sf T_1})
-\sin( \theta_{\sf T_1})) ({\bf m_{\sf T_{1 A'}}} \times {\bf m_{\sf T_2}})_x
=0 \nonumber \\
&&-m_{\sf T_2}^x m_{\sf T_2}^z + 
(\sin( \theta_{\sf T_1})^2
+\sqrt{2}\sin(2  \theta_{\sf T_1})) m_{\sf T_{1 A'}}^x  m_{\sf T_{1 A'}}^z  \nonumber \\
&& \ \ \ + (\sqrt{2} \cos( \theta_{\sf T_1})-\sin( \theta_{\sf T_1})) 
({\bf m_{\sf T_{1 A'}}} \times {\bf m_{\sf T_2}})_y
=0 \nonumber \\
&&-m_{\sf T_2}^x m_{\sf T_2}^y + 
(\sin( \theta_{\sf T_1})^2
+\sqrt{2}\sin(2  \theta_{\sf T_1})) m_{\sf T_{1 A'}}^x  m_{\sf T_{1 A'}}^y  \nonumber \\
&& \ \ \ + (\sqrt{2} \cos( \theta_{\sf T_1})-\sin( \theta_{\sf T_1})) 
({\bf m_{\sf T_{1 A'}}} \times {\bf m_{\sf T_2}})_z
=0 \nonumber \\
\label{eq:FMPCconstraints}
\end{eqnarray}
where $\theta_{\sf T_1}$ is defined in Eq. (\ref{eq:FMangle}).


In general, the ground state manifold on the boundary of the Palmer-Chalker phase 
is two-dimensional.
To establish this, we consider small deviations from a given solution 
\begin{eqnarray}
{\bf m}_{\sf T_2} &=& {\bf m}_{\sf T_2}^0 + {\bf \delta m}_{\sf T_2} \nonumber \\
{\bf m}_{\sf T_{1 A^{\prime}}} &=& {\bf m}_{\sf T_{1 A^{\prime}}}^0 + {\bf \delta m}_{\sf T_{1 A^{\prime}}}
\end{eqnarray}
and expand the constraint Eq.~(\ref{eq:FMPCconstraints}) to linear order 
in ${\bf \delta m}$.  
Generally, we find two linearly-independent solutions for 
$({\bf \delta m}_{\sf T_2},{\bf \delta m}_{\sf T_{1 A^{\prime}}})$, 
and the manifold in the vicinity of 
$({\bf m}_{\sf T_2}^0, {\bf m}_{\sf T_{1 A^{\prime}}}^0)$
is locally two-dimensional.


However if we expand around a state 
$(\tilde{\bf m}_{\sf T_2}^0, \tilde{\bf m}_{\sf T_{1 A^{\prime}}}^0)$
where {\it both} order parameters are aligned with the same cubic axis, e.g.
\begin{eqnarray}
\tilde{m}_{\sf T_2}^{0y} 
= \tilde{m}_{\sf T_2}^{0z} 
= \tilde{m}_{\sf T_{1 A^{\prime}}}^{0y}
= \tilde{m}_{\sf T_{1 A^{\prime}}}^{0z}
= 0
\end{eqnarray}
one of the Eqs.~(\ref{eq:FMPCconstraints}) is satisfied trivially, 
leaving only three constraints on six variables.
It follows that the manifold is locally three-dimensional in the vicinity of 
$(\tilde{\bf m}_{\sf T_2}^0, \tilde{\bf m}_{\sf T_{1 A^{\prime}}}^0)$.\\

On a final note, the emergent degeneracies observed in presence of inverse Dzyaloshinskii-Moriya
interactions~\cite{canals08,chern-arxiv} can be described in the same way when turning $J_4$ positive.

\section{Classical low-temperature expansion}
\label{low-T}


\begin{figure*}
\includegraphics[width=0.6\textwidth,height=9cm]{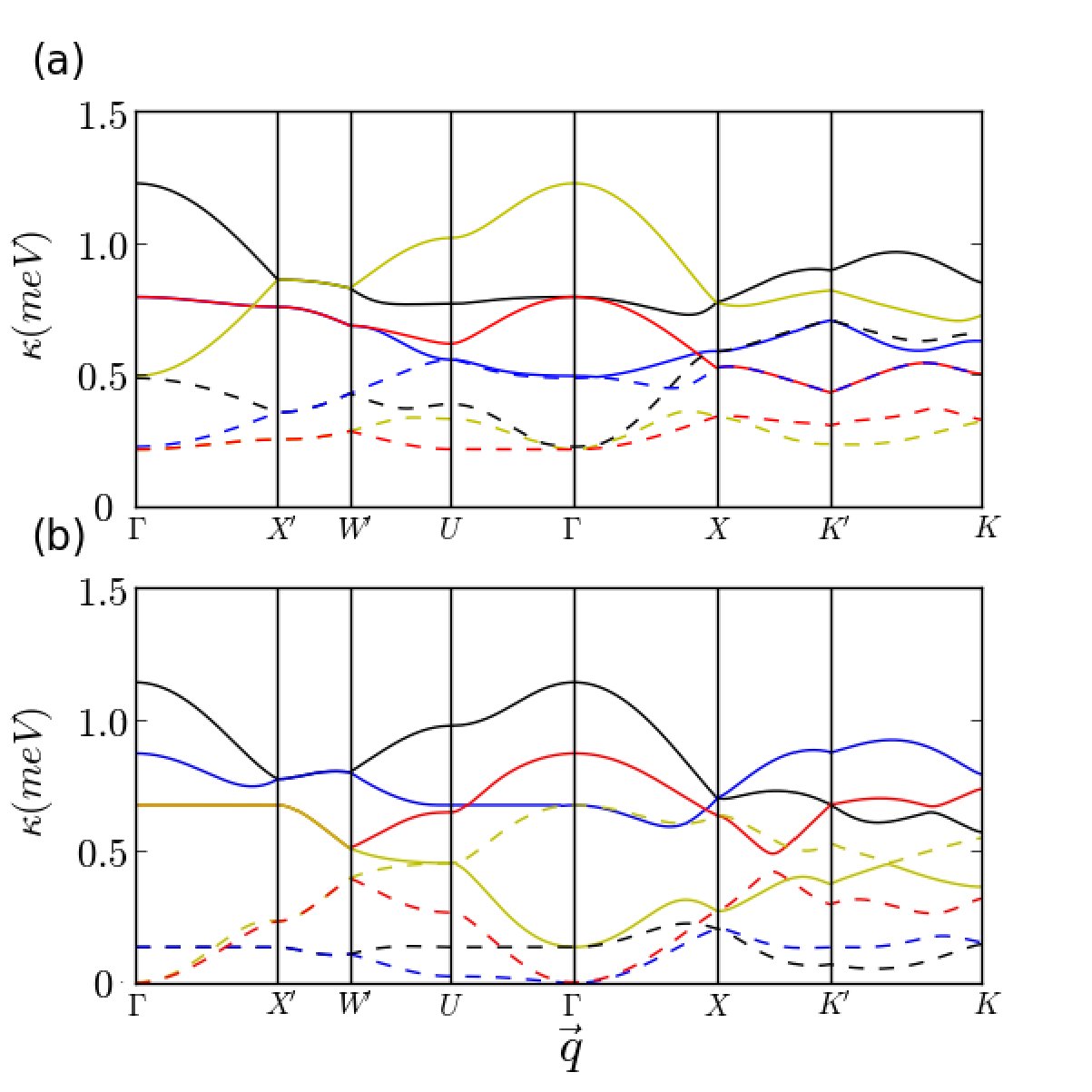}
\caption{
Spin-wave dispersion calculated within a classical, low-temperature 
expansion, showing dimensional reduction of a subset of excitations.
(a) Excitations of the FM ground state, 
for exchange parameters appropriate to Yb$_2$Ti$_2$O$_7$, 
i.e. 
\mbox{$J_1=-0.09 \text{meV}$}, 
\mbox{$J_2=-0.22 \text{meV}$}, 
\mbox{$J_3=-0.29 \text{meV}$},
setting 
$J_4=0$.  
The ferromagnet possesses a flat band in the $(h, h, h)$ ($\Gamma \to U$) 
direction at energy $\Delta=0.22$ meV, which gives rise to rods in the equal 
time structure factor (cf. Fig.~5, main text).
(b) Excitations of the $\Psi_3$ ground state, for exchange parameters 
on the boundary between the $\Psi_3$ and FM phases,  
i.e. 
\mbox{$J_1=-0.029 \text{meV}$}, 
\mbox{$J_2=-0.22 \text{meV}$}, 
\mbox{$J_3=-0.29 \text{meV}$}
with
$J_4=0$.  
The $\Psi_3$ phase, on the phase boundary also possesses 
a quasi-flat band along $(h, h, h)$, which in this case is gapless 
at the $\Gamma$ point of the Brillouin zone.
This leads us to suggest that the low energy rod like 
features observed in the paramagnetic
phase of Yb$_2$Ti$_2$O$_7$ arise from its proximity
in parameter space to the $\Psi_3$ phase
and the low energy modes 
which are present on the phase boundary.
\label{fig:dispersionsYTO}
}
\end{figure*}


We have performed a classical, low-temperature \mbox{(low-T)} spin-wave expansion 
around each of the three different ordered phases identified in the main text.
This has a number of applications.
Firstly, by calculating the entropy associated with each of the ${\sf E}$ symmetry 
states, we can determine the phase boundary between the  $\Psi_2$ and $\Psi_3$ 
ground states in the limit $T \to 0$ [cf. Fig.~1, main text].
Secondly, knowledge of the spin wave dispersion provides further insight into the 
nature of the degeneracies where phases with different symmetries meet.
And thirdly, by using the low-T expansion to calculate the structure factor
$S({\bf q})$, we can link correlation functions measured in experiment 
explicitly to the (classical) spin-wave spectrum.
Finally, the low-T expansion also provides a useful benchmark for classical Monte Carlo 
simulations, particularly at low temperatures, where simulations are hard to equilibrate.


We define a local co-ordinate system by introducing a set of orthogonal 
unit vectors $\{ {\bf u}_i,  {\bf v}_i,  {\bf w}_i \}$ for each of the four
sublattices $i=0,1,2,3$ [cf. Fig.~\ref{fig:spindef}].
The local ``$z$-axis'', ${\bf w}_i$, is chosen to be aligned with the
spins in a given four-sublattice ground state
\begin{eqnarray}
{\bf S}_i= S {\bf w}_i \quad \forall \ i
\end{eqnarray}
The remaining unit vectors, ${\bf u}_i$ and  ${\bf v}_i$,  are only determined
up to a rotation about ${\bf w}_i$, and any convenient choice can be made.


Using this basis, the fluctuations on site $i$ of tetrahedron $k$ can be 
parameterized as
\begin{eqnarray}
{\bf S}_{ik}&=& 
\begin{pmatrix}
\sqrt{S} \delta u_{ik} \\
\sqrt{S} \delta v_{ik} \\
\sqrt{S^2- S \delta u_{ik}^2-S \delta v_{ik}^2}
\end{pmatrix} \nonumber \\
&\approx&
\begin{pmatrix}
\sqrt{S} \delta u_{ik} \\
\sqrt{S} \delta v_{\vec{ik}} \\
S - \frac{1}{2}  \delta u_{ik}^2- \frac{1}{2}  \delta v_{ik}^2
\end{pmatrix}.
\label{eq:fluc}
\end{eqnarray}
Substituting Eq.~(\ref{eq:fluc}) into $\mathcal{H}_{\sf ex}$ [Eq.~(\ref{eq:Hex})]
we obtain
\begin{eqnarray}
\mathcal{H}_{\sf ex}  
   &=& \sum_{{\sf tet} \  k} \sum_{i < j} {\bf S}_{ik} \cdot {\bf J}_{ij}\cdot {\bf S}_{jk} \nonumber \\
   &=& {\mathcal E}_0 +  \mathcal{H}_{\sf ex}^{CSW} + \ldots
 \end{eqnarray}
where
\begin{eqnarray}
{\mathcal E}_0  &=& \frac{N S^2}{4} \sum_{i,j=0}^{3} {\bf w}_i \cdot {\bf J}_{ij} \cdot {\bf w}_j 
\label{eq:E0}
\end{eqnarray}
is the classical ground-state energy of the chosen \mbox{4-sublattice} state, 
and %
\begin{eqnarray}
\mathcal{H}_{\sf ex}^{\sf CSW}
 &=&  \frac{S}{2} \sum_k \sum_{i, j=0}^3 \nonumber \\
&&   \bigg[
 - \frac{1}{2} (\delta u_{ik}^2 +\delta u_{jk}^2 +\delta v_{ik}^2 +\delta v_{jk}^2)
  \left( {\bf w}_i \cdot {\bf J}_{ij} \cdot {\bf w}_j\right) \nonumber \\
 &+& \ \delta u_{i k} \delta u_{j k} \left( {\bf u}_i \cdot {\bf J}_{ij} \cdot {\bf u}_j \right)
 +  \delta v_{i k} \delta v_{j k} \left( {\bf v}_i \cdot {\bf J}_{ij} \cdot {\bf v}_j \right) \nonumber \\
 &+&  \delta u_{i k} \delta v_{j k} \left( {\bf u}_i \cdot {\bf J}_{ij} \cdot {\bf v}_j \right) 
 +  \delta v_{i k} \delta u_{j k} \left( {\bf v}_i \cdot {\bf J}_{ij} \cdot {\bf u}_j \right)
  \bigg] \nonumber \\
  \label{eq:Hfluc}
\end{eqnarray}
describes the leading effect of (classical) fluctuations about this state.  
Performing Fourier transformation, we find 
\begin{eqnarray}
\mathcal{H}_{\sf ex}^{\sf CSW}
   &=& \frac{N S^2}{4} \sum_{i,j=0}^{3} 
         {\bf w}_i \cdot {\bf J}_{ij} \cdot {\bf w}_j 
\nonumber \\
                  &+& \frac{1}{2}  \sum_{{\bf q}} \tilde{u}(-{\bf q})^T \cdot {\bf M}({\bf q})
                 \cdot \tilde{u}({\bf q})
\end{eqnarray}
Here $ \tilde{u}({\bf q})$ is the vector 
\begin{eqnarray}
\tilde{u}({\bf q})=\bigg( \delta \tilde{u}_0 ({\bf q}), 
\delta u_1 ({\bf q}),
\delta u_2 ({\bf q}),
\delta u_3 ({\bf q}), \nonumber \\
\delta v_0 ({\bf q}),
\delta v_1 ({\bf q}),
\delta v_2 ({\bf q}),
\delta v_3 ({\bf q}) \bigg)^T,
\end{eqnarray}
and ${\bf M}({\bf q})$ the $8 \times 8$ matrix 
\begin{eqnarray}
{\bf M}({\bf q})=
2 S \begin{pmatrix}
{\bf M}^{11}({\bf q})
& {\bf M}^{12}({\bf q}) \\
{\bf M}^{21}({\bf q})
& {\bf M}^{22}({\bf q})  \\
  \end{pmatrix}  
  \label{eq:M}
\end{eqnarray}
built $4 \times 4$ blocks 
\begin{eqnarray}
{\bf M}^{11}_{ij}({\bf q})
   &=&\cos({\bf q} \cdot {\bf r}_{ij} )  \nonumber \\
&&\bigg( {\bf u}_i \cdot {\bf J}_{ij} \cdot {\bf u}_j 
 - \delta_{ij}  \sum_{l} \left(  {\bf w}_l \cdot {\bf J}_{lj} \cdot {\bf w}_j   \right)  
\bigg) \nonumber \\
\\
{\bf M}^{12}_{ij}({\bf q})
   &=&{\bf M}^{21}_{ji}({\bf q})=\cos({\bf q} \cdot {\bf r}_{ij} ) 
\bigg( {\bf v}_i \cdot {\bf J}_{ij} \cdot {\bf u}_j  
\bigg) \\
{\bf M}^{22}_{ij}({\bf q})
   &=&\cos({\bf q} \cdot {\bf r}_{ij} ) \nonumber \\
&& \bigg( {\bf v}_i \cdot {\bf J}_{ij} \cdot {\bf v}_j 
 - \delta_{ij}  \sum_{l} \left(  {\bf w}_l \cdot {\bf J}_{lj} \cdot {\bf w}_j   \right)  
\bigg) \nonumber \\
\end{eqnarray}
where \mbox{$i, j  \in \{0, 1, 2,  3\}$} and \mbox{${\bf r}_{ij} = {\bf r}_j - {\bf r}_i$} 
[cf. Eq.~(\ref{eq:r})].


The matrix ${\bf M}({\bf q})$ [Eq.~(\ref{eq:M})] can be diagonalized 
by a suitable orthogonal transformation, \mbox{${\bf U}=({\bf U}^T)^{-1}$} to give
\begin{equation}
{\mathcal H}_{\sf ex}^{\sf CSW}
           =  \frac{1}{2}  \sum_{{\bf q}} \sum_{\nu=1}^{8}
           \kappa_{\nu {\bf q}} \upsilon_{\nu {\bf q}}  
           \upsilon_{\nu -{\bf q}}
           \label{eq:HlowT}
\end{equation}
where the eight normal modes of the system are given by 
$\upsilon({\bf q})={\bf U} \cdot \tilde{u}({\bf q})$ with associated eigenvalues 
$\kappa_{\nu} ({\bf q})$.
Since ${\mathcal H}_{\sf ex}^{\sf CSW}$ [Eq.~(\ref{eq:HlowT})] is quadratic in 
$\upsilon_{\nu {\bf q}}$, the associated partition function can be calculated
exactly
\begin{eqnarray}
\mathcal{Z}^{\sf CSW}_{\sf ex} &=& \frac{1}{\sqrt{2 \pi}}\int \prod_{\nu=1}^{8} \prod_{{\bf q}}
d\upsilon_{\nu {\bf q}}  \nonumber \\
&&\exp{\left( -\frac{1}{2}
  \frac{\sum_{\nu=1}^{8} \sum_{{\bf q}} \kappa_{\nu {\bf q}} \upsilon_{\nu {\bf q}} 
   \upsilon_{i -{\bf q}}}{T}
 \right)} \nonumber \\
 & =& \prod_{\nu=1}^{8} \prod_{{\bf q}} \left( \sqrt{\frac{T}{\kappa_{\nu {\bf q}}}}  \right). 
  \label{eq:partitionfunction}
\end{eqnarray}
It follows that, for $T \to 0$, the free energy of the system is given by
\begin{eqnarray}
\mathcal{F}_{\sf ex}^{\sf low-T} &=& 
   {\mathcal E}_0 
   - \frac{T}{2} \sum_{\nu {\bf q}}  \ln \kappa_{\nu {\bf q}} 
   + N (T \ln T + T)
   + {\mathcal O}(T^2)   \nonumber \\
   \label{eq:FlowT}
\end{eqnarray}


Within this classical, low-T expansion, the eigenvalues $\kappa_{\nu} ({\bf q})$ 
correspond to independent, low energy modes, which determine the 
physical properties of the states, and have the interpretation of a classical 
spin-wave spectrum.
This is illustrated for the non-collinear FM, with parameters appropriate for
Yb$_2$Ti$_2$O$_7$, in Fig.~\ref{fig:dispersionsYTO}.
However the classical spectrum $\kappa_{\nu} ({\bf q})$ should {\it not} be confused 
with the semi-classical spin-wave dispersion $\omega_{\nu} ({\bf q})$ found in 
linear spin-wave theory, where quantum effects are included.

\section{Ground-state selection within the one-dimensional manifold 
                   of states with ${\sf E}$ symmetry}

Knowledge of the free energy $\mathcal{F}_{\sf ex}^{\sf low-T}$ [Eq.~(\ref{eq:FlowT})] makes 
it possible to determine which of possible ${\sf E}$ symmetry ground states
is selected by thermal fluctuations in the limit $T \to 0$.


Expanding the free energy in components of 
${\bf m}_{\sf E}$ [cf.~Eq.~(\ref{eq:thetaE})], we find 
\begin{eqnarray}
{\mathcal F}_{\sf E} 
   &=& {\mathcal F}_0
    +  \frac{1}{2} \; a \; m_{\sf E}^2 
    + \frac{1}{4} \; b \; m_{\sf E}^4 
    + \frac{1}{6} \; c \; m_{\sf E}^6 \nonumber\\   
  &&  + \frac{1}{6} \; d \; m_{\sf E}^6 \; \cos(6\,\theta_{\sf E})
    + {\mathcal O} (m_{\sf E}^8)
\label{eq:Landau}
\end{eqnarray}
where ${\mathcal F}_0$ is an unimportant constant.
It follows that (1) a suitable (secondary) order parameter for 
symmetry breaking within this manifold is $c_{\sf E} = \cos 6 \theta_{\sf E}$ 
[cf.~Eq.~(\ref{eq:ctheta})], and that 
(2) the two states spanning ${\bf m}_{\sf E}$, $\Psi_2$ and  $\Psi_3$, 
are distinguished only at sixth-order in 
$m_{\sf E}$ [\onlinecite{oitmaa-arXiv}].
These facts have important consequences for the finite temperature
phase transition into the paramagnet, as discussed below.


For $T \to 0$, we can parameterise ${\mathcal F}_{\sf E}$ [Eq.~(\ref{eq:Landau})]
from ${\mathcal F}_{\sf ex}^{\sf low-T}$ [Eq.~(\ref{eq:FlowT})].  
Since ${\mathcal H'}_{\sf ex}^{[{\sf T_d}]}$ ~[Eq.~(\ref{eq:HTd})] is
quadratic in ${\bf m}_{\sf E}$, all other terms in the free energy 
must be of purely entropic origin.
Moreover, from the form of ${\mathcal F}_{\sf E}$, we anticipate that the entropy 
associated with the ${\sf E}$--symmetry states will vary as 
\begin{eqnarray}
{\mathcal S}_{\sf E} (\theta_{\sf E})  
   &=& N \sum_{n=0, 1, 2\ldots} s_n \cos (6n \theta_{\sf E})
\end{eqnarray}
%


\begin{figure*}
\includegraphics[width=0.95\textwidth]{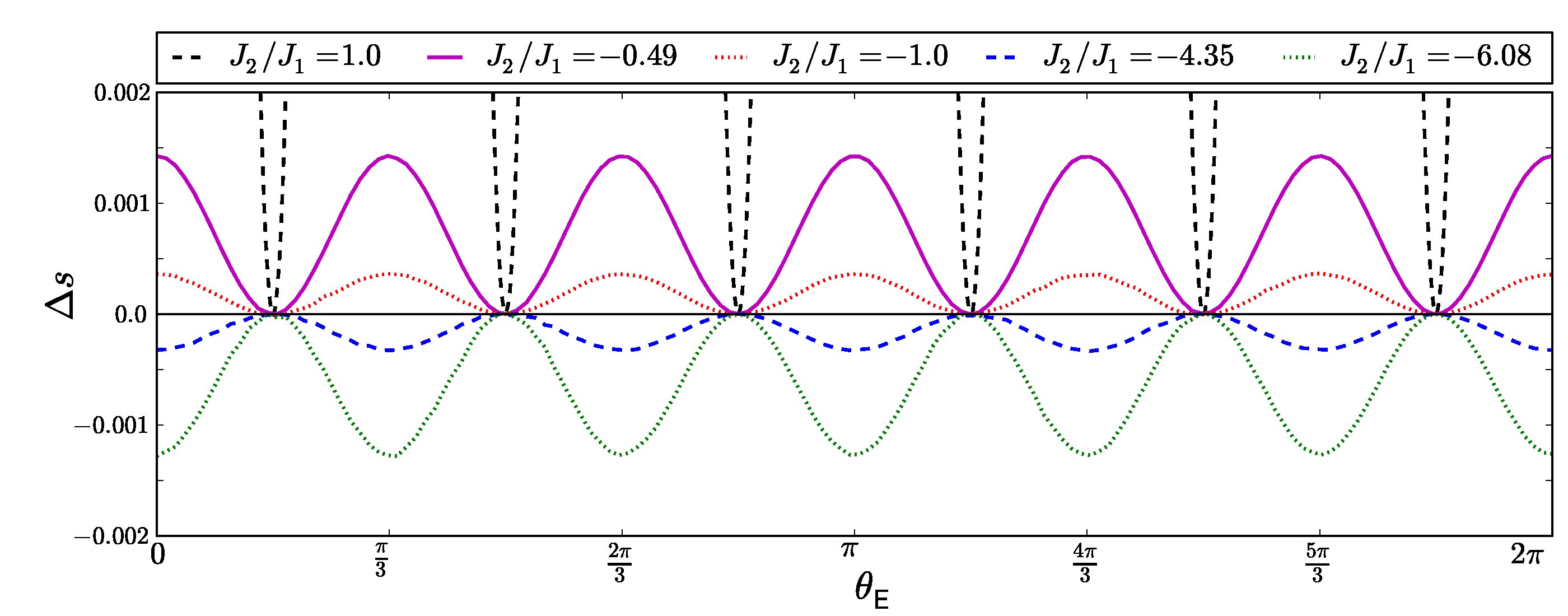}
\caption{
Variation of entropy per spin within the one-dimensional manifold of 
states with symmetry ${\sf E}$. 
Entropy ${\mathcal S}(\theta_{\sf E})$ has been estimated using the low-temperature 
expansion [Eq.~(\ref{eq:S})], for a range of values of $J_2$, with the entropy 
of the $\Psi_3$ state subtracted as a reference, i.e. 
$\Delta s_{\theta_{\sf E}} = [{\mathcal S}(\theta_{\sf E}) - {\mathcal S}(\pi/6)]/ N$.
The parameters $J_1=0.115 \text{meV}$ and $J_3=-0.099 \text{meV}$ were fixed 
at values appropriate to Er$_2$Ti$_2$O$_7$ [\onlinecite{savary12-PRL109}], 
setting $J_4 \equiv 0$.
In all cases, $\Delta s_{\theta_{\sf E}}$ repeats with period $2\pi/6$.  
For a choice of $J_2$ appropriate to Er$_2$Ti$_2$O$_7$  
[$J_2/J_1=-0.49$ --- solid purple line],  entropy takes on its maximum value for 
$\theta_{\sf E}=\frac{n \pi}{3}$, with $n=0,1,2,3,4,5$, corresponding to the six 
$\Psi_2$ ground states.  
The extreme variation in entropy at the boundary of the Palmer-Chalker phase 
[$J_1=J_2$ --- dashed black line], reflects the presence of an $\mathcal{O}(L^2)$ 
set of zero modes in the spectrum of $\Psi_2$ ground state.
None the less, the entropy difference between $\Psi_2$ and $\Psi_3$,  
$\Delta s_{\pi/3} \approx 0.18$ remains finite.
For sufficiently negative $J_2$ (dashed blue line, dotted yellow line) 
$\Delta  s_{\pi/3} < 0$, and fluctuations select the $\Psi_3$ state.
All results have been calculated from Eq.~(\ref{eq:S}), 
with the sum evaluated numerically by a Monte Carlo method.
Statistical errors are smaller than the point size.
}
\label{fig:entropy}
\end{figure*}


The sign of the coefficients $s_n$ then determines the ground state 
selected by fluctuations.
This expectation is confirmed by explicit calculation of 
\begin{eqnarray}
\frac{{\mathcal S}_{\sf E} (\theta_{\sf E})}{N} 
    = \ln{T} + 1 - \frac{1}{2 N} \sum_{{\bf q}} 
\ln{\left( \det({\bf M} ({\bf q})) \right)}
\label{eq:S}
\end{eqnarray}
[cf.~Ref.~\onlinecite{shannon10}], where ${\bf M}({\bf q})$ is the $8 \times 8$ matrix 
defined in Eq.~(\ref{eq:M}).
These results are illustrated in Fig.~\ref{fig:entropy}.
Equivalent calculations, carried out numerically for all parameters associated
with ${\sf E}$--symmetry ground states, lead to the phase boundary between 
$\Psi_2$ and  $\Psi_3$ shown in Fig.~1 of the main text.
For parameters appropriate to Er$_2$Ti$_2$O$_7$ [\onlinecite{savary12-PRL109}], 
we find that fluctuations select a $\Psi_2$ ground state, in keeping with earlier
published work on quantum fluctuations~\cite{mcclarty09,savary12-PRL109,zhitomirsky12}


We can now learn more about how ground state selection works by realising
that, for some choices of parameters, the 
operation connecting different ${\sf E}$--symmetry ground states 
becomes an {\it exact} symmetry of the Hamiltonian.
This is most easily seen by writing $\mathcal{H}_{\sf ex}$ 
[Eq.~(\protect\ref{eq:Hex})] in a coordinate frame tied to 
the local [111] axes.
Following the notation of [\onlinecite{ross11-PRX1}], 
\begin{eqnarray}
{\mathcal H}^\prime_{\sf ex} 
    &=& \sum_{\langle ij\rangle} 
               \Big\{ J_{zz} \mathsf{S}_i^z \mathsf{S}_j^z - J_{\pm}
                (\mathsf{S}_i^+ \mathsf{S}_j^- + \mathsf{S}_i^- \mathsf{S}_j^+) 
                \nonumber \\ 
       && + J_{\pm\pm} \left[\gamma_{ij} \mathsf{S}_i^+ \mathsf{S}_j^+ + \gamma_{ij}^*
                 \mathsf{S}_i^-\mathsf{S}_j^-\right] 
                 \nonumber \\
       && + J_{z\pm}\left[ 
                           \mathsf{S}_i^z (\zeta_{ij} \mathsf{S}_j^+ + \zeta^*_{ij} \mathsf{S}_j^-) 
                           + {i\leftrightarrow j}
                 \right]\Big\}
\label{eq:Hross}                 
\end{eqnarray}
where $\mathsf{S}_i^\alpha$ are the transformed spins, 
\begin{eqnarray}
J_{zz} &=& -\frac{1}{3} (2 J_1 - J_2 + 2(J_3 + 2J_4)) \\
J_{\pm} &=& \frac{1}{6} (2 J_1 - J_2 -J_3 - 2 J_4) \\
J_{\pm \pm} &=& \frac{1}{6} (J_1 + J_2 - 2 J_3 + 2 J_4) \\
J_{z \pm} &=& \frac{1}{3 \sqrt{2}} (J_1 + J_2 + J_3-J_4)
\end{eqnarray}
and the matrices 
\begin{equation}
\zeta 
    = \left(\begin{array}{cccc}
            0 & -1 & e^{i\frac{\pi}{3}} & e^{-i\frac{\pi}{3}}\\
           -1 & 0 & e^{-i\frac{\pi}{3}} & e^{i\frac{\pi}{3}}\\
            e^{i\frac{\pi}{3}} & e^{-i\frac{\pi}{3}} & 0 & -1\\
            e^{-i\frac{\pi}{3}} & e^{i\frac{\pi}{3}} & -1 & 0
      \end{array}\right)
    \quad 
       \gamma=-\zeta^*
\end{equation}
encode the rotations in co-ordinate frame between different sublattices.


For the simple choice 
$$
(J_{zz},\ J_{\pm},\ J_{\pm \pm},\ J_{z \pm}) 
    = (0,\  J,\ 0,\ 0) 
    \quad \quad J > 0
$$
the ground state belongs to ${\sf E}$ and
${\mathcal H}^\prime_{\sf ex}$~[Eq.~(\ref{eq:Hross})] reduces
to an $XY$ ferromagnet.  
In this case the entire one-dimensional manifold of 
${\sf E}$--symmetry states are connected by an explicit symmetry 
of the Hamiltonian (rotation around the local $\langle 111 \rangle $ axes).
It follows that order-by-disorder is ineffective, and 
the ground state retains its $U(1)$ symmetry 
--- for a related discussion, see [\onlinecite{wong13}].


To gain insight into the phase diagram for \mbox{$J_3 < 0,\ J_4 \equiv 0$}  
[cf.~Fig.~1, main text], we expand about a point in parameter space
\begin{eqnarray}
(J_{zz},\ J_{\pm},\ J_{\pm \pm},\ J_{z \pm}) 
   & =& (-2J,\ J,\ 0,\ 0)
    \quad \quad J > 0 \nonumber \\
\implies (J_1, J_2, J_3, J_4)  &=& (2J, -2J, 0, 0)\nonumber
\end{eqnarray}
where  ${\mathcal H}^\prime_{\sf ex}$~[Eq.~(\ref{eq:Hross})]
reduces to a Heisenberg ferromagnet.
At this point the ground state manifold is formed from linear
combinations of $E$ and $A_2$ symmetry states and all
the entire ground state manifold is connected by an exact symmetry
of the Hamiltonian, so one again there is no order by disorder.
For $J_3<0$ states with a finite value of $m_{\sf A_2}$
are removed from the ground state manifold and fluctuations select
a ground state from amongst the ${\sf E}$ states.
%
%
It follows that, for \mbox{$J_3 \to 0^{-},\ J_4 \equiv 0$},  the phase boundary 
between the $\Psi_2$ and $\Psi_3$ states should tend to the 
line \mbox{$J_2/|J_3| = - J_1/|J_3|$} [cf. Fig.~1, main text].  


To see which phase is preferred for finite $J_3$, we expand the 
difference in entropy $\mathcal{S}_{\sf E} (\theta_{\sf E})$ 
between the $\Psi_2$ and $\Psi_3$ ground states
\begin{eqnarray}
\Delta s_{\pi/3}
   = \frac{\mathcal{S}_{\sf E}(\pi/3) 
     - \mathcal{S}_{\sf E} (\pi/6)}{N} 
\end{eqnarray}
in powers of $J_{\pm \pm}$ and $J_{z \pm}$.  
We do this by writing the matrix ${\bf M}({\bf q})$ [Eq.~\ref{eq:M}] as
\begin{eqnarray}
{\bf M}({\bf q}) = {\bf M}_0({\bf q}) + \epsilon {\bf X}({\bf q})
\end{eqnarray}
where ${\bf M}_0({\bf q})$ is the matrix associated with the high-symmetry point,
and ${\bf X}({\bf q})$ that associated with the perturbation, and noting that 
\begin{eqnarray}
&& \ln( \det( {\bf M}_0 + \epsilon {\bf X})) 
   =  \ln( \det({\bf M}_0) ) \nonumber \\
&&
+ \sum_{n=1}^{\infty}
 (-1)^{(n+1)} \frac{\epsilon^n}{n}  
 \Tr \bigg[ \left(  {\bf X} \cdot {\bf M}_0^{-1} \right)^n \bigg].
 \end{eqnarray}
We then expand in powers of $J_{\pm\pm}$ and $J_{z\pm}$.


We find that the leading correction to $\Delta S$ is
\begin{eqnarray}
\Delta s_{\pi/3}
    \approx a \left( \frac{J_{\pm\pm}}{J_{\pm}} \right)^3
\end{eqnarray}
where $a=0.0045$.
%
%
It follows that, for sufficiently small $J_3$, the phase boundary between 
$\Psi_2$ and $\Psi_3$ should tend to the line $J_{\pm\pm} = 0$, with the 
$\Psi_2$ phase favoured for $J_{\pm\pm}>0$ and $\Psi_3$ favoured 
for $J_{\pm\pm}<0$.
Numerical evaluation of Eq. \ref{eq:S}, in the limit $J_3 \to 0$, 
yields results in agreement with these arguments [cf.~Fig.~1, main text].


On the line $J_{\pm\pm} = 0$ itself, we find that the leading correction 
to the difference in entropy is 
\begin{eqnarray}
\Delta s_{\pi/3} \approx b \left( \frac{J_{z\pm}}{J_{\pm}} \right)^6
\end{eqnarray}
with $b=-5.3 \times 10^{-5}$. 
Hence the $\Psi_3$ state is weakly preferred, and the phase boundary will bend 
towards positive $J_2/|J_3|$, as observed in Fig.~1 of the main text.
Since $J_{z\pm}$ is a term which drives out of plane fluctuations, a negative sign 
for $b$ is consistent with the argument that $\Psi_3$ is better connected to the 
ferromagnetic phase, and hence has a softer spectrum for out-of-plane fluctuations.


In the limit $|J_3| \gtrsim (|J_1|, |J_2|)$, numerical evaluation of Eq.~\ref{eq:S} 
yields the more complex, reentrant behaviour, as seen in Fig.~1 of the main text. 
This behaviour occurs over a very narrow region of parameter space, and is discussed 
in detail (for the case of quantum, as opposed to thermal order by disorder) 
in [\onlinecite{wong13}].

\section{Semiclassical spin wave theory}

The effect of quantum fluctuations on ordered states can be estimated with 
a conventional large-$S$ expansion.
The sublattice dependent basis $\{ {\bf u}_i, {\bf v}_i, {\bf w}_i\}$, 
previously introduced for classical spins [Eq.~(\ref{eq:fluc})], 
again provides a convenient starting point.
Working to leading order in Holstein-Primakoff bosons  
\mbox{$\big[ a_i^{\phantom \dagger} , a_j^{\dagger} \big] =  \delta_{ij}$}, 
we write 
\begin{eqnarray}
\label{eq:HPsw}
S^w_i 
   &=& S-a^{\dagger}_i a^{\phantom \dagger}_i 
   \\
    \label{eq:HPs+}
S^+_i 
    &=& S^u_i+iS^v_i 
    = (2 S - a^{\dagger}_i a^{\phantom \dagger}_i)^{1/2} a^{\phantom \dagger}_i 
\approx \sqrt{2S} 
       a_i^{\phantom \dagger} 
      \\
      \label{eq:HPs-}
S^-_i 
   &=& S^u_i-iS^v_i
   = a^{\dagger}_i (2 S - a^{\dagger}_i a_i^{\phantom \dagger})^{1/2} 
   \approx \sqrt{2S} a^{\dagger}_i  
\end{eqnarray}
Substituting these expressions in $\mathcal{H}_{\sf ex}$ [Eq.~(\ref{eq:Hex})] 
and Fourier transforming them, we obtain
\begin{eqnarray}
{\mathcal H}_{\sf ex} 
   &=& {\mathcal E}_0 + {\mathcal H}^{\sf LSW}_{\sf ex} + \ldots
\end{eqnarray}
where ${\mathcal E}_0$ is the classical ground state energy 
defined in Eq.~(\ref{eq:E0}), and 
\begin{eqnarray}
{\mathcal H}^{\sf LSW}_{\sf ex}   
   &=&  \frac{1}{2} \sum_{{\bf q}} 
       \tilde{A}^{\dagger}_i({\bf q}) \cdot {\bf X}({\bf q}) \cdot
       \tilde{A}({\bf q}) 
\label{eq:HLSW}
\end{eqnarray}
describes quantum fluctuations at the level of linear spin wave theory.
Here $\tilde{A}^{\dagger}({\bf q}), \tilde{A}({\bf q})$ 
are eight-component vectors of operators
\begin{eqnarray}
\tilde{A}^{\dagger}({\bf q}) 
   = (a_0^{\dagger}({\bf q}), 
       a_1^{\dagger}({\bf q}), 
       a_2^{\dagger}({\bf q}), 
       a_3^{\dagger}({\bf q}), 
       \nonumber \\
       a_0^{\phantom \dagger}(-{\bf q}),
       a_1^{\phantom \dagger}(-{\bf q}),
       a_2^{\phantom \dagger}(-{\bf q}), 
        a_3^{\phantom\dagger}(-{\bf q}))
\end{eqnarray}
and $X({\bf q})$ is an $8 \times 8$ matrix written in block form as
\begin{eqnarray}
{\bf X}({\bf q})
   &=& 2 S \begin{pmatrix}
            {\bf X}^{11}({\bf q})
         & {\bf X}^{12}({\bf q}) \\
            {\bf X}^{21}({\bf q})
         & {\bf X}^{22}({\bf q})  \\
             \end{pmatrix} 
           \label{eq:X}  \\
{\bf X}^{11}_{ij}({\bf q}) 
   &=&
    \cos({\bf q} \cdot {\bf r}_{ij} ) \nonumber \\
   && \bigg(
     {\bf c}_i \cdot {\bf J}^{ij} \cdot  {\bf c}_j^{\ast}
      - \delta_{ij}  \sum_{l}   {\bf w}_l \cdot {\bf J}^{lj} \cdot {\bf w}_j  
     \bigg)  \\
{\bf X}^{12}_{ij}({\bf q})
   &=& {\bf X}^{21 \ast}_{ji}=\cos({\bf q} \cdot {\bf r}_{ij} ) 
       \bigg(
         {\bf c}_i \cdot {\bf J}^{ij} \cdot {\bf c}_j  
       \bigg) \\
{\bf X}^{22}_{ij}({\bf q})
   &=& \cos({\bf q} \cdot {\bf r}_{ij} ) \nonumber \\
    && \bigg( 
         {\bf c}_i^{\ast} \cdot {\bf J}^{ij} \cdot {\bf c}_j
        - \delta_{ij}  \sum_{l}  {\bf w}_l \cdot {\bf J}^{lj} \cdot {\bf w}_j   
        \bigg)  
\end{eqnarray}
where
\begin{eqnarray}
{\bf c}_i=\frac{1}{\sqrt{2}} \left( {\bf u}_i + i {\bf v}_i \right).
\end{eqnarray}


The spin-wave Hamiltonian ${\mathcal H}^{\sf LSW}_{\sf ex}$ [Eq.~(\ref{eq:HLSW})] 
can be diagonalized by a suitable Bogoliubov transformation.
We accomplish this following the method outlined in Ref.~\onlinecite{roger83}
by introducing new Bose operators
\mbox{$\big[ b_i^{\phantom \dagger} , b_j^{\dagger} \big] =  \delta_{ij}$}, 
such that
\begin{eqnarray}
\label{eq:bogoliubov}
B^{\dagger}({\bf q}) 
   &=&  (b_0^{\dagger}({\bf q}), 
             b_1^{\dagger}({\bf q}), 
             b_2^{\dagger}({\bf q}), 
             b_3^{\dagger}({\bf q}), 
             \nonumber \\
     &&   \qquad 
             b_0^{\phantom \dagger}(-{\bf q}),
             b_1^{\phantom \dagger}(-{\bf q}),
             b_2^{\phantom \dagger}(-{\bf q}), 
             b_3^{\phantom\dagger}(-{\bf q})) 
             \nonumber \\
   &=& A^{\dagger}({\bf q}) \cdot 
           {\bf U}^{ \dagger}({\bf q}) 
\end{eqnarray}
The condition that these operators are Bosonic 
\begin{eqnarray}
\left[ B_i^{\phantom \dagger} ({\bf q}) ,  B_j^{\dagger} ({\bf q'}) \right] 
     &=& \sigma_{ij} \delta_{{\bf q} {\bf q}'}
\end{eqnarray}
where
\begin{eqnarray}
\hat{\sigma} = 
\begin{pmatrix}
{\bf 1} & {\bf 0} \\
{\bf 0} & -{\bf 1}
\end{pmatrix}.
\end{eqnarray}
is an $8 \times 8$ matrix (written in block form), and 
\begin{eqnarray}
\left[ B_i^{\dagger} ({\bf q}) ,  B_j^{\dagger} ({\bf q'}) \right] 
    &=& \left[ B_i^{\phantom \dagger} ({\bf q}) ,  B_j^{\phantom \dagger} ({\bf q'}) \right] = 0
\end{eqnarray}
leads to a pseudo-unitary condition  on ${\bf U}({\bf q})$
\begin{eqnarray}
{\bf U}^{-1}({\bf q})=\hat{\sigma} \cdot {\bf U}^{\dagger}({\bf q}) \cdot \hat{\sigma}.
\end{eqnarray}
Substituting in Eq. \ref{eq:HLSW}, we obtain
\begin{eqnarray}
\label{eq:MatrixH}
{\mathcal H}^{\sf LSW}_{\sf ex}   
   &=& \frac{1}{2} \sum_{{\bf q}} B^{\dagger} ({\bf q}) \cdot  
           {\bf U}^{ -1 \dagger}  ({\bf q})
            \cdot \vec{X}({\bf q}) \cdot  {\bf U}^{ -1}  ({\bf q}) 
            \cdot B^{\phantom \dagger} ({\bf q}) \nonumber\\
   &=&  \frac{1}{2} 
           \sum_{{\bf q}} B^{\dagger} ({\bf q}) \cdot  \sigma  \cdot 
           {\bf U}({\bf q}) \cdot
            \sigma \cdot \vec{X}({\bf q}) \cdot  {\bf U}^{ -1} ({\bf q})  
            \cdot  B^{\phantom \dagger} ({\bf q}) . 
            \nonumber \\
\end{eqnarray}
The object 
\mbox{$ {\bf U}({\bf q}) \cdot  \sigma \cdot \vec{X}({\bf q}) \cdot  {\bf U}^{ -1}  ({\bf q})$} 
is a similarity transformation on the matrix \mbox{$\sigma \cdot \vec{X}({\bf q})$}, 
and for correctly chosen ${\bf U}({\bf q})$, will be a diagonal matrix containing the 
eigenvalues of \mbox{$\sigma \cdot \vec{X}({\bf q})$}.
We then arrive at
\begin{eqnarray}
{\mathcal H}^{\sf LSW}_{\sf ex} 
   &=& \frac{1}{2} \sum_{{\bf q}} B^{\dagger} ({\bf q}) \cdot  \sigma  \cdot 
           \begin{pmatrix}
                \omega_{\nu}(\mathbf{k}) & 0 \\
                0 & -\omega_{\nu}(\mathbf{k})
           \end{pmatrix}
           \cdot B^{\phantom \dagger} ({\bf q}). \nonumber \\
\end{eqnarray}
Collecting all terms and reordering operators, we have
\begin{eqnarray}
{\mathcal H}_{\sf ex} 
    &=& {\mathcal E}_0 \left(1 + \frac{1}{S} \right)
            \nonumber \\
      && \;  + \sum_{{\bf q}} 
             \sum_{\nu=0}^{3} \omega_{\nu}({\bf q})
             \left(
                  b_{\nu}^{\dagger}({\bf q})
                  b^{\phantom \dagger}_{\nu}({\bf q}) +\frac{1}{2} 
             \right) 
             + \ldots  
             \nonumber \\
\end{eqnarray}
The dispersion $\omega_{\nu}({\bf q})$ of the four branches of spin waves can be 
found by numerical diagonalization of \mbox{$\sigma \cdot \vec{X}({\bf q})$}.


The effect of quantum fluctuations on classical order may be estimated by 
calculating the correction to the ordered moment on sublattice $i$
\begin{eqnarray}
\langle S^{i}_w \rangle
   &=& S -  \langle a_i^{\dagger} a_i \rangle 
          \nonumber \\
   &=&   S -\frac{4}{N} \sum_{{\bf q}} 
           \sum_{m=4}^{7} |{\bf U}_{i m}({\bf q})|^2
\label{eq:momentcorrection}
\end{eqnarray}
where $\sum_{m=4}^{7} $ implies a sum
over the off-diagonal
$4 \times 4$ block of ${\bf U}({\bf q})$.
In all of the 4-sublattice phases described in this text, 
$\langle S^{i}_w \rangle$ is the same for 
all sublattices $i=0,1,2,3$.


The results of this analysis are shown in Fig. 6 of the main text, where we show 
that divergences in the ordered moment correction approaching the high degeneracy limits
of the model lead to regions where the conventional magnetic order is completely eliminated.

\section{Classical Monte Carlo simulation}


The Monte Carlo simulations in this paper are based on the Metropolis algorithm with 
parallel tempering~\cite{swendsen86,geyer91} and  over-relaxation~\cite{creutz87}. 
The spins are modelled as classical vectors of length $|S_{i}|=1/2$ and locally updated 
using the standard Marsaglia method~\cite{marsaglia72}.   
We consider cubic clusters of linear dimension $L$, based on the 16-site cubic unit cell 
of the pyrochlore lattice, and containing $N = 16 L^3$ sites.
A Monte Carlo step (MCs) is defined as $N$ attempts to locally update a randomly 
chosen spin, and $t_{max}$ (measured in MCs) is the total Monte Carlo time over 
which data are collected.


Equilibration is performed for each temperature in two successive steps.   
First the system is slowly cooled down from high temperature (random initial spin configuration) 
to the temperature of measurement $T$ during $t_{max}/10$ MCs.   
Then, the system is equilibrated at temperature $T$ during additional $t_{max}/10$ MCs.   
After equilibration, Monte Carlo time is set to zero and measurements start and go on 
for $t_{max} \sim 10^{5}-10^{7}$ MCs.


All thermodynamical observables have been averaged over Monte Carlo time every 10 MCs, 
except for calculations of the equal-time structure factor $S({\bf q})$, where data points were 
taken every 100 MCs for efficiency.  
The parallel tempering method implies simultaneously simulating a large number of replicas
of the system in parallel, with each replica held at a different temperature.
The program then regularly attempts to swap the spin configurations of replicas
with neighbouring temperatures, in such a way as to maintain detailed 
balance~\cite{swendsen86,geyer91}.  
Simulating $\sim 100$ replicas, with swaps attempted every 100 MCs appears to offer 
a good compromise between efficiency and decorrelation for L=6. 


In the case of the over-relaxation method, after each Monte Carlo step, 
two further sweeps are made of the entire lattice.
Each spin feels an effective field due to the interaction with its six neighbours; any 
rotation around this axis conserves the energy and is thus an acceptable move respecting 
detailed balance. 
To avoid rotating successive neighbouring spins, we first update all spins of sublattice 0, 
then sublattice 1, 2 and finally 3.  
The first iteration of all $N$ spins is deterministic, \textit{i.e.} we rotate them by the maximum 
allowed angle; while for the second iteration, a random angle of rotation is chosen for each spin. 
The generation of so many random numbers is of course time consuming but is 
recommended for better equilibration~\cite{kanki05}.
We note that convergence of the specific heat $c_h \to 1$ for $T \rightarrow 0$ 
is a good indication of the equilibration at low temperatures.

\section{Finite temperature phase diagram}

\subsection{Details of simulations}

In Fig.~3 of the main text we show a finite temperature phase diagram 
spanning all four of the ordered phases discussed in the article.
This phase diagram was determined from simulations for 64 different parameter 
sets, equally spaced on the circle defined by 
\mbox{$\sqrt{J_{1}^{2}+J_{2}^{2}} = 3\,|J_{3}|$} 
with \mbox{$J_3=-0.1\ \text{meV}$} and \mbox{$J_4=0$} 
[cf. white circle in Fig. 1 of main text]. 
Transition temperatures for each phase were extracted from the relevant 
order-parameter susceptibilities.
This is described in turn for each of the four ordered phases, below.


\begin{figure}[t]
\centering\includegraphics[width=0.95\columnwidth]{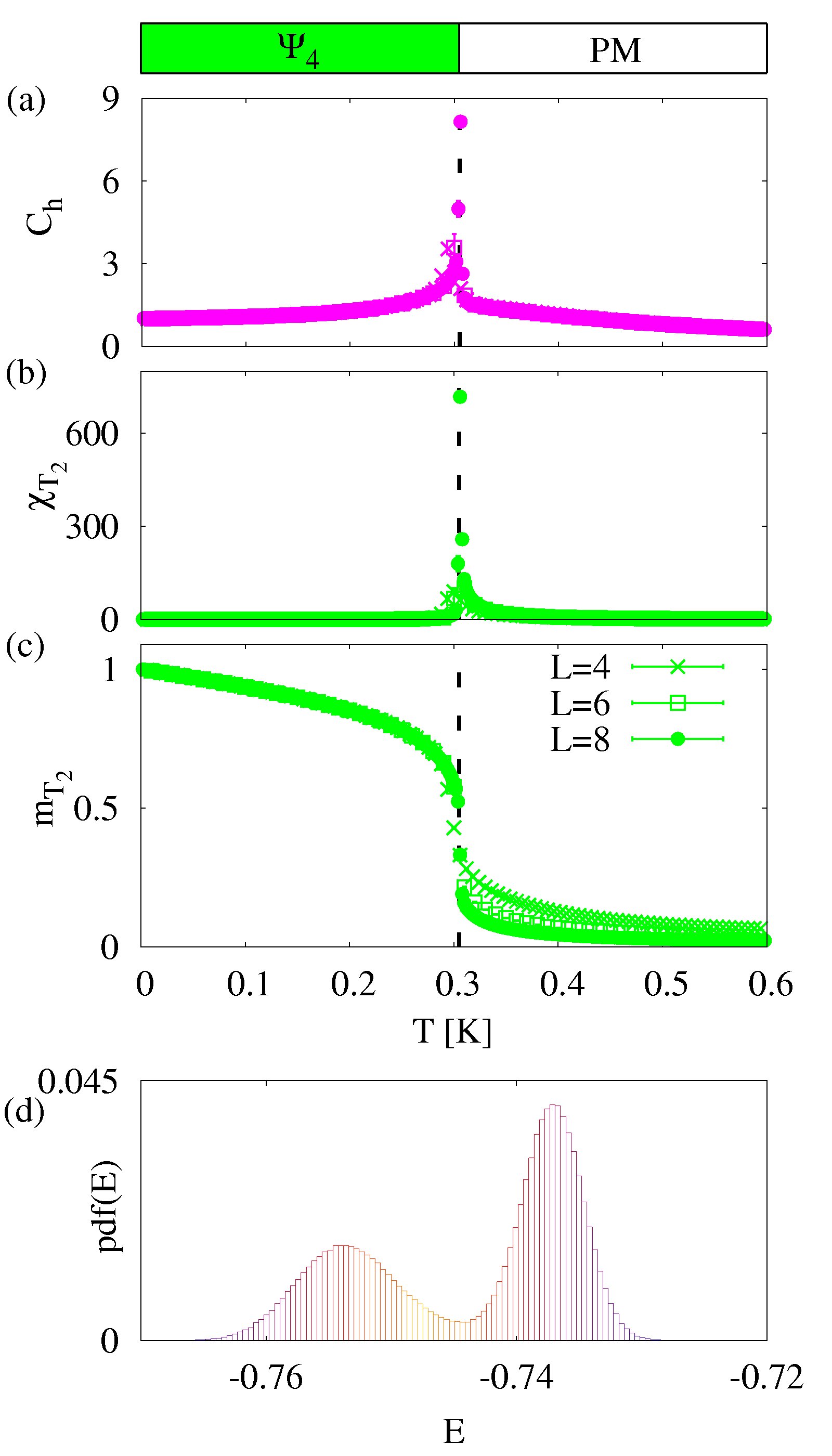}
\caption{Finite-temperature phase transition from the paramagnet into the 
Palmer-Chalker phase [$\Psi_4$], as determined by classical Monte Carlo 
simulation of $\mathcal{H}_{\sf ex}$~[Eq.~(\ref{eq:Hex})], for parameters 
\mbox{$J_1 = 0$\ meV}, 
\mbox{$J_2 = 0.3$\ meV},
\mbox{$J_3 = -0.1$\ meV},
\mbox{$J_4 = 0$\ meV}. 
a) Temperature dependence of the specific heat $c_h(T)$.
b) Temperature dependence of the order-parameter susceptibility, $\chi_{\sf T_2}(T)$.
c) Temperature dependence of the order parameter, $|{\bf m}_{\sf T_2}(T)|$.    
d) Probability distribution of the energy $E$ evaluated at the transition temperature $T_c=306.5~ \text{mK}$
    for a cluster of size $L=12$.
The black dashed line in (a)-(c) indicates a first-order phase transition  
at $T_{\sf T_2} = 305 \pm 5$ mK. 
Simulations were performed for clusters of $N=16L^3$ spins, 
with $L=4,6,8,12$.  
\label{fig:PMtoPC}}
\end{figure}


Simulations were performed for a cluster of $N=3456$ spins ($L=6$), and 
data averaged over 10 independent runs during $t_{max}=10^{6}$ MCs. 
Parallel tempering was used, typically with 121 replicas, at temperatures 
equally-spaced from 0 to 1.2 K.
However, close to the boundaries between phases with different symmetries, the 
large number of competing ground states makes simulations difficult to equilibrate.
Here, additional data points with better statistics were sometimes necessary, typically with 
201 temperatures on a smaller temperature window, with $t_{max}=10^{7}$ MCs 
and $N=8192$ \mbox{(\textit{i.e.} L=8)}. 
Under such conditions, over-relaxation was usually not necessary to determine the 
transition temperature.


\begin{figure}[t]
\centering\includegraphics[width=0.95\columnwidth]{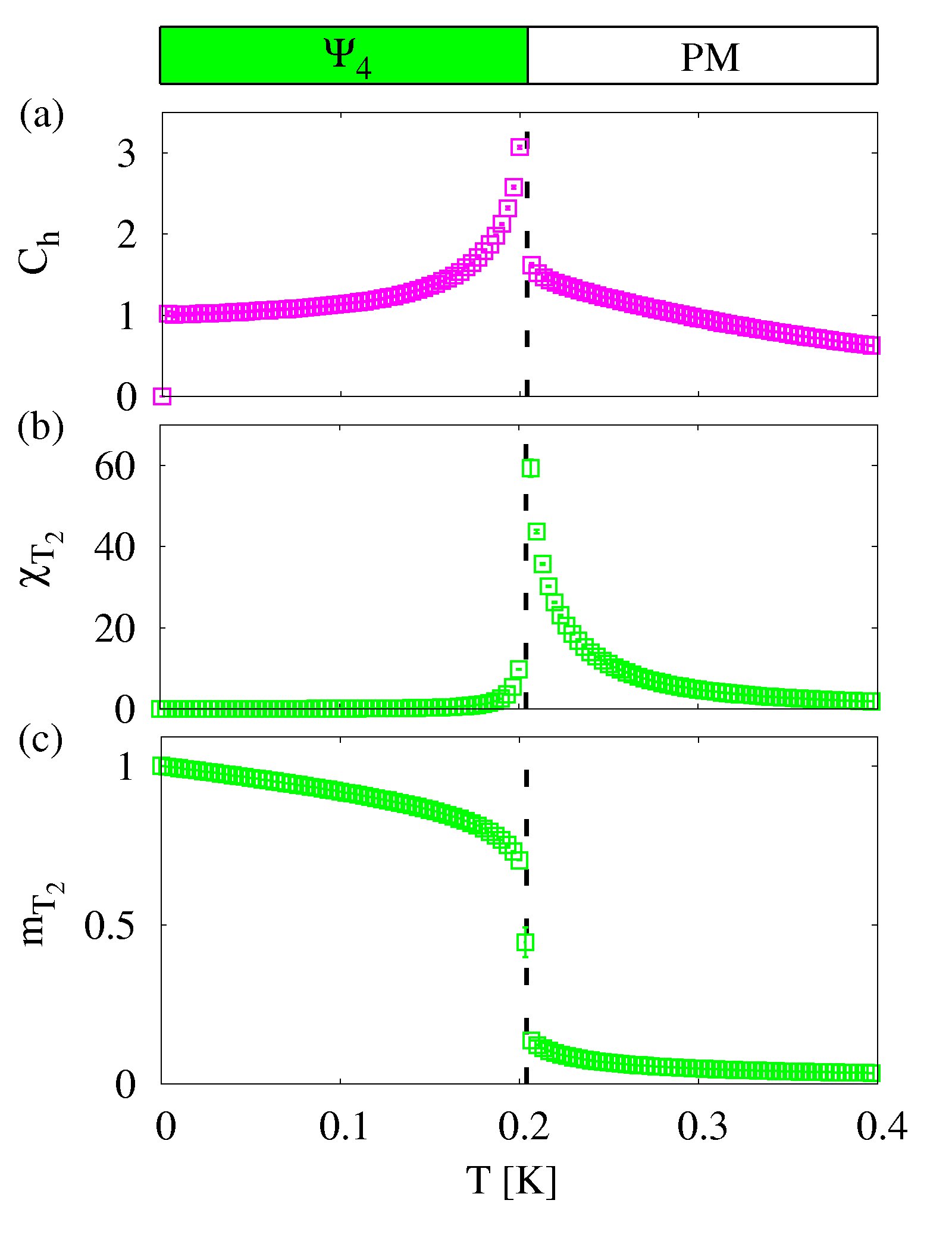}
\caption{Finite-temperature phase transition from the paramagnet into the 
Palmer-Chalker phase [$\Psi_4$], as determined by classical Monte Carlo 
simulation of $\mathcal{H}_{\sf ex}$~[Eq.~(\ref{eq:Hex})], for parameters 
appropriate to Er$_2$Sn$_2$O$_7$ [\onlinecite{guitteny13}], i.e.
\mbox{$J_1 = 0.07$\ meV}, 
\mbox{$J_2 = 0.08$\ meV},
\mbox{$J_3 = -0.11$\ meV},
\mbox{$J_4 = 0.04$\ meV}. 
a) Temperature dependence of the specific heat $c_h(T)$.
b) Temperature dependence of the order-parameter susceptibility, $\chi_{\sf T_2}(T)$.
c) Temperature dependence of the order parameter, $|{\bf m}_{\sf T_2}(T)|$.    
The black dashed line in (a)-(c) indicates a first-order phase transition  
at $T_{\sf T_2} = 204 \pm 5\ \text{mK}$.
Simulations were performed for clusters of $N=16L^3$ spins, 
with $L=6$. 
\label{fig:PMtoPC-Er2Sn2O7}}
\end{figure}


\begin{figure}[h]
\centering\includegraphics[width=0.95\columnwidth]{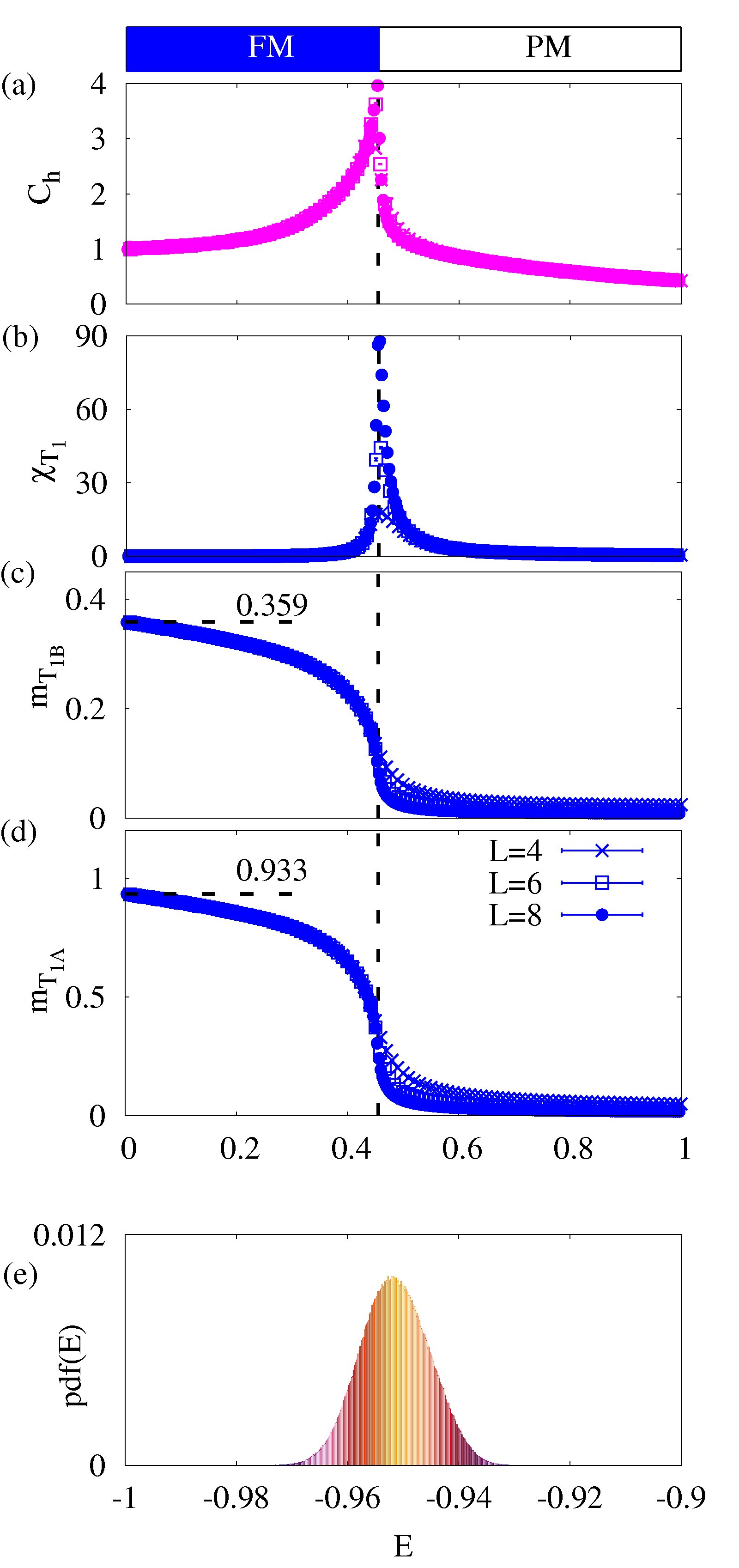}
\caption{
Finite-temperature phase transition from the paramagnet into the 
non-collinear ferromagnet (FM), as determined by classical Monte Carlo 
simulation of $\mathcal{H}_{\sf ex}$~[Eq.~(\ref{eq:Hex})], for parameters 
appropriate to Yb$_2$Ti$_2$O$_7$ \cite{ross11-PRX1}, i.e.
\mbox{$J_1 = -0.09$\ meV}, 
\mbox{$J_2 = -0.22$\ meV},
\mbox{$J_3 = -0.29$\ meV}
setting
\mbox{$J_4 = 0$\ meV}. 
a)  Temperature dependence of the specific heat $c_h(T)$.
b) Temperature dependence of the order-parameter susceptibility, $\chi_{\sf T_2}(T)$.
c) Temperature dependence of the order parameter, $|{\bf m}_{\sf T_{1, A}}(T)|$.
d) Temperature dependence of the order parameter, $|{\bf m}_{\sf T_{1, B}}(T)|$.
e) Probability distribution of the energy $E$ evaluated at the transition 
    for a system of size $L=12$.
The black dashed line in (a)--(d) indicates a continuous phase transition 
at \mbox{$T_{\sf T_1} = 455 \pm 5\ \text{mK}$}.  
Simulations were performed for clusters of $N=16L^3$ spins, 
with $L=4,6,8,12$. 
}
\label{fig:PMtoFM}
\end{figure}


\begin{figure}[h]
\centering\includegraphics[width=0.95\columnwidth]{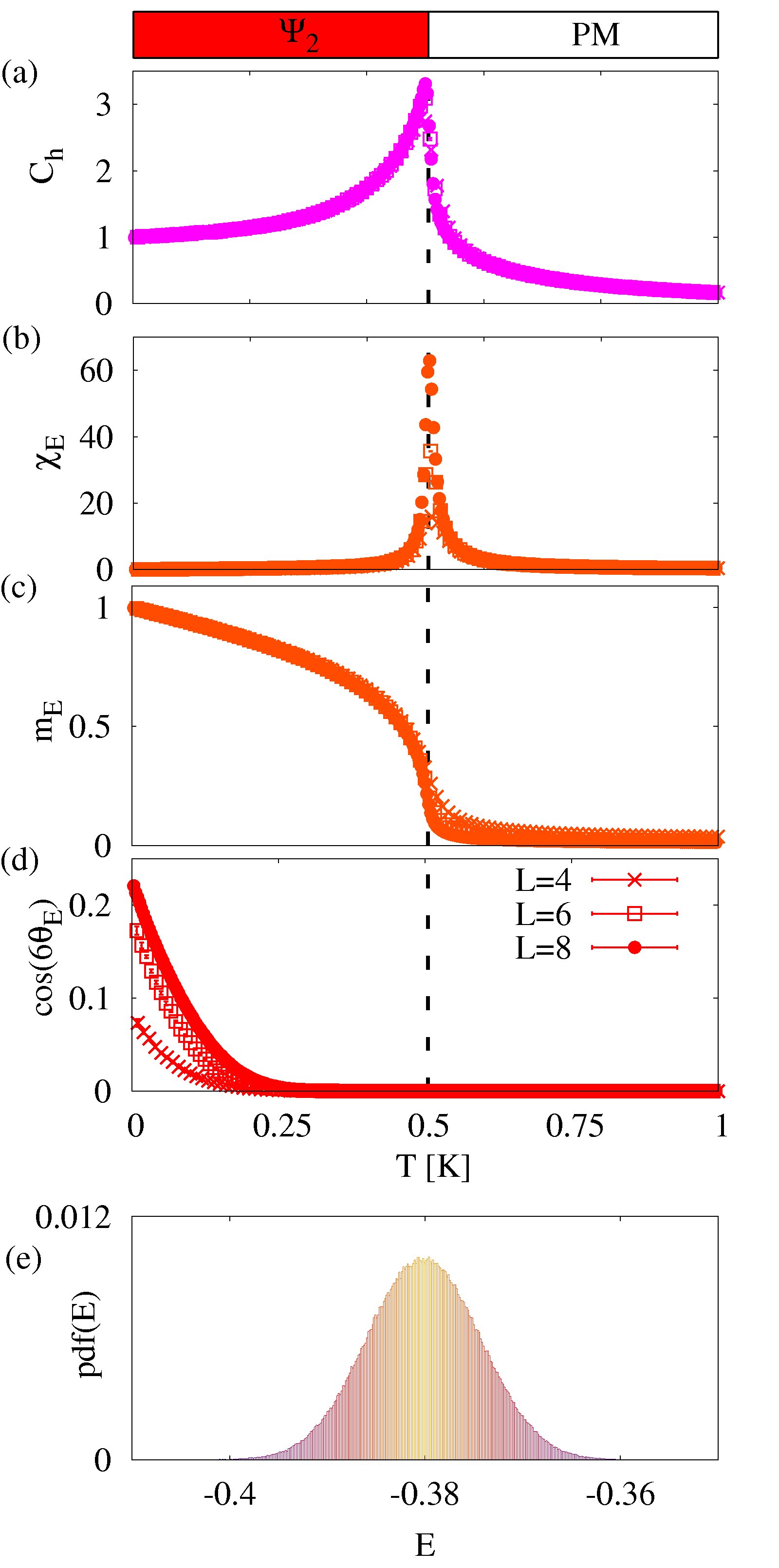}
\caption{
Finite-temperature phase transition from the paramagnet into the non-coplanar 
antiferromagnet $\Psi_2$, as determined by classical Monte Carlo 
simulation of $\mathcal{H}_{\sf ex}$~[Eq.~(\ref{eq:Hex})], for parameters 
appropriate to Er$_2$Ti$_2$O$_7$ \cite{savary12-PRL109}, i.e.
\mbox{$J_1 = 0.11$\ meV}, 
\mbox{$J_2 = -0.06$\ meV},
\mbox{$J_3 = -0.1$\ meV}
setting
\mbox{$J_4 = 0$\ meV}. 
a) Temperature dependence of the specific heat $c_h(T)$.
b) Temperature dependence of the order-parameter susceptibility, $\chi_{\sf E}(T)$.
c) Temperature dependence of the order parameter, $|{\bf m}_{\sf E}(T)|$.
d) Temperature dependence of the order parameter, $\cos 6\theta_{\sf E}$.
e) Probability distribution of the energy $E$ evaluated at the transition 
    temperature $T_c=502~ \text{mK}$ for a system of size $L=12$.
The black dashed line indicates a continuous phase transition 
at \mbox{$T_N=505 \pm 5$\ mK}.  
Simulations were performed for clusters of $N=16L^3$ spins, 
with $L=4,6,8,12$. 
}
\label{fig:PMtoPsi2}
\end{figure}

\subsection{Transition from the paramagnet into the Palmer-Chalker phase, $J_4 \equiv 0$}

In Fig.~\ref{fig:PMtoPC} we show simulation results for the finite-temperature
phase transition from the paramagnet into Palmer-Chalker phase for parameters 
$$ (J_1,\  J_2,\ J_3,\ J_4) = (0, 0.3, -0.1,  0)\quad \text{meV} $$ 
deep within the Palmer-Chalker phase.
Anomalies in both the specific heat $c_h(T)$ [Fig.~\ref{fig:PMtoPC}(a)] 
and order-parameter susceptibility $\chi_{\sf T_2}(T)$ [Fig.~\ref{fig:PMtoPC}(b)]
at \mbox{$T_{\sf T_2} = 305  \pm 5$\ mK}, provide clear evidence of a phase transition.


For this parameter set, the transition is first order, 
as is evident from the discontinuity in the value of the order parameter 
${\bf m}_{\sf T_2}$ for $T=T_{\sf T_2}$ [Fig.~\ref{fig:PMtoPC}(c)],
and double peak in the probability distribution for the energy [Fig.~\ref{fig:PMtoPC}(d)].

\subsection{Transition from the paramagnet into the Palmer-Chalker phase, for parameters
                     appropriate to Er$_2$Sn$_2$O$_7$, $J_4 \ne 0$}

In Fig.~\ref{fig:PMtoPC-Er2Sn2O7} we show simulation results for the finite-temperature
phase transition from the paramagnet into Palmer-Chalker phase for parameters 
appropriate to Er$_2$Sn$_2$O$_7$
$$ (J_1,\  J_2,\ J_3,\ J_4) = (0.07, 0.08, -0.11,  0.04)\quad \text{meV} $$ 
near to the boundary of the Palmer-Chalker phase.
Anomalies in both the specific heat $c_h(T)$ [Fig.~\ref{fig:PMtoPC-Er2Sn2O7}(a)] 
and order-parameter susceptibility $\chi_{\sf T_2}(T)$ [Fig.~\ref{fig:PMtoPC-Er2Sn2O7}(b)]
at \mbox{$T_{\sf T_2} = 200  \pm 5$\ mK}, provide clear evidence of a phase transition.
No ordering transition has ever been observed in experiment on Er$_2$Sn$_2$O$_7$, 
although anomalies in the magnetic susceptibility below \mbox{$T = 200\ \text{mK}$} have 
been interpreted as evidence of spin freezing.


For this parameter set, the transition is first order, 
as is evident from the discontinuity in the value of the order parameter 
${\bf m}_{\sf T_2}$ for $T=T_{\sf T_2}$ [Fig.~\ref{fig:PMtoPC-Er2Sn2O7}(c)].
%
We have confirmed by repeating simulations with $J_4 \equiv 0$ that Dzyaloshinskii-Moriya 
interactions no not have any qualitative effect on the thermodynamics of Er$_2$Sn$_2$O$_7$.
However the finite value of $J_4$ does have an effect on the transition temperature, 
which drops to $T_{\sf T_2} \approx 70\ \text{mK}$ for $J_4 = 0$.

\subsection{Transition from the paramagnet into the ferromagnetic phase, 
                    for parameters appropriate to Yb$_2$Ti$_2$O$_7$}

In Fig.~\ref{fig:PMtoFM} we show simulation results for the finite-temperature
phase transition from the paramagnet into the non-colinear ferromagnet (FM), 
for parameters appropriate to Yb$_2$Ti$_2$O$_7$ \cite{ross11-PRX1}, 
setting $J_4 =0$  
$$ (J_1,\  J_2,\ J_3,\ J_4) = (-0.09, -0.22, -0.29,  0)\quad \text{meV} $$ 
Anomalies in both the specific heat $c_h(T)$ [Fig.~\ref{fig:PMtoFM}(a)] 
and order-parameter susceptibility $\chi_{\sf T_1}(T)$ [Fig.~\ref{fig:PMtoFM}(b)]
at \mbox{$T_{\sf T_1} = 455  \pm 5\ \text{mK}$}, provide clear evidence of a phase transition.


This estimate of the transition temperature compares reasonably well with experiment, 
where those samples which order undergo a phase transition at 
\mbox{$T_c^{\sf Yb_2Ti_2O_7} = 240 \pm 30 \text{mK}$}~\cite{bloete69,ross11-PRB84,chang12}.
At low temperatures, the temperature-dependence of the order parameters
${\bf m}_{\sf T_{1,A}}$ and ${\bf m}_{\sf T_{1,B}}$ [Fig.~\ref{fig:PMtoFM}(c)--(d)] 
converges on the values expected from a low-temperature expansion about the 
FM ground state (not shown).


The single peak in the probability distribution for the energy [Fig.~\ref{fig:PMtoFM}(e)]
suggests that, for parameters appropriate to Yb$_{2}$Ti$_{2}$O$_{7}$, the thermal 
phase transition from paramagnet to non-collinear FM in a classical model is at most 
very weakly first order.
This contrasts with experiment, where the phase transition in those samples which order 
is believed to be strongly first order~\cite{chang12}.


It is also interesting to note that classical Monte Carlo simulations for parameter sets close 
to the border with the Palmer-Chalker phase --- where fluctuation effects are more 
pronounced --- reveal a strongly first-order transition.
We have confirmed by repeating simulations with $J_4 = 0.01\ \text{mK}$ (cf. \cite{ross11-PRX1}) 
that Dzyaloshinskii-Moriya interactions have a negligible effect on the thermodynamics 
of Yb$_2$Ti$_2$O$_7$, changing the transition temperature to 
\mbox{$T_{\sf T_1} = 452 \pm 10\ \text{mK}$}

\subsection{Transition from the paramagnet into the $\Psi_2$ phase, 
                    for parameters appropriate to Er$_2$Ti$_2$O$_7$}

In Fig.~\ref{fig:PMtoPsi2} we show simulation results for the finite-temperature
phase transition from the paramagnet into the $\Psi_2$ phase, 
for parameters appropriate to Er$_2$Ti$_2$O$_7$~[\onlinecite{savary12-PRL109}], 
setting $J_4 =0$  
$$ (J_1,\  J_2,\ J_3,\ J_4) = (0.11, -0.06, -0.1,  0)\quad \text{meV} $$ 
This shows a number of interesting features.


Anomalies in both the specific heat $c_h(T)$ [Fig.~\ref{fig:PMtoPsi2}(a)] 
and order-parameter susceptibility $\chi_{\sf E}(T)$ [Fig.~\ref{fig:PMtoPsi2}(b)]
 at \mbox{$T_{\sf E} = 505  \pm 5\ \text{mK}$} offer clear evidence of a phase transition. 
Surprisingly, this transition occurs at a significantly {\it lower} temperature in 
simulation than experiment, where a transition is observed at
\mbox{$T_{\sf N}^{\sf Er_2Ti_2O_7} = 1.2 \pm 0.1 \text{K}$} 
[\onlinecite{bloete69,champion03,oitmaa-arXiv}]. 
We have confirmed by repeating simulations with $J_4 = -0.003\ \text{mK}$ 
(cf. \cite{ross11-PRX1}) that Dzyaloshinskii-Moriya interactions have a 
negligible effect on the thermodynamics of Er$_2$Ti$_2$O$_7$, changing
the transition temperature by only a few degrees 
to \mbox{$T_{\sf E} = 506  \pm 10\ \text{mK}$}


Both the smooth evolution of the primary order parameter, 
${\bf m}_{\sf E}$~[Fig.~\ref{fig:PMtoPsi2}(c)], and the 
single peak in the probability distribution for the energy [Fig.~\ref{fig:PMtoPsi2}(e)]
suggests that the phase transition seen in simulation is at most weakly first-order.
For the clusters simulated, we find that it is possible to obtain a fairly good collapse of 
data for $\chi_{\sf E}(T)$ [Fig.~\ref{fig:PMtoPsi2}(b)] using 3D XY exponents.


However there are only a discrete number of $\Psi_2$ ground states, and 
a finite value of  $|{\bf m}_{\sf E}|$ alone does not imply $\Psi_2$ order.
Evidence for the $\Psi_2$ ground state comes from the secondary 
order parameter $c_{\sf E} = \cos 6 \theta_{\sf E} > 0$ [Fig.~\ref{fig:PMtoFM}(d)].
Here simulation results are strongly size-dependent, but suggest a slow 
crossover into the $\Psi_2$ state, occurring at a $T^* \ll T_{\sf E}$, without any 
accompanying feature in $c_h(T)$ [Fig.~\ref{fig:PMtoPsi2}(a)].


\begin{figure}[h!]
\centering\includegraphics[width=0.95\columnwidth]{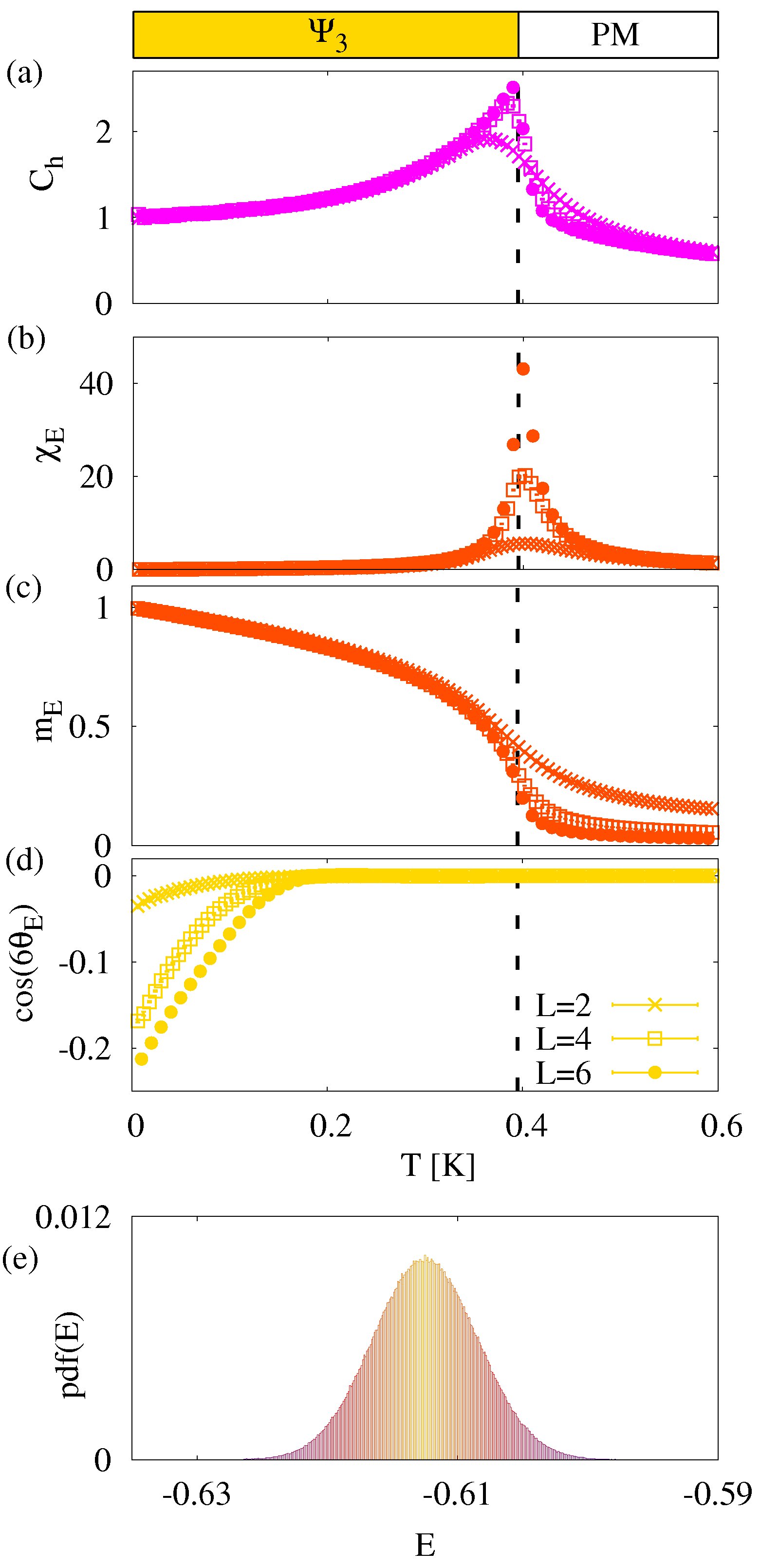}
\caption{
Finite-temperature phase transition from the paramagnet into the coplanar 
antiferromagnet $\Psi_3$, as determined by classical Monte Carlo 
simulation of $\mathcal{H}_{\sf ex}$~[Eq.~(\ref{eq:Hex})], for parameters 
\mbox{$J_1 = 0$\ meV}, 
\mbox{$J_2 = -0.3$\ meV},
\mbox{$J_3 = -0.1$\ meV},
\mbox{$J_4 = 0$\ meV}. 
a) Temperature dependence of the specific heat $c_h(T)$.
b) Temperature dependence of the order-parameter susceptibility, $\chi_{\sf E}(T)$.
c) Temperature dependence of the order parameter, $|{\bf m}_{\sf E}(T)|$.
d) Temperature dependence of the order parameter, $\cos 6\theta_{\sf E}$.
e) Probability distribution of the energy $E$ evaluated at the transition 
    temperature $T_c=395 ~\text{mK}$ for a system of size $L=12$.
The black dashed line indicates a continuous phase transition 
at \mbox{$T_{\sf E}=395 \pm 5$\ mK}.  
Simulations were performed for clusters of $N=16L^3$ spins, 
with $L=2,4,6,12$. 
}
\label{fig:PMtoPsi3}
\end{figure}


On the basis of the Landau theory ${\mathcal F}_{\sf E}$ [Eq.~(\ref{eq:Landau})], 
we anticipate that {\it any} finite value of $m_{\sf E}  = |{\bf m}_{\sf E}|$ will induce 
symmetry breaking in $\theta_{\sf E}$, and that both symmetries should 
therefore be broken at the same temperature.
Depending on the sign of the relevant coupling, 
\begin{eqnarray}
\delta {\mathcal F}_{\sf E} = \frac{1}{6}\ d\  m_{\sf E}^6\ \cos 6 \theta_{\sf E}
\end{eqnarray}
the system will then enter either a $\Psi_2$ or a $\Psi_3$ ground state.


However, the free-energy barrier separating the $\Psi_2$ and $\Psi_3$ 
ground states is very small, and this in turn sets a very large  
length-scale for the selection of the $\Psi_2$ ground state.
Based on the low-temperature expansion $\mathcal{F}_{\sf ex}^{\sf low-T}$ 
[Eq.~(\ref{eq:FlowT})], we estimate that clusters with linear dimension $L \sim 1000$
may be needed to resolve this as a single transition.

\subsection{Transition from the paramagnet into the $\Psi_3$ phase}

In Fig.~\ref{fig:PMtoPsi3} we show simulation results for the finite-temperature
phase transition from the paramagnet into the $\Psi_3$ phase, 
for parameters 
$$ (J_1,\  J_2,\ J_3,\ J_4) = (0, -0.3, -0.1,  0) \quad \text{meV} $$ 
close to the border with the non-collinear ferromagnet. 
Anomalies in both the specific heat $c_h(T)$ [Fig.~\ref{fig:PMtoPsi3}(a)] 
and order-parameter susceptibility $\chi_{\sf E}(T)$ [Fig.~\ref{fig:PMtoPsi3}(b)]
at \mbox{$T_{\sf E} = 395  \pm 5 \text{mK}$} offer clear evidence of a phase transition. 
Both the smooth evolution of the primary order parameter, 
${\bf m}_{\sf E}$~[Fig.~\ref{fig:PMtoPsi3}(c)], and the 
single peak in the probability distribution for the energy [Fig.~\ref{fig:PMtoPsi3}(e)]
suggest that this phase transition is continuous.


Evidence for the $\Psi_3$ ground state comes from the finite value of the secondary 
order parameter \mbox{$c_{\sf E} = \cos 6 \theta_{\sf E} < 0$} [Fig.~\ref{fig:PMtoPsi3}(d)].
This secondary order parameter shows only a slow onset, consistent with a crossover
into the $\Psi_3$ state, and is {\it very} strongly size-dependent.
As with the $\Psi_2$ state considered above, we infer that, with increasing system 
size, the temperature associated with this crossover scales towards $T = T_N$, 
and that in the thermodynamic limit, a single phase transition from takes place from 
the paramagnet into the $\Psi_3$ state.

\section{Equal-time structure factors $S({\bf q})$}

\subsection{Definitions}

The equal-time structure factor $S({\bf q})$ is defined by
\begin{eqnarray}
   S({\bf q}) = 
   \sum_{\alpha, \beta=1}^{3} \sum_{i, j=1}^{4} 
   \left( \delta_{\alpha \beta}- \frac{q_{\alpha} {q_\beta}}{q^2} \right)
   \langle 
      m_{\alpha}^{i}(-{\bf q})  m_{\beta}^{j}({\bf q}) 
   \rangle\nonumber\\
   \label{eq:defSq}
\end{eqnarray}
Here the magnetic moment $m_{\alpha}^{i}$ is related to the pseudospin-$1/2$ 
via the g-tensor $S^{\beta}_{i} (\mathbf{R}_i)$
\begin{eqnarray}
m^{i}_{\alpha}({\bf q}) 
   = \sqrt{\frac{4}{N}} \sum_{\beta=1}^3
         g^i_{\alpha \beta}  
         \left( 
            \sum_{\vec{R}_i} e^{i {\bf q}.\vec{R}_i} S_{\beta}(\vec{R}_i) 
         \right).
\end{eqnarray}


\begin{figure*}
\includegraphics[width=0.8\textwidth]{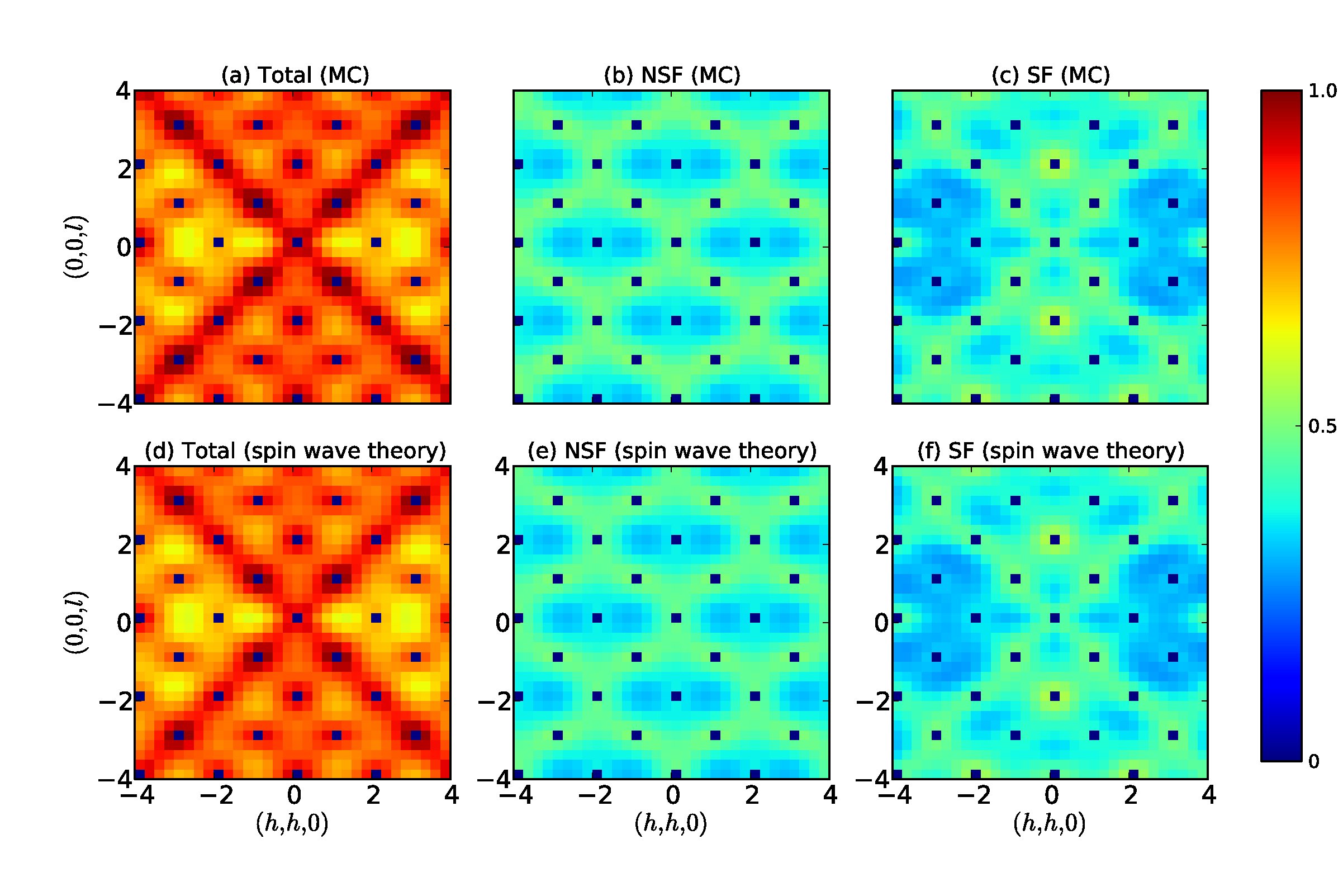}
\caption
{Comparison between results for equal-time structure factor $S({\bf q})$ obtained
in classical Monte Carlo (MC) simulation and classical low-temperature expansion
(spin wave theory) for parameters appropriate to Yb$_2$Ti$_2$O$_7$ \cite{ross11-PRX1}.   
a) Total scattering in the $(h, h, l)$ plane within MC simulation.
b) Associated scattering in the non spin-flip ({\sf NSF}) channel.
c) Associated scattering in the spin-flip ({\sf SF}) channel.
d) Total scattering in the $(h, h, l)$ plane within a spin-wave expansion
about the ferromagnetic ground state.
e) Associated scattering in the {\sf NSF} channel.
f) Associated scattering in the {\sf SF} channel.
Rods of scattering in the $(h, h, h)$ direction, associated with a low-energy 
spin-wave excitation, are visible in both {\sf SF} and {\sf NSF} channels.
All results were obtained at $T=0.05$\ K, for exchange parameters 
\mbox{$J_1=-0.09 \text{meV}$},
\mbox{$J_2=-0.22 \text{meV}$}, 
\mbox{$J_3=-0.29 \text{meV}$}, 
setting \mbox{$J_4=0$}.
{\sf SF} and {\sf NSF} channels are defined with respect to a neutron with polarisation 
in the $(1, -1, 0)$ direction, as in Ref. \onlinecite{fennell09}. 
$S({\bf q})$ has been calculated using the experimentally measured g-tensor 
for Yb$_2$Ti$_2$O$_7$ \cite{hodges01, ross11-PRX1}, with $g_z=1.77, g_{xy}=4.18$.
In order to avoid saturating the colour scale, the spectral weight associated 
with Bragg peaks at reciprocal lattice vectors has been subtracted.
\label{fig:YTOsq}}
\end{figure*}


\begin{figure*}
\centering
\includegraphics[width=0.8\textwidth]{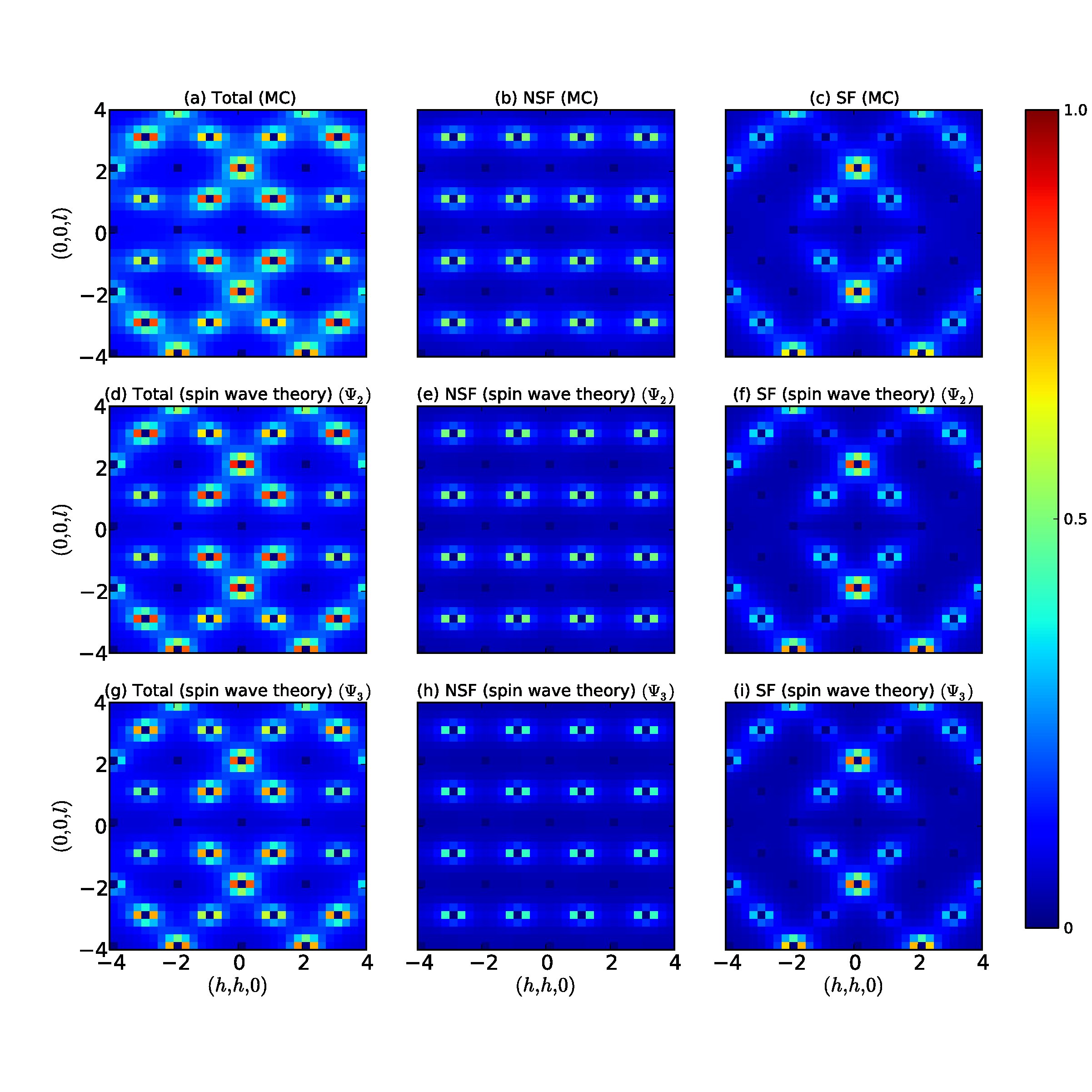}
\caption{
Comparison between results for equal-time structure factor $S({\bf q})$ obtained
in classical Monte Carlo (MC) simulation and low-temperature expansion
(classical spin wave theory) for parameters appropriate to Er$_2$Ti$_2$O$_7$.   
a) Total scattering in the $(h, h, l)$ plane within MC simulation.
b) Associated scattering in the non spin-flip ({\sf NSF}) channel.
c) Associated scattering in the spin-flip ({\sf SF}) channel.
d) Total scattering in the $(h, h, l)$ plane within a spin-wave expansion 
about a $\Psi_2$ ground state.
e) Associated scattering in the {\sf NSF} channel.
f) Associated scattering in the {\sf SF} channel.
g) Total scattering in the $(h, h, l)$ plane within a spin-wave expansion 
about a $\Psi_3$ ground state.
h) Associated scattering in the {\sf NSF} channel.
i) Associated scattering in the {\sf SF} channel.
Careful comparison of the distribution of scattering in the vicinity of the $(1, 1, 1)$, 
$(3, 3, 3)$ and $(1, 1, 3)$ reciprocal lattice vectors supports the conclusion 
that the $\Psi_2$ state is preferred for these exchange parameters, in agreement 
with experiment and the calculations described in the text.
All results were obtained at $T=0.36$\ K, for exchange parameters 
\mbox{$J_1=0.11\ \text{meV}$},
\mbox{$J_2=-0.06\ \text{meV}$},
\mbox{$J_3=-0.10\ \text{meV}$}, 
setting  \mbox{$J_4 \equiv 0$} [\onlinecite{savary12-PRL109}].
For clarity, spectral weight associated with Bragg peaks at reciprocal 
lattice vectors has been subtracted.
\label{fig:ETOsq}
}
\end{figure*}


In the local co-ordinate frame $\{ {\bf x}_i^{\sf \ local},  {\bf y}_i^{\sf \ local},
{\bf z}_i^{\sf \ local} \}$ in which the $ {\bf z}_i^{\sf \ local}$ axis is 
the local $\langle 111 \rangle$ $C_3$ symmetry axis, the $g$-tensor is diagonal
 \begin{eqnarray}
 {\bf g}^{\sf local}=
    \begin{pmatrix}
       g_{xy} & 0 & 0 \\
       0 & g_{xy} & 0\\
       0 & 0 & g_z \\
    \end{pmatrix}
 \end{eqnarray}
rotating back into the global co-ordinate frame the $g$-tensor is sublattice dependent
\begin{eqnarray}
{\bf g}_{0} &=&
    \begin{pmatrix}
        g_{1} & g_{2} & g_{2} \\
        g_{2} & g_{1} & g_{2}\\
        g_{2} & g_{2} & g_{1} \\
    \end{pmatrix} \quad
{\bf g}_{1} =
   \begin{pmatrix}
       g_{1} & -g_{2} &- g_{2} \\
      -g_{2} & g_{1} & g_{2}\\
      -g_{2} & g_{2} & g_{1} \\
   \end{pmatrix} \nonumber \\
 {\bf g}_{2}&=&
   \begin{pmatrix}
       g_{1} & -g_{2} & g_{2} \\
      -g_{2} & g_{1} & -g_{2}\\
       g_{2} & -g_{2} & g_{1} \\
   \end{pmatrix} \quad
 {\bf g}_{3} =
   \begin{pmatrix}
      g_{1} & g_{2} & -g_{2} \\
      g_{2} & g_{1} & -g_{2}\\
     -g_{2} & -g_{2} & g_{1} \\
   \end{pmatrix} \nonumber \\
\end{eqnarray}
where
\begin{eqnarray}
g_1 &=& \frac{2}{3} g_{xy} +\frac{1}{3} g_z \quad
g_2 = -\frac{1}{3} g_{xy} + \frac{1}{3} g_z.
\end{eqnarray}
 
The structure factor $S({\bf q})$ can also be resolved into spin flip ({\sf SF}) 
and non-spin flip ({\sf NSF}) components, for comparison with experiments 
carried out using polarised neutrons.
For neutrons with polarisation along \mbox{${\bf \hat{n}}$},
these are given by
\begin{eqnarray}
S^{\sf NSF}({\bf q})
   &=& \sum_{\alpha, \beta=1}^{3} \sum_{i, j=1}^{4} 
          \langle 
               ( {\bf m}^{i}(-{\bf q}) \cdot {\bf \hat{n}} )  
               ({\bf m}^{j}({\bf q})  \cdot {\bf \hat{n}} )
          \rangle \nonumber\\
          \label{eq:sqnsf}\\
S^{\sf SF}({\bf q})
   &=& \sum_{\alpha, \beta=1}^{3} \sum_{i, j=1}^{4} 
           \frac{1}{q^2}
                    \langle 
                          ( {\bf m}^{i}(-{\bf q}) 
                          \cdot 
                          \left( 
                               {\bf \hat{n}} \times {\bf q}) 
                           \right) 
                   \nonumber \\
        &&   \qquad  \qquad \qquad \times 
                    \left( 
                        {\bf m}^{j}({\bf q})  \cdot ({\bf \hat{n}} \times {\bf q})
                   \right)
                   \rangle
                   \nonumber \\
                    \label{eq:sqsf}
\end{eqnarray}
Where we quote results for {\sf SF} and {\sf NSF} components of $S({\bf q})$ 
below, we consider \mbox{${\bf \hat{n}}=(1, -1, 0)/\sqrt{2}$}.


The correlation function 
$\langle m_{\alpha}^{i}(-{\bf q})  m_{\beta}^{j}({\bf q}) \rangle$ 
needed to evaluate $S({\bf q})$~[Eq.~(\ref{eq:defSq})] can be 
calculated directly from correlations of the spins $S_{\beta}(\vec{R}_i)$
in a classical Monte Carlo simulation.
For ordered phases, it can also be calculated analytically within either 
the classical (low-T)  or semi-classical ({\sf LSW}) spin-wave approximations.
In the case of the low-T expansion, discussed below, this makes use of the 
fact that
$\langle m_{\alpha}^{i}(-{\bf q})  m_{\beta}^{j}({\bf q}) \rangle$ 
can be expressed in terms of 
\begin{eqnarray}
\langle \upsilon_{i {\bf q}} \tilde{\upsilon}_{j -{\bf q}} \rangle 
    = \delta_{ij} \frac{T}{\kappa_{i {\bf q}}}
\end{eqnarray}
(c.f. Eq. (\ref{eq:partitionfunction})).

\subsection{Details of simulations}

In Fig.~4 of the main text we show classical Monte Carlo simulation results 
for the equal-time structure factor $S({\bf q})$, for a range of parameters 
associated with the non-collinear ferromagnet.
These simulations were carried out for a cluster of $N=27 648$ spins ($L=12$), 
with $t_{max}=10^{5}$ MCs, and averaged over 10 independent samples. 
The figure is composed of 9216 pixels, each corresponding to one of the 
allowed ${\mathbf q}$-vectors in the $[hhl]$ plane for a cluster of this size.
$S({\bf q})$ was calculated following the definition Eq.~(\ref{eq:defSq}), 
using the measured g-tensor for Yb$_{2}$Ti$_{2}$O$_{7}$ 
\cite{hodges01,ross11-PRX1}, with $g_{xy}=4.18$ and $g_{z}=1.77$.
Since simulations were performed in the paramagnetic phase, at relatively 
high temperatures, neither parallel tempering nor over-relaxation were needed 
to obtain well-equilibriated results.   


In Fig.~5 of the main text we show classical Monte Carlo simulation results
for the equal-time structure factor $S({\bf q})$, for a range of parameters associated 
with the $\Psi_2$ phase.
Details of these simulations were exactly as for Fig.~4, described above.
However in this case, the structure factor was calculated using the measured g-tensor 
for Er$_{2}$Ti$_{2}$O$_{7}$ \cite{savary12-PRL109}, with $g_{xy}=5.97$ and $g_{z}=2.45$. 

\subsection{Comparison of Monte Carlo simulation and spin wave theory}

Here, to demonstrate the quality of our simulation data,
we compare the structure factors, as calculated from
the classical spin wave theory ${\mathcal H}_{\sf ex}^{\sf CSW}$ [Eq.~(\ref{eq:HlowT})] 
and Monte Carlo simulation, for three different parameter sets:
the parameters of Yb$_2$Ti$_2$O$_7$ as found in Ref. \cite{ross11-PRX1}
where the classical ground state is ferromagnetic,
the parameters of Er$_2$Ti$_2$O$_7$ as found in Ref. \cite{savary12-PRL109}
where we expect the order by disorder mechanism to favour the $\Psi_2$
states
and one set of parameters where the order by disorder mechanism favours
the $\Psi_3$ states.
We find excellent, quantitative agreement between the two methods.


\begin{figure*}
\includegraphics[width=0.8\textwidth]{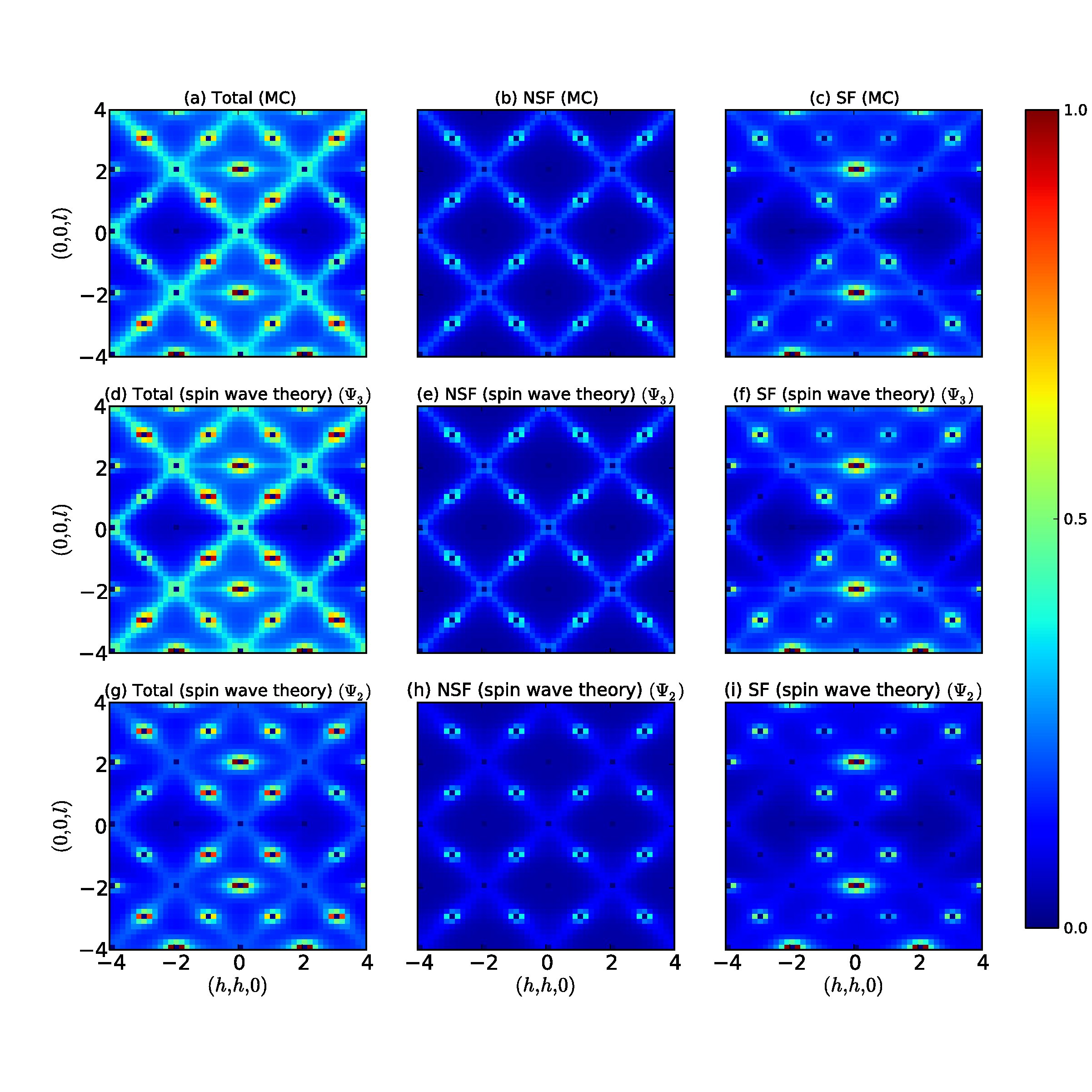}
\caption{
Comparison between results for equal-time structure factor $S({\bf q})$ obtained
in classical Monte Carlo (MC) simulation and low-temperature expansion
(classical spin wave theory) in the ordered phase for parameters 
\mbox{$J_1 = 0$},
\mbox{$J_2=-1.0\ \text{meV}$}, 
\mbox{$J_3 = -0.10\ \text{meV}$} , 
\mbox{$J_4 \equiv 0$},
approaching the non-collinear ferromagnet from within the ${\sf E}$--symmetry phase.  
a) Total scattering in the $(h, h, l)$ plane within MC simulation showing strong rod-like 
features in $[111]$ directions.
b) Associated scattering in the non spin-flip ({\sf NSF}) channel.
c) Associated scattering in the spin-flip ({\sf SF}) channel.
d) Total scattering in the $(h, h, l)$ plane within a classical spin-wave expansion 
about a $\Psi_3$ ground state, showing strong rod-like features in $[111]$ directions.
e) Associated scattering in the {\sf NSF} channel.
f) Associated scattering in the {\sf SF} channel.
g) Total scattering in the $(h, h, l)$ plane within a classical spin-wave expansion 
about a $\Psi_2$ ground state.
h) Associated scattering in the {\sf NSF} channel.
i) Associated scattering in the {\sf SF} channel.
Comparison of the scattering supports the conclusion that the $\Psi_3$
ground state is found in simulation, in agreement with the results 
of the low-T expansion [cf. Fig.~\ref{fig:entropy}].
An isotropic g-tensor $g_z=1, g_{xy}=1$ has been assumed.
For clarity, spectral weight associated with Bragg peaks at reciprocal 
lattice vectors has been subtracted.
}
\label{fig:J1=0sq}
\end{figure*}


In Fig. \ref{fig:YTOsq} we show the structure factor $S({\bf q})$ calculated both
from classical spin wave theory and from Monte Carlo simulation at $T=0.05 $K, in the
NSF, {\sf SF} and total scattering channels (see Eqs. (\ref{eq:defSq}), ({\ref{eq:sqnsf}}) and ({\ref{eq:sqsf}})).
We have used the experimentally determined parameters for the g-tensor \cite{hodges01}
$g_z=1.77$, $g_{xy}=4.18$  and exchange integrals \cite{ross11-PRX1} 
$J_1=-0.09 \text{meV}$, $J_2=-0.22 \text{meV}$ and $J_3=-0.29 \text{meV}$, 
setting $J_4 \equiv 0$.
Rod-like features are clearly visible in the total scattering along $[111]$ directions.
These are associated with a low-energy spin-wave mode which disperses 
very weakly in the $[111]$ direction [cf.~Fig.~\ref{fig:dispersionsYTO}].
The excellent, quantitative agreement between spin wave theory and simulation demonstrates
the excellent equilibration of the simulations down to $0.05$K for the parameters
of Yb$_2$Ti$_2$O$_7$, and strongly supports our understanding of the origin
of the rod-like features seen in neutron scattering 
[\onlinecite{bonville04,ross11-PRB84,ross09,thompson11,chang12}]. 


$S({\bf q})$ is also useful for studying the entropic ground state selection 
within the one-dimensional manifold of state with ${\sf E}$ symmetry.
For a given set of parameters we may compare the diffuse scattering 
calculated in spin wave theory in expansions around the $\Psi_3$ and $\Psi_2$ 
phases with the diffuse scattering calculated in simulations.


 
\begin{figure*}[h!]
\centering\includegraphics[width=18cm]{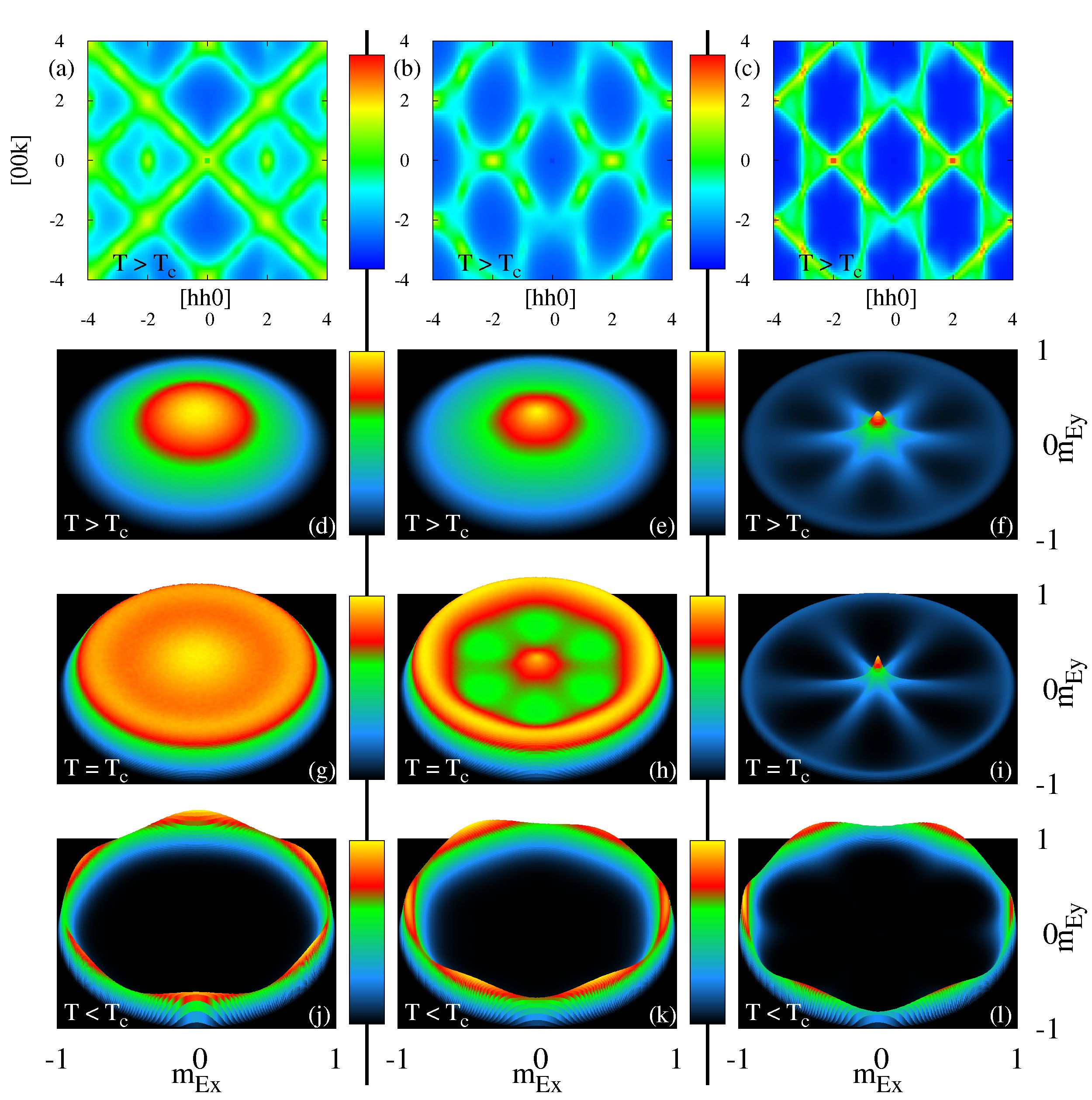}
\caption{
Influence of ground-state degeneracy on finite-temperature phase transitions, 
as revealed by the probability distribution of the order parameter 
${\bf m}_{\sf E} = m_{\sf E}\ (\cos\theta_{\sf E},\ \sin\theta_{\sf E})$~[Eq.~(\ref{eq:thetaE})].
Results are taken from simulation of $\mathcal{H}_{\sf ex}$ [Eq.~(\protect\ref{eq:Hex})],
with parameters chosen correspond to 
(1)~a $\Psi_3$ ground state, approaching the non-collinear FM (a,\ d,\ g,\ j) with $T_c=0.39$ K;
(2)~ a $\Psi_2$ ground state (b,\ e,\ h,\ k) with $T_c=0.26$ K;
(3)~a $\Psi_2$ ground state, on the border of the Palmer-Chalker phase (c,\ f,\ i,\ l) with $T_c=0.065$ K.  
(a)-(c) quasi-elastic scattering $S({\bf q})$ in the paramagnetic phase $T > T_c$.
(d)-(f) corresponding results for the probability density function, ${\sf P}({\bf m}_{\sf E})$.
(g)-(i) ${\sf P}({\bf m}_{\sf E})$ at the transition temperature $T = T_c$.
(j)-(l) ${\sf P}({\bf m}_{\sf E})$ in the ordered phase $T < T_c$.
(1)~For a finite-size system, the onset of $\Psi_3$ occurs progressively, 
through (g) the emergence of a one-dimensional manifold of states with 
finite $|{\bf m}_{\sf E}|$, and then (j) the entropic selection of $\theta_{\sf E}$ 
corresponding to one of six distinct $\Psi_3$ ground states.
(a) The connection with the non-collinear FM is evident in $S({\bf q})$, with 
rods of scattering strongly reminiscent of those seen in Yb$_2$Ti$_2$O$_7$.
(2)~The same process occurs
, but in this case ${\sf P}({\bf m}_{\sf E})$ shows $\Psi_2$ ground states are favoured 
at low temperatures (k) and even at the transition (h).
(3)~On the boundary of the Palmer-Chalker phase, the ground state manifold includes 
additional manifolds of states which mix ${\bf m}_{\sf E}$ and ${\bf m}_{{\sf T}_2}$.
These are evident (i) in the ``spoked wheel'' seen in ${\sf P}({\bf m}_{\sf E})$ at 
$T = T_c$, and drive the entropic selection of the $\Psi_2$ ground state.
(c) The high degeneracy at this phase boundary is also evident in the ``bow-tie''
structure in $S({\bf q})$.
Further details of simulations and parameters for (1), (2) and (3) are given in the text.
}
\label{fig:figure-of-doom}
\end{figure*}
 

Such a comparison is shown in Fig. \ref{fig:ETOsq}, 
for exchange parameters appropriate to Er$_2$Ti$_2$O$_7$ 
(\mbox{$J_1=0.11\ \text{meV}$},
\mbox{$J_2=-0.06\ \text{meV}$}
and \mbox{$J_3=-0.10\ \text{meV}$}, 
setting  \mbox{$J_4 \equiv 0$})  
and temperature \mbox{$T=0.36 \text{K}$}.
From the entropy calculations shown in Fig. \ref{fig:entropy} 
we expect the $\Psi_2$ state to be preferred for these
values of the exchange parameters.
Comparison of of the distribution of weight in the vicinity of
the $(1, 1, 1)$, $(3, 3, 3)$ and $(1, 1, 3)$ reciprocal 
lattice vectors between the Monte Carlo data
and the spin wave expansions around the $\Psi_2$ and $\Psi_3$ 
phases supports this conclusion.


Similarly, in Fig. \ref{fig:J1=0sq} we show a comparison of the diffuse scattering between 
Monte Carlo simulations and spin wave expansions around the $\Psi_2$ and $\Psi_3$ phases
for exchange parameters approaching the non-collinear ferromagnetic 
phase ($J_1=0$, $J_2 = -1.0\ \text{meV}$ and $J_3 = -0.1\ \text{meV}$, $J_4=0$), at $T=0.4$K.
Calculations of the entropy within spin wave theory show that the $\Psi_3$ state should be preferred by
fluctuations for these parameters, and this is confirmed by the comparison of the structure factors,
in particular by the presence of bright rods in the $[111]$ direction.

\section{Living on the edge : the influence of ground state manifolds on finite-temperature 
             phase transitions}
\label{s:MC3}

The major assertion of this article is that many of the interesting properties of pyrochlore 
magnets  --- for example the rods of scattering observed in 
Yb$_2$Ti$_2$O$_7$, and the order-by-disorder selection of a $\Psi_2$ ground state in 
Er$_2$Ti$_2$O$_7$ --- are the direct consequence of the high ground-state 
degeneracy where phases with different symmetry meet.
While the arguments for enlarged ground state manifolds at $T=0$ are easy
to understand, it is far less obvious that this degeneracy should make itself
felt at finite temperature, especially where it is not protected by symmetry.


We can test the internal consistency of these ideas by using the probability 
distribution of the order parameter 
$${\bf m}_{\sf E} = m_{\sf E}\ (\cos\theta_{\sf E},\ \sin\theta_{\sf E})$$
[cf. Eq.~(\ref{eq:thetaE})] to deconstruct the order-by-disorder selection 
of $\Psi_2$ and $\Psi_3$ ground states in finite-temperatures simulations of 
$\mathcal{H}_{\sf ex}$ [Eq.~(\protect\ref{eq:Hex})].
The probability density function $P({\bf m}_{\sf E})$ is sensitive both to the 
formation of a one-dimensional manifold of states with ${\sf E}$ symmetry
--- which manifests itself as a ring in $P({\bf m}_{\sf E})$ --- and to
the selection of an ordered ground state within this manifold --- which will
appear as six degenerate maxima within the ring.


$P({\bf m}_{\sf E})$ also enables us to study the evolution of the ground state
manifolds at the boundaries between phases with competing symmetry --- in this case 
${\bf m}_{\sf T_2}$ and ${\bf m}_{\sf T_{1,A'}}$.
At these phase boundaries, ${\bf m}_{\sf E}$ takes on a new, constrained 
set of values, characteristic of the way in which different manifolds connect.
For example Eq.~(\ref{eq:spoke1}--\ref{eq:spoke3}) predicts that, on the boundary
with the Palmer-Chalker phase, the one-dimensional manifold of states with 
$|{\bf m}_{\sf E}| = 1$ acquires ``spokes'' in the directions  
$$\theta_{\sf E} = \Big\{ 0,\ \frac{\pi}{3},\ \frac{2\pi}{3},\ \pi,\ \frac{4\pi}{3},\ \frac{5\pi}{3} \Big\}$$
connecting ${\bf m}_{\sf E} = 0$ with the six $\Psi_2$ ground states.
Observation of such a ``spinning wheel'' pattern in $P({\bf m}_{\sf E})$ at finite 
temperature would therefore confirm that the zero-temperature degeneracies
were still operative.


In Fig.~\ref{fig:figure-of-doom} we present results for $P({\bf m}_{\sf E})$ and $S({\bf q})$ 
taken from simulations of $\mathcal{H}_{\sf ex}$ [Eq.~(\protect\ref{eq:Hex})] for three sets of 
parameters
$$(1) \quad (J_1,\ J_2,\ J_3,\ J_4) =  (0, -0.3, -0.1, 0) \quad \text{meV} $$
where we expect a $\Psi_3$ ground state, but are approaching the border with the 
non-collinear FM [Fig.~\ref{fig:figure-of-doom} \mbox{(a,\ d,\ g,\ j)}];
$$(2) \quad (J_1,\ J_2,\ J_3,\ J_4) =  (0.11, 0.06, -0.1, 0) \quad \text{meV} $$
where we expect a $\Psi_2$ ground state, but are approaching the border with the 
Palmer-Chalker phase [Fig.~\ref{fig:figure-of-doom} \mbox{(b,\ e,\ h,\ k)}]; and
$$(3) \quad (J_1,\ J_2,\ J_3,\ J_4) =  (0.11, 0.11, -0.1,0) \quad \text{meV} $$
exactly on the $T=0$ border of the Palmer-Chalker phase 
[Fig.~\ref{fig:figure-of-doom} \mbox{(c,\ f,\ i,\ l)}].


The results for $S({\bf q})$ shown in Fig.~\ref{fig:figure-of-doom}(a-c), demonstrate 
the diffuse structure expected in the paramagnet in each case~: 
(1) Fig.~\ref{fig:figure-of-doom}(a) --- rods of scattering, 
reminiscent of those observed in Yb$_2$Ti$_2$O$_7$ 
[\onlinecite{bonville04,ross11-PRB84,ross09,thompson11,chang12}]; 
(2) Fig.~\ref{fig:figure-of-doom}(b) --- a diffuse web of rings, 
reminiscent to that observed in experiments on Er$_2$Ti$_2$O$_7$ 
[\onlinecite{dalmas12}], also ordering in $\Psi_2$;
(3) Fig.~\ref{fig:figure-of-doom}(c) --- ``bow-tie'' patterns reminiscent of the 
pinch points observed in the Heisenberg antiferromagnet on a pyrochlore lattice 
[\onlinecite{moessner98-PRB58,conlon10}],


The corresponding results for $P({\bf m}_{\sf E})$ in the paramagnet 
show a broad distribution of ${\bf m}_{\sf E}$, consistent with fluctuations in the absence of order, 
for both parameter sets (1) [Fig.~\ref{fig:figure-of-doom}(d)].
However on the border of the Palmer-Chalker phase (3) [Fig.~\ref{fig:figure-of-doom}(f)] 
$P({\bf m}_{\sf E})$ shows a diffuse spoked wheel, confirming that the connection 
implied by the $T=0$ ground state manifold survives at finite temperature. 
The $\mathbb{Z}_6$ symmetry actually also transpires at finite temperature for parameter sets (2) [Fig.~\ref{fig:figure-of-doom}(h)],
indicating that the $T=0$ ground state degeneracy on the boundaries can be felt even in the paramagnetic regime
away from the boundaries. The reason why we see it at $T_c=0.26$ K for parameter set (2)
and not at $T_c=0.39$ K for parameter set (1) is probably a consequence of the strong finite size dependence of
the entropic selection between $\Psi_2$ and $\Psi_3$.



\end{document}